\documentclass[11pt, preprint]{aastex}
\usepackage{natbib}
\newcommand{\flux}{ergs\,s$^{-1}$\,cm$^{-2}$}

\begin{document}
\title{Optical Spectral Properties of Swift BAT Hard X-ray Selected\\ Active Galactic Nuclei Sources}
\author{Lisa M. Winter\altaffilmark{1,*}, Karen T. Lewis\altaffilmark{2}, Michael Koss\altaffilmark{3}, Sylvain Veilleux\altaffilmark{3,4}, Brian Keeney\altaffilmark{1},\\ Richard F. Mushotzky\altaffilmark{3,5}}

\altaffiltext{*}{Hubble Fellow}
\altaffiltext{1}{Center for Astrophysics and Space Astronomy, University of Colorado, Boulder, CO}
\altaffiltext{2}{Department of Physics \& Astronomy, Dickinson College, Carlisle, PA}
\altaffiltext{3}{Department of Astronomy, University of Maryland, College Park, MD}
\altaffiltext{4}{Max-Planck-Institut f\"ur extraterrestrische Physik, Postfach 1312, D-85741 Garching, Germany}
\altaffiltext{5}{NASA Goddard Space Flight Center, Greenbelt, MD}

\begin{abstract}
The Swift Burst Alert Telescope (BAT) survey of Active Galactic Nuclei (AGN) is providing
an unprecedented view of local AGNs ($<z> \approx 0.03$) and their host galaxy properties.
In this paper, we present an analysis of the optical spectra of a sample of 64 AGNs from the 9-month survey, detected solely based on their 14-195\,keV flux.  Our analysis includes both archived spectra from the Sloan Digital Sky Survey and our own observations from the 2.1-m Kitt Peak National Observatory telescope.
Among our results, we include line ratio classifications utilizing standard emission line diagnostic plots, [\ion{O}{3}] 5007\AA~ luminosities, and H$\beta$ derived black hole masses.  As in our X-ray study, we find the type 2 sources to be less luminous (in [\ion{O}{3}] 5007\AA~and 14--195\,keV luminosities) with lower accretion rates than the type 1 sources.  We find that the optically classified LINERs, \ion{H}{2}/composite galaxies, and ambiguous sources have the lowest luminosities, while both broad line and narrow line Seyferts have similar luminosities.  From a comparison of the hard X-ray (14--195\,keV) and [\ion{O}{3}] luminosities, we find that both the observed and extinction-corrected [\ion{O}{3}] luminosities are weakly correlated with X-ray luminosity.  In a study of the host galaxy properties from both continuum fits and measurements of the stellar absorption indices, we find that the hosts of the narrow line sources have properties consistent with late type galaxies.
\end{abstract}
\keywords{X-rays: galaxies, galaxies:active}

\section{Introduction}
The Swift Burst Alert Telescope (BAT) provides an unprecedented opportunity to study
the optical properties of an unbiased sample of AGN.  Conducting an all-sky mission in
the 14--195\,keV band, the BAT survey has detected 153 AGN in the first 9-months\footnotemark{} 
\citep{2008ApJ...681..113T, 2008ATel.1429....1B}.  Since the sources were
detected based on 14--195\,keV flux, with a flux limit of 2--3 $\times 10^{-11}$\,\flux, selection effects due to obscuring material are minimal.  Due to the unbiased nature of the Swift BAT survey, Suzaku follow-ups of Swift-detected sources led to the identification of a new class of ``hidden'' AGN \citep{2007ApJ...664L..79U}, heavily obscured ($N_{\rm H} > 10^{23}$\,cm$^{-2}$) sources that would not likely be identified as AGN based on their optical or soft X-ray ($E < 3$\,keV) properties alone.  This class of ``hidden'' sources was found to comprise 24\% of the 9-month BAT AGNs \citep{2009ApJ...690.1322W}, making an analysis of the collective optical properties an important piece in understanding the properties of the Swift BAT-detected AGN.

\footnotetext{\url{http://heasarc.gsfc.nasa.gov/docs/swift/results/bs9mon/}}

Currently, great progress is being made in collecting and analyzing the multi-wavelength properties of this uniquely selected, very hard X-ray, 9-month Swift BAT AGN sample. The collective properties of the 0.3--10\,keV X-ray band have been analyzed and presented in  \citet{2009ApJ...690.1322W}.  A comparison of the IR [\ion{O}{4}], optical [\ion{O}{3}], and X-ray 2--10\,keV luminosity are presented in \citet{2008ApJ...682...94M} for a sample of 40 BAT AGNs.  Simultaneous optical-to-X-ray spectral energy distributions are analyzed for 26 of the BAT AGNs in \citet{2009arXiv0907.2272V}.  Additionally, some details of the optical host properties are presented in \citet{2009ApJ...690.1322W} as well as 
\citet{2009ApJ...692L..19S}.  Further, the results a full analysis of the optical colors and morphologies are being compiled in Koss {\it et al.} (in prep) and the Spitzer-based IR properties will be presented in Weaver {\it et al.} (submitted).
In this paper, we present an analysis of the optical spectral properties of a sub-sample of the AGN from the BAT 9-month catalog.

Since the BAT-detected sources are bright ($m_V < 16$) and nearby ($<z> = 0.03$), they are easily observable with ground-based facilities.  Between published optical spectral analyses, the publicly available Sloan Digital Sky Survey (SDSS) spectra, and our own follow-up observations with the Kitt Peak National Observatory's (KPNO) 2.1-m telescope of sources for which optical spectra/analyses were not available, we present the optical emission line properties of 64/153 of the SWIFT BAT AGNs.   This sample includes 35 broad line (55\%) and 29 narrow line (45\%) sources, the same ratio as in the total sample.  All selected sources were chosen based on positions viewable from the Kitt Peak Observatory.  In this way, our sample represents 81\% of the non-blazar ``northern'' BAT AGN sources.  As in our X-ray study \citep{2009ApJ...690.1322W}, we exclude the beamed sources due to the different physical mechanisms producing their spectra (i.e. jets).  The missing ``northern'' sources were missed purely due to observation scheduling and poor weather conditions.  In the following sections, we describe the observations, data analysis, and finally our results.

\section{Observations and Data Reduction}
For our analysis of the optical spectra of the Swift BAT-detected AGN, we first obtained spectra of our sources that were publicly available from the Sloan Digital Sky Survey (SDSS).  We supplemented this data set with our own observations at the Kitt Peak Observatory.  Additionally, we included several of the SDSS observed sources as Kitt Peak targets, in order to compare the results of our analysis from each observatory.

Our Kitt Peak observations were obtained on the 2.1-m telescope as part of MD-TAC time.  Over the course of 5 observing trips, from August 2006 -- April 2009, we used the GoldCam spectrograph to observe the central region of $\approx 50$ objects, including AGN and template galaxies.  The AGN observed were sources for which we could not find archived optical spectra or analyses of optical lines in the literature.  The template galaxies (10) were chosen from non-active templates listed in \citet{1997ApJS..112..315H}.  A majority of the sources
were observed with two 30-minute exposures in both the red (grating 35, which covers 4760--7240\AA) and blue (grating 26new, which covers 3660--6140\AA), through a 2\arcsec\,slit.  Both of these gratings have a spectral resolution of 3.3\AA, corresponding to a velocity width of 200 km\,s$^{-1}$ at 5007\AA.  The exposure times were chosen in order to achieve a S/N of $\approx$ 70 per pixel for the AGN sources and the dispersion relation for both gratings corresponds to $\approx 1.25$\,\AA~pixel$^{-1}$.  However, for some of the faintest sources we used a lower dispersion grating, grating 32, which covered a larger wavelength range than the higher dispersion gratings (4280--9220\AA, at  2.25\,\AA~pixel$^{-1}$ and which has a spectral resolution of 6.7\AA).  We used this grating because of the unknown redshift of many of these sources.

Initial processing of the data proceeded using the standard tasks in IRAF to extract the spectra and remove cosmic rays.  The spectra were dispersion corrected using comparison observations of the HeNeAr lamp taken at each telescope position.  They were then flux calibrated using standard stars, from the spectrophotometric standards compiled by \citet{1988ApJ...328..315M}, observed on the same night as the template/AGN.  We then added the medium resolution red and blue spectra together to obtain a single medium resolution spectrum for each source.

In addition to the Kitt Peak observations, we include spectra from the SDSS data release 7 \citep{2009ApJS..182..543A}.  Such spectra were publicly available for 24 of the non-blazar BAT AGN sources.  A list of the BAT AGN 9-month sources for which we analyzed Kitt Peak/Sloan spectra are listed in Table~\ref{tbl-1}.  The KPNO observations (with typical total exposure times of 1\,hr in each medium resolution grating) were planned such that we would obtain similar signal-to-noise spectra as the SDSS spectra (S/N $\approx 75$), to provide an easy comparison between both sets of spectra.  In total, our sample consists of 64 sources, including 40 spectra from our KPNO observations, 24 with SDSS spectra (4 sources having both a KPNO and SDSS spectrum), and 13 with emission line properties listed in the literature (9 of which also have either a KPNO or SDSS spectrum).  Details of the KPNO observations, including the extraction aperture along the slit, are listed in Table~\ref{tbl-kpno} for the target AGN sources and in Table~\ref{tbl-kpnotemplates} for the galaxy template sources.  Details of the SDSS observations are listed in Table~\ref{tbl-sdss}.  Based on visual inspections of the AGN spectra, we indicate in the tables which sources display broad lines with a `B'.  In total, 33 sources (including 3 with emission line properties available in the literature), 55\% of the sample, exhibit clear broad lines (i.e. broad H$\alpha$ and H$\beta$).

\section{Data Analysis}
Analysis of the spectra consisted of three steps: de-reddening the spectrum to correct for reddening from the Milky Way, continuum subtraction, and fitting the emission lines.  The spectra were de-reddenned using the IDL procedure {\tt CCM\_UNRED} from the Goddard IDL Astronomy User's Library.  This procedure uses the reddening curve of \citet{1989ApJ...345..245C}, with R$_V = 3.1$, and the input value of E$_{B-V}$.  The Milky Way E$_{B-V}$ values (listed in Table~\ref{tbl-1}) were obtained for the Kitt Peak and SDSS observed sources from the NASA Extragalactic Database (NED).  Following this step, the spectra were de-redshifted to their restframe wavelengths, using the NED redshifts or measured redshift from the [OIII] 5007 \AA\,line.  Following the continuum fits (\S~\ref{continuum}), we measured the emission (\S~\ref{emission}) and absorption (\S~\ref{gal-absorption}) line parameters for prominent spectral features.  We then tested for aperture effects by comparing emission line and stellar absorption line measurements with redshift, finding no correlations.

\subsection{Continuum Modeling}
\label{continuum}
In order to fit the emission lines as correctly as possible, great care must be taken in modeling the continuum.  
For an AGN source, we expect the continuum to be a combination of non-thermal emission from the AGN 
and stellar light from the host galaxy.  To model the contribution from stellar light, we used the population synthesis models from the GALAXEV package\footnotemark{} \citep{2003MNRAS.344.1000B} in the 3200\AA--9300\AA~ range.  The spectral resolution of these models ($\approx 3$\AA) is directly comparable to that in the SDSS and KPNO samples.  We assume that the galaxy light is the sum of bursts of formation at different ages, using stellar populations at 3 different ages (25, 2500, and 10000 Myr) to determine whether the host is consistent with a young, intermediate, or old population (or any combination of these three).  While a three component stellar model (young, intermediate, old) does not fully describe the spectra of all galaxies, we have found (see the appendix) that adding more components results in degenerate solutions with different sums of the 10 spectral models in the Bruzual \& Charlot instantaneous burst models.  We thus use the 3 component models adopted and recognize that this may not be a fully accurate description of the stellar components of the host galaxies.  Additionally, we used 3 metallicity levels: 0.05\,Z ($2.5 Z_{\sun}$), 0.02\,Z ($Z_{\sun}$),  and 0.004\,Z ($\frac{1}{5} Z_{\sun}$).  We use the same code described in  \citet{2004ApJ...613..898T}, which was used to measure the continuum in a sample of 53,000 SDSS galaxies.  As described in \citet{2004ApJ...613..898T}, the best fit is obtained using a nonnegative least squares fit using the same metallicity for all 3 of the `age' groups, attenuated by dust (which is modeled as a free parameter).  The $\chi^2$ values, using different metallicity populations, are compared to find the best fit metallicity range.  We note, however, that these models depend upon necessary assumptions, such as stellar populations created in an instantaneous burst of star formation (see \citet{2009ApJ...699..486C} for a discussion of many of the associated uncertainties in single stellar population models), which are not physical.

\footnotetext{\url{http://www2.iap.fr/users/charlot/bc2003/ }}

\begin{figure}
\includegraphics[height=6cm]{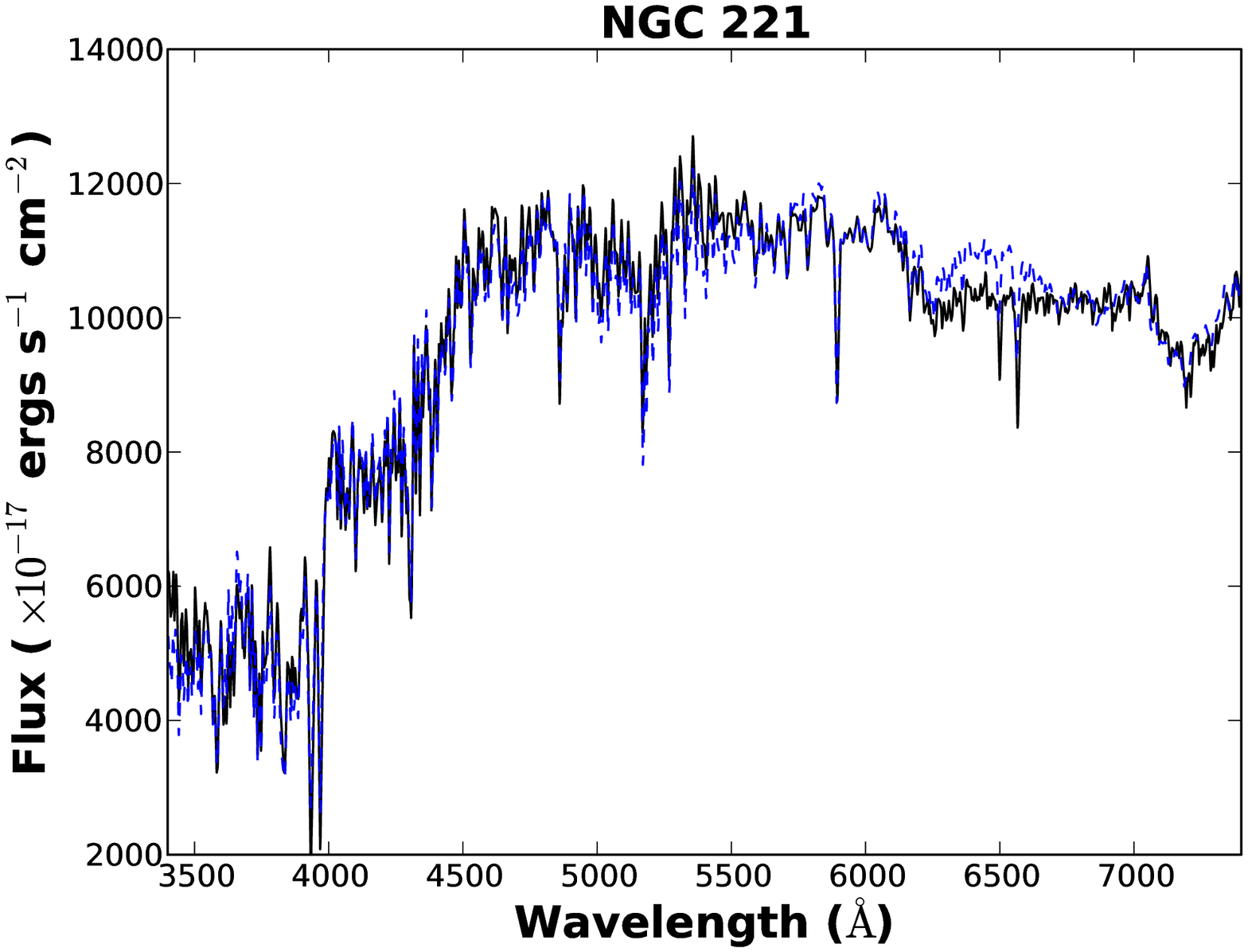}
\includegraphics[height=6cm]{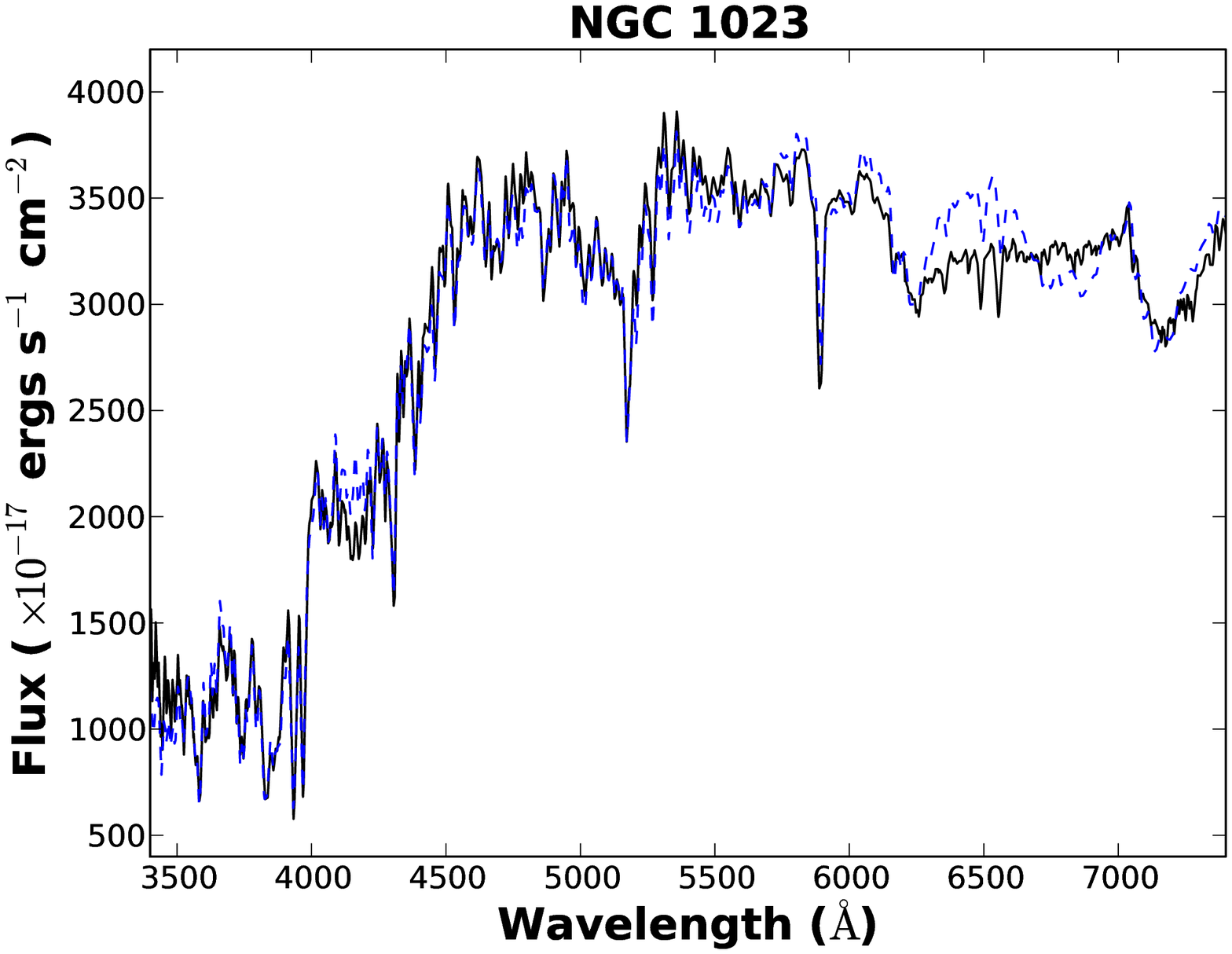}\\

\includegraphics[height=6cm]{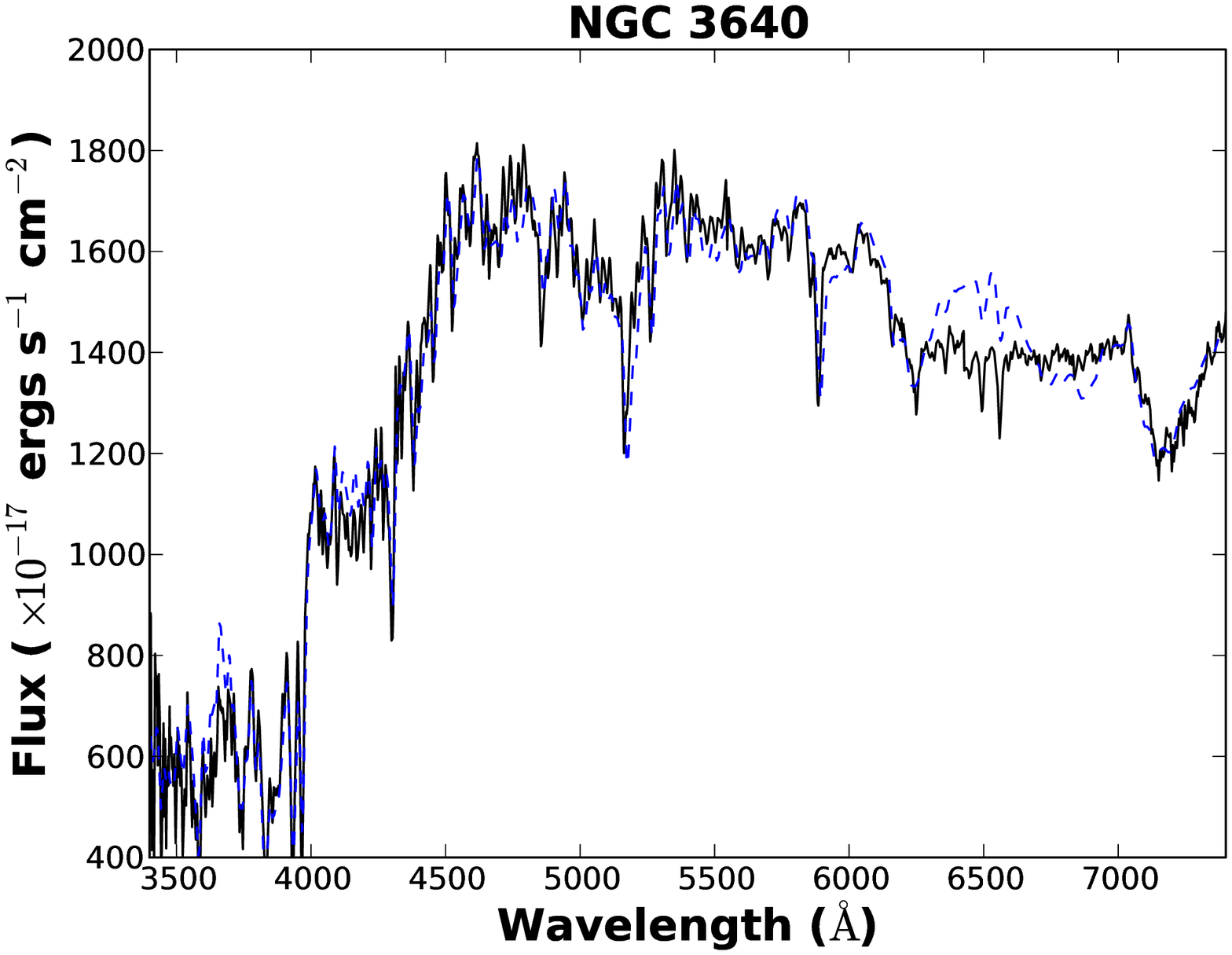}
\includegraphics[height=6cm]{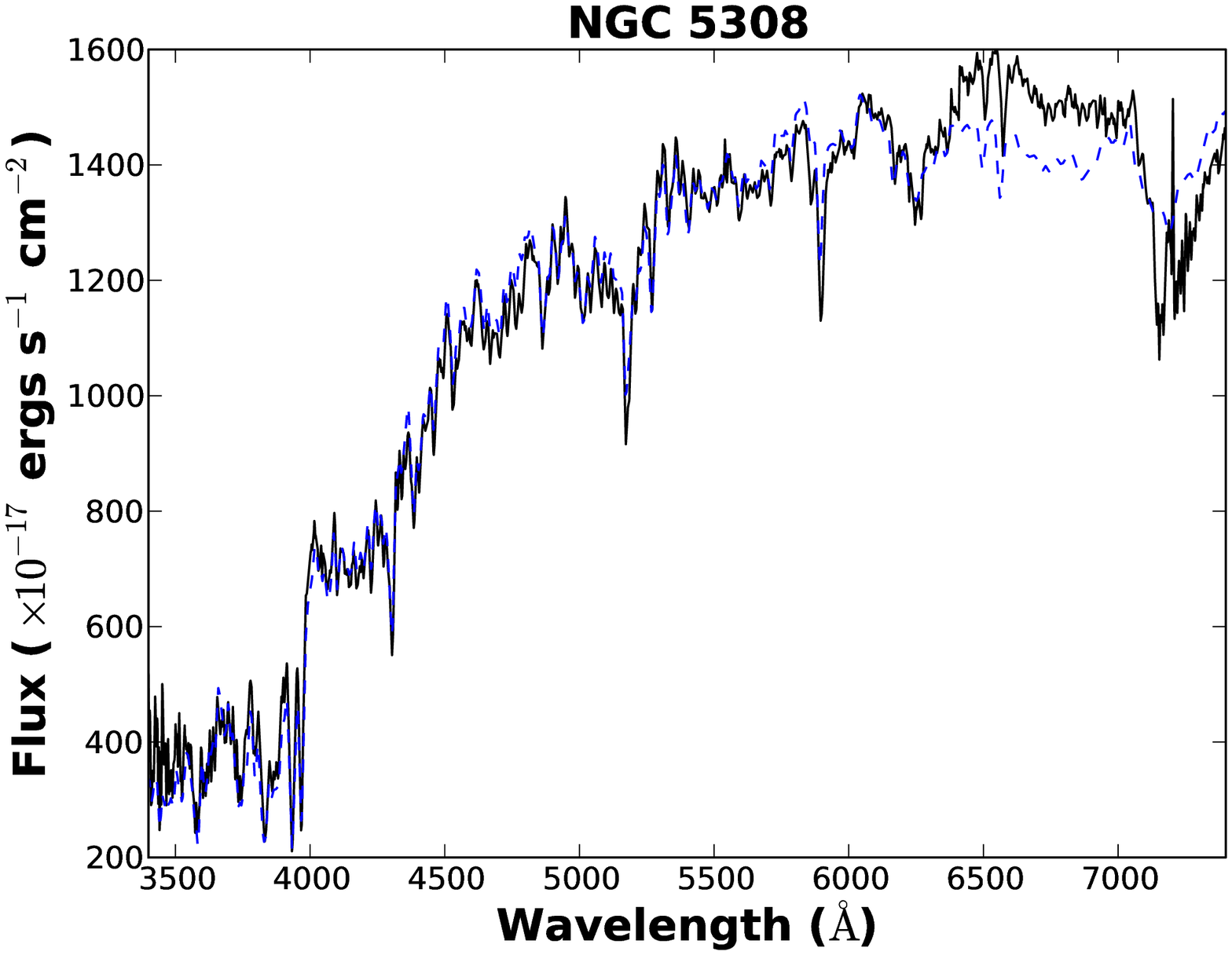}

\caption{Plotted are the KPNO spectra of 4 template galaxies (black) with the best-fit continuum model (blue) described in Table~\ref{tbl-templates}.  Using three simple stellar population models (young, intermediate, old), we find that we can replicate the spectra well, particularly in the blue end of the spectrum.  Using additional populations at intervening ages, we could better replicate the spectra.  However, such fits are degenerate 
\citep{2004ApJ...613..898T} and we would lose information about the host galaxy, when fitting to our AGN sources (for which the host properties are not well-defined).
\label{fig-templates}}
\end{figure}

To test the effectiveness of the galaxy continuum fits, we first applied the models to our set of template galaxies obtained at KPNO.  The final input to the \citet{2004ApJ...613..898T} code is the galaxy's velocity dispersion, a quantity that is unknown for many of our AGN host galaxies.  Therefore, we fit each of the templates with a range of dispersion values to obtain the best-fit.  These values were then compared to the known galaxy parameters, listed in LEDA\footnote{\url{http://leda.univ-lyon1.fr}} \citep{2003AA...412...45P}.  The galaxy type and velocity dispersion, as well as the fitted values, are listed in Table~\ref{tbl-templates}.  On average, we find that the fitted dispersion velocities (for a Gaussian, $FWHM = 2.35 \times \sigma$) are in agreement with the central velocity dispersions listed in LEDA ($<\sigma_M> = 132$\,km\,s$^{-1}$, while the LEDA values give $<v_{disp}> = 159$\,km\,s$^{-1}$).  From a comparison of the galaxy type to the light fraction (at 5500\AA) from the young, intermediate, and old stellar populations, there are no obvious contradictions.  Our sample includes late spirals through ellipticals and we find that the models suggest the light is dominated by intermediate to old stellar populations in most of the galaxies (this is consistent with the color analysis of the images found by Koss {\it et al.} (in preparation)).  In Figure~\ref{fig-templates}, we plot examples of the results of the stellar continuum fitting.  We find that the models are particularly accurate at fitting the blue end (below 5000\AA) of the spectra.  While the addition of more stellar populations (at different ages) would provide better fits to the spectra, \citet{2004ApJ...613..898T} point out that the fits are often degenerate (they use 10 different population ages).  Therefore, in an effort to get a broad understanding of the stellar properties of the AGN host galaxy properties, we confine our fits to the young, intermediate, and old populations indicated above.

Additionally, we created a grid of test spectra using different combinations of the three stellar populations indicated.  Random noise was added to the test spectra, which were then broadened with FWHM $ = 300$\,km\,s$^{-1}$ and an instrumental resolution of 75\,km\,s$^{-1}$, and reddened using the Charlot \& Fall law
\citep{2000ApJ...539..718C}.  The results of continuum fits to these test spectra are presented in \S~\ref{testspectra}.  As shown, we find that the velocity dispersion is well-determined for our test spectra while the metallicities are not.  We can clearly distinguish young stellar populations from the intermediate/old populations, however, there is a degeneracy between the intermediate and old populations when they are combined with the young populations.  These degeneracies are taken into account in the following discussions.  We also created a grid of test spectra including a power law contribution similar to that of our sample along with the stellar populations, from which we found no degeneracy between the power law and stellar components (see appendix).  

In order to subtract a continuum from the KPNO and the SDSS spectra, we modified the galaxy modeling code to include a non-thermal power law contribution from the AGN ($p_0 \times \lambda^{p_1}$, where $p_0$ is constrained to range from 0 to 1 and $p_1 > 0$).  In our model, separate reddening values were fitted for both the power law component and the stellar component.  Additionally, in our fits we masked out regions near prominent emission line positions (i.e. H$\beta$, [\ion{O}{3}] $\lambda 5007$\AA) at a standard width of 500\,km\,s$^{-1}$ and used a larger width of 7000\,km\,s$^{-1}$ around H$\alpha$.  For the broad line sources (identified as such by visual inspection of the optical spectra), we masked a larger region with a width of 10000\,km\,s$^{-1}$ around prominent hydrogen and helium emission features (H$\gamma$, H$\delta$, H$\beta$, H$\alpha$, \ion{He}{1}, and \ion{He}{2}).  The results of these fits are presented in Table~\ref{tbl-continuum}.  Average values of $p_1$ for our sources were 0.67, very similar to the power law slopes found for luminous quasars by \citet{2006ApJS..166..470R}, with a range of fitted values from 0 to 2.89.  The average value for $p_0$ is 0.47, with values ranging from 0 to 1.  As listed in the Table~\ref{tbl-continuum}, $p_0$ was calculated for the specific flux at 1\AA~and has units of $10^{-17}$\,\flux\,\AA$^{-1}$.

As we show in the appendix, we found no statistical degeneracies between the power law component and stellar continua based upon our simulations.  However, the issue of separating stellar and non-thermal AGN continua is complex.  In order to assess the degree of degeneracy in our models, we carefully analyzed the results of our model fits.  From our models, 37\% of the narrow line sources (sources in this category tend to be classified as Sy 1.8/1.9 sources by other authors) and 38\% of the broad line sources have contributions of 50\% or greater from a power law.  We examined the spectra of these sources in the region from 3800--4200\AA, which includes the important stellar diagnostic lines of Ca H and K as well as the H$\delta$ absorption.  For broad line sources with high power law contributions, we find that absorption lines tend to be weak, while [\ion{Ne}{3}] (at 3869 and 3968\AA) and occasionally weaker hydrogen Balmer (H$\zeta$, H$\epsilon$, H$\delta$) emission lines  are comparatively strong. For the narrow line sources, sources with strong power law contributions tend to have weak to no clearly evident absorption features.  Nearly half of these narrow line spectra have either poor fits to the data ($\chi^2 >> 1$) or no spectral coverage at the blue wavelengths which include important stellar lines like Ca H and K (making the fits less reliable).  Therefore, the effects of any degeneracies between power law and stellar population models are likely small for our purposes (i.e. rough estimates of the continuum).

\begin{figure}
\includegraphics[height=6cm]{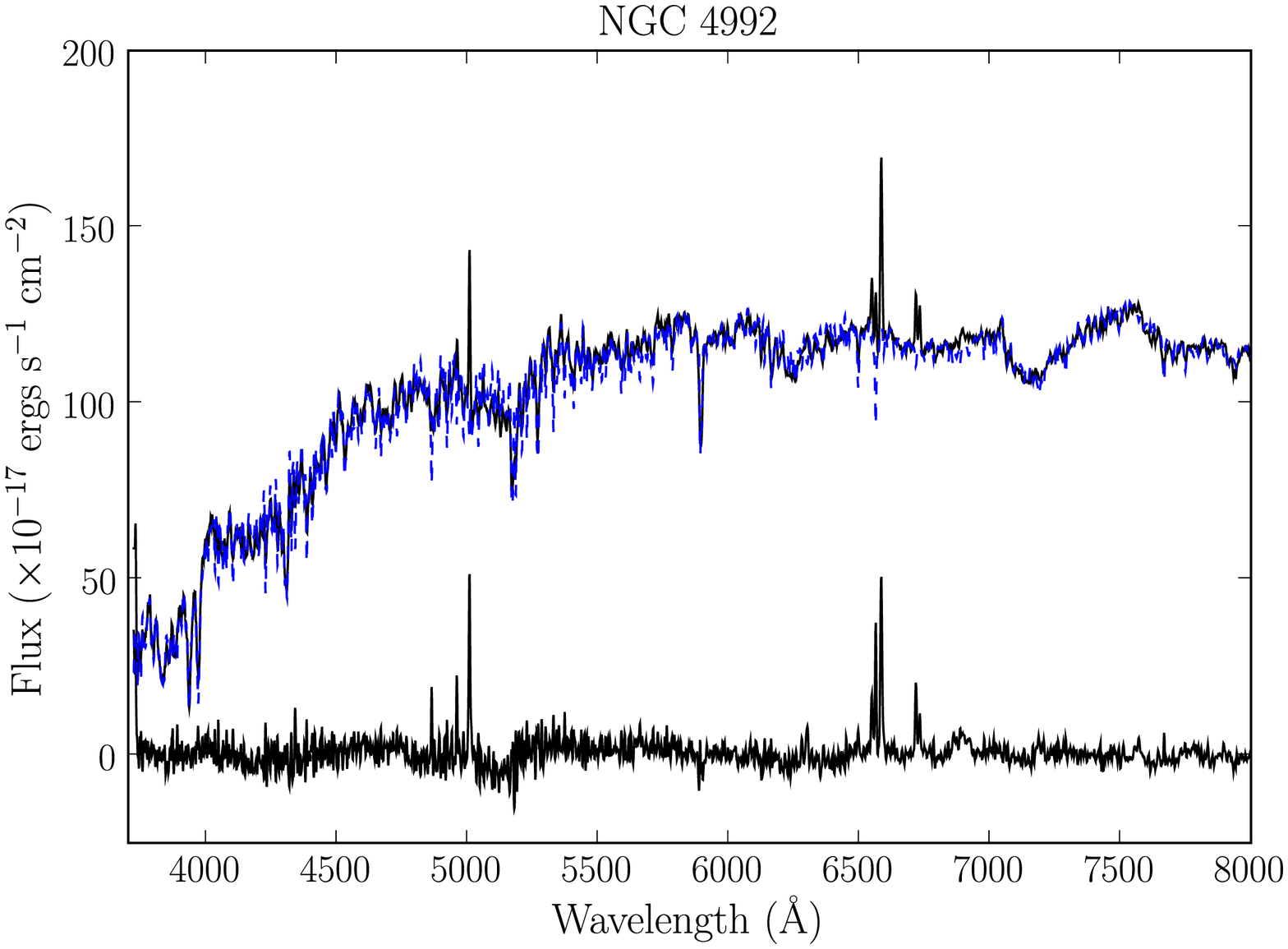}
\includegraphics[height=6cm]{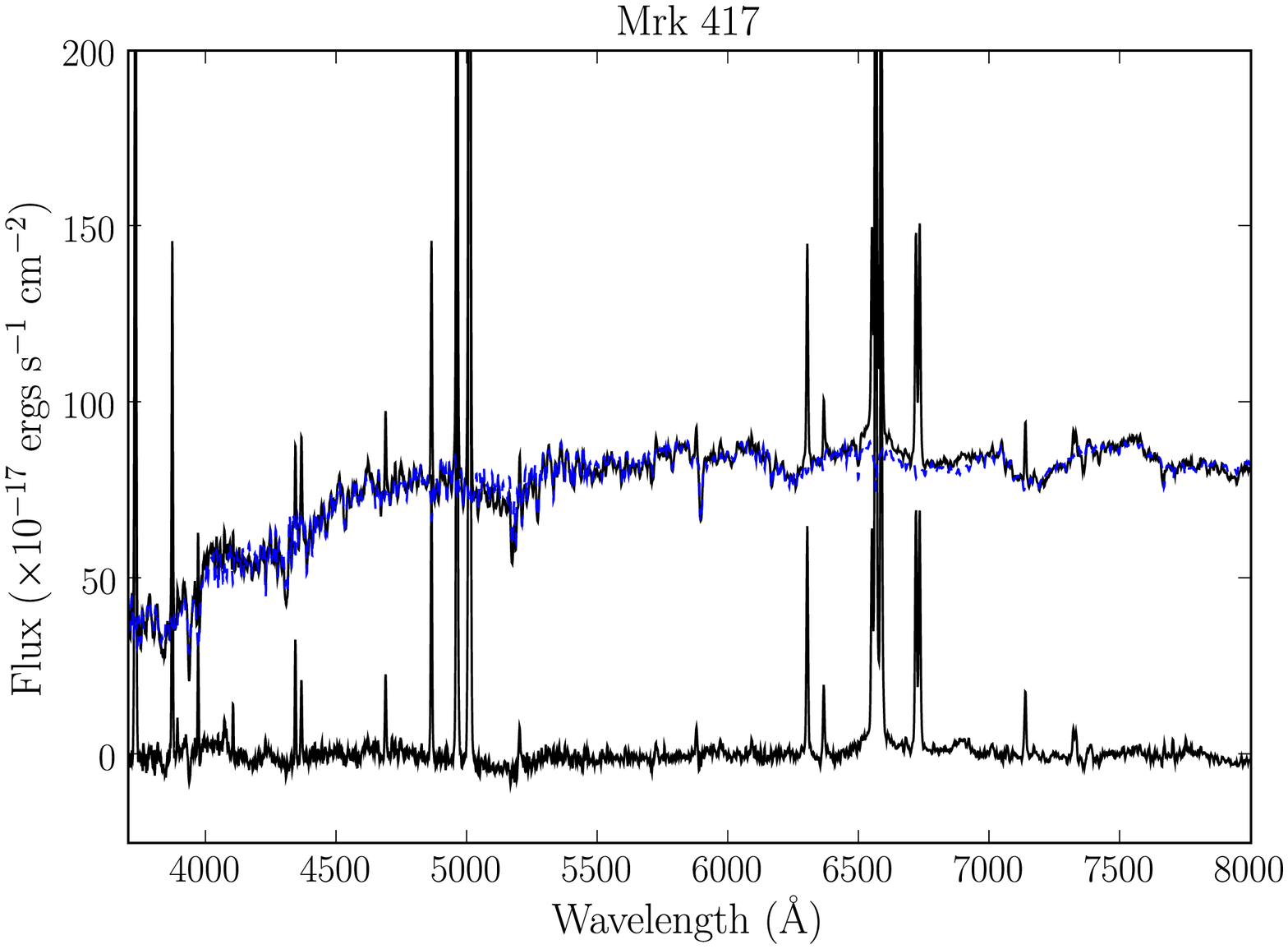}
\includegraphics[height=6cm]{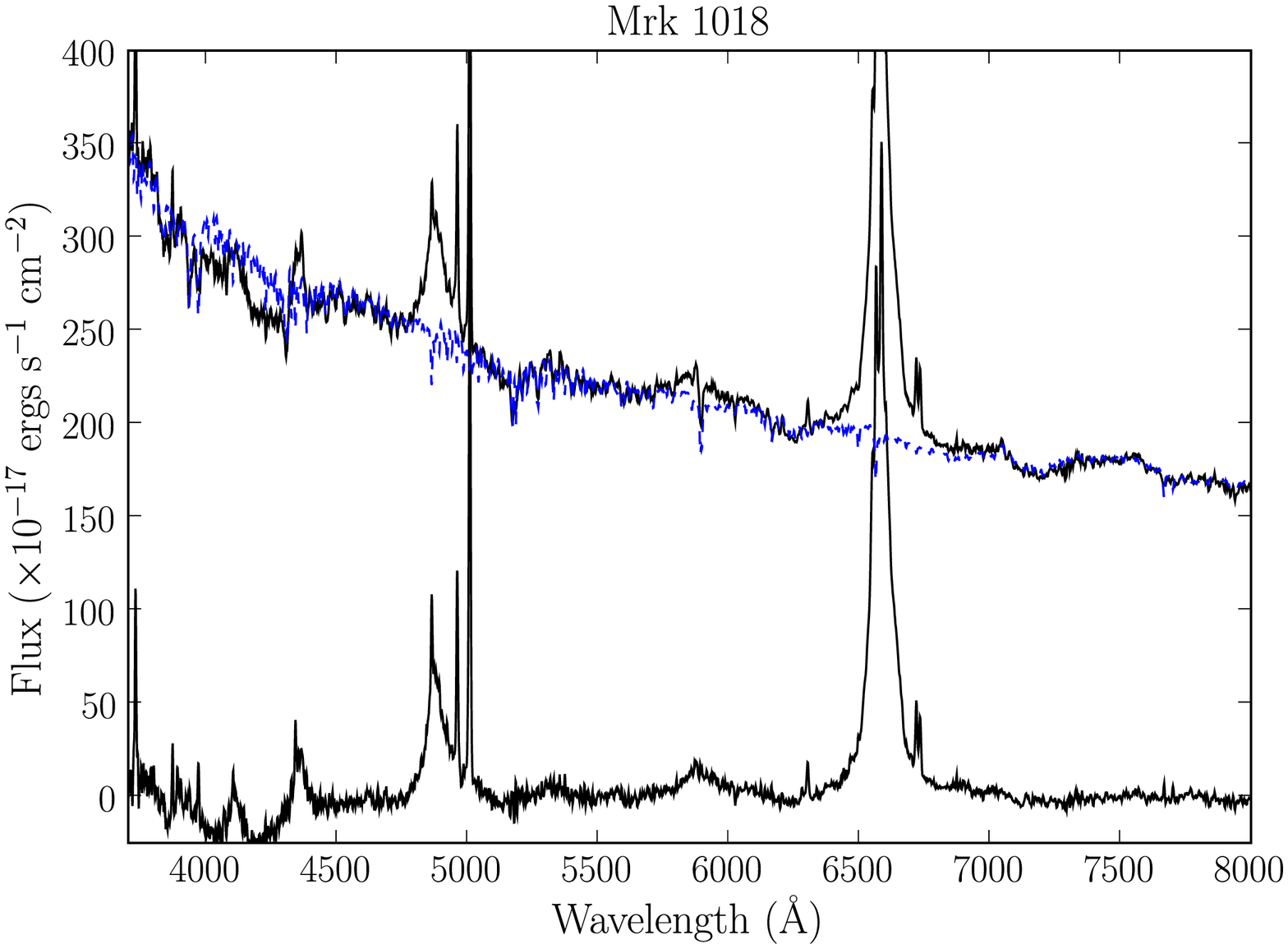}
\includegraphics[height=6cm]{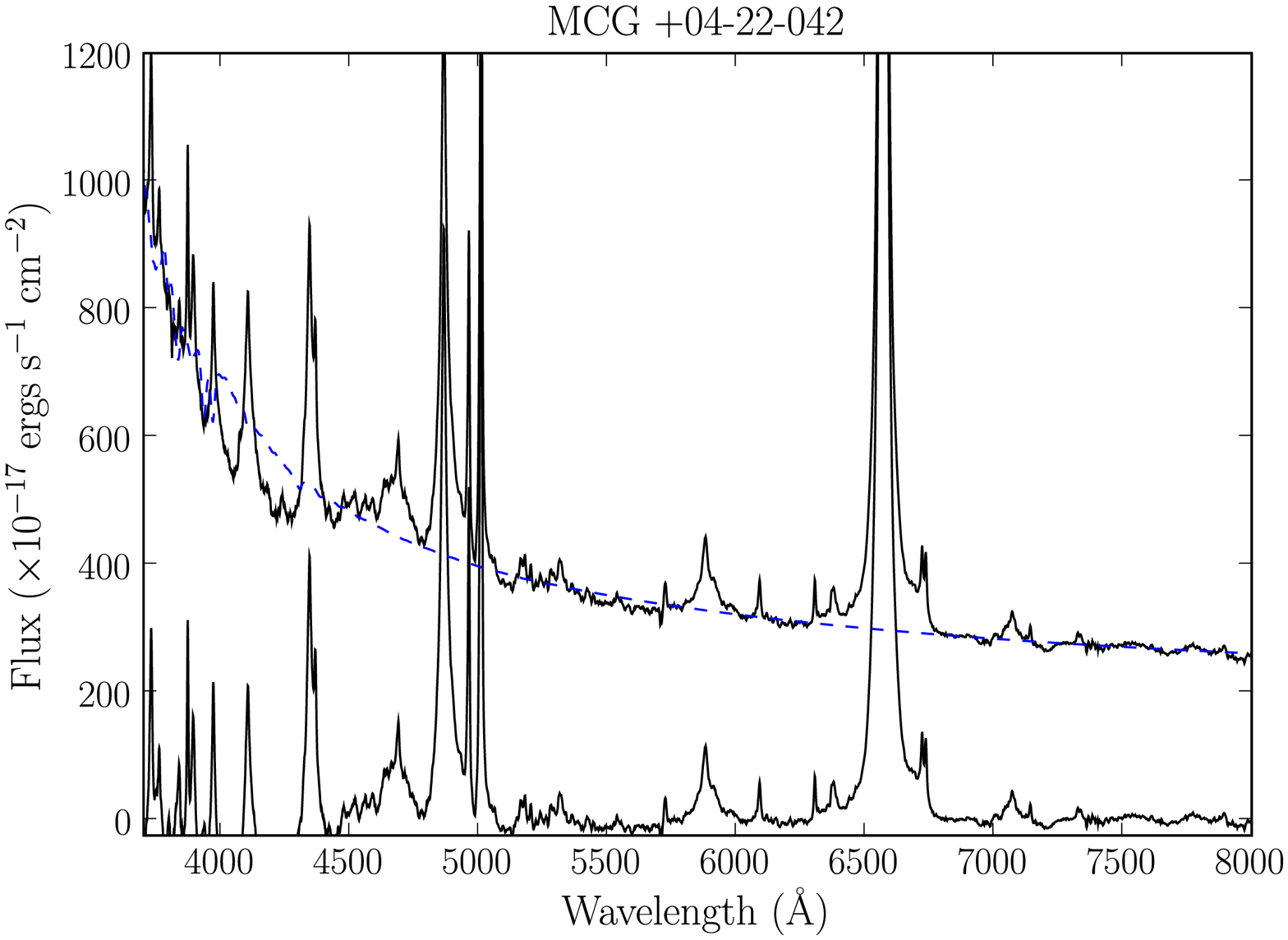}

\caption{Plotted are the SDSS spectra of 4 AGN, two narrow line sources (top) and two broad line sources (bottom), before and after the continnum subtraction (black).  The best-fit continuum model is plotted in blue (described in Table~\ref{tbl-continuum}).  The continuum model utilizes three simple stellar population models (young, intermediate, old) along with a power law model to account for AGN emission.
\label{fig-sloansubtr}}
\end{figure}
\begin{figure}
\includegraphics[height=6cm]{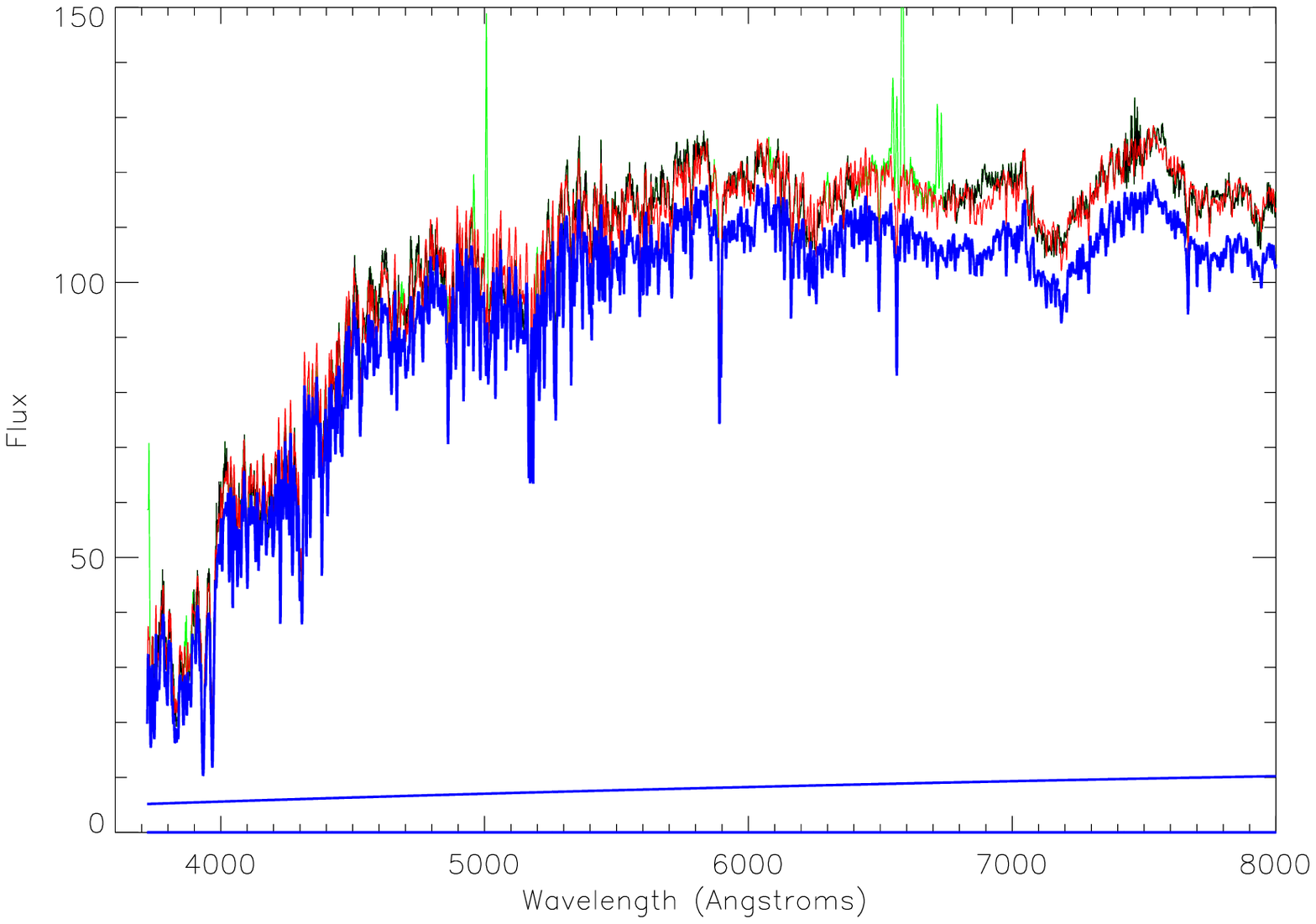}
\includegraphics[height=6cm]{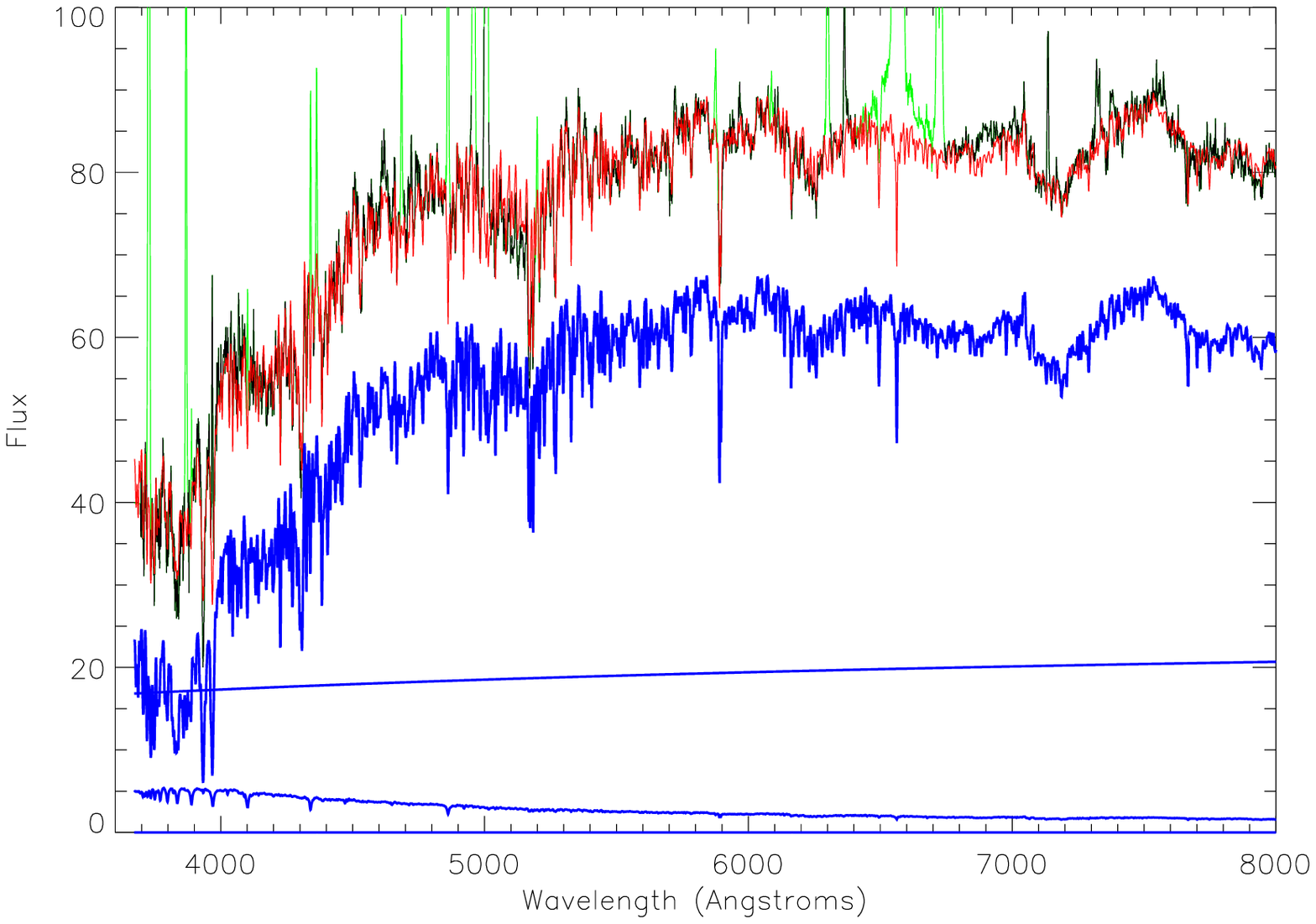}
\includegraphics[height=6cm]{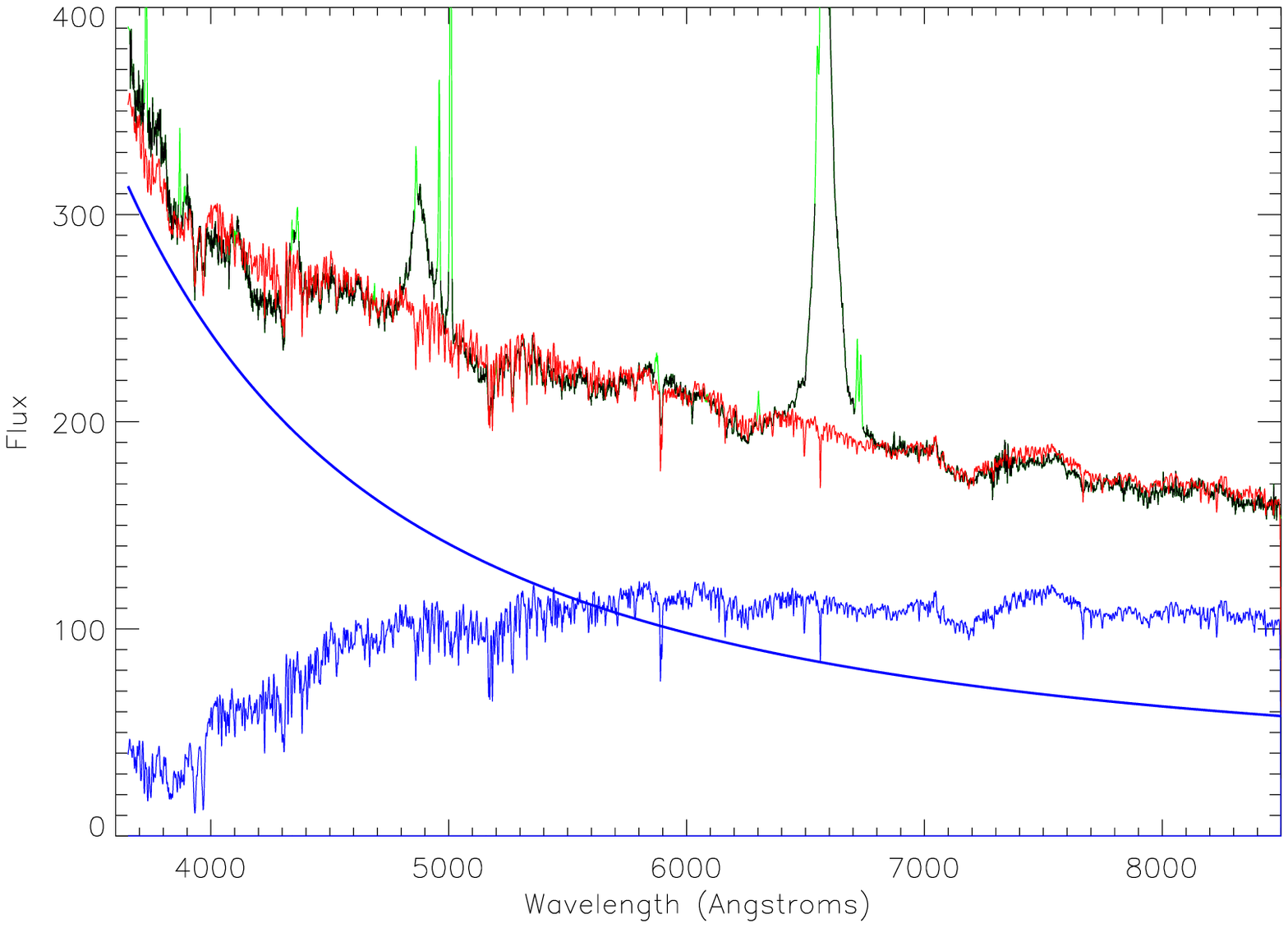}
\includegraphics[height=6cm]{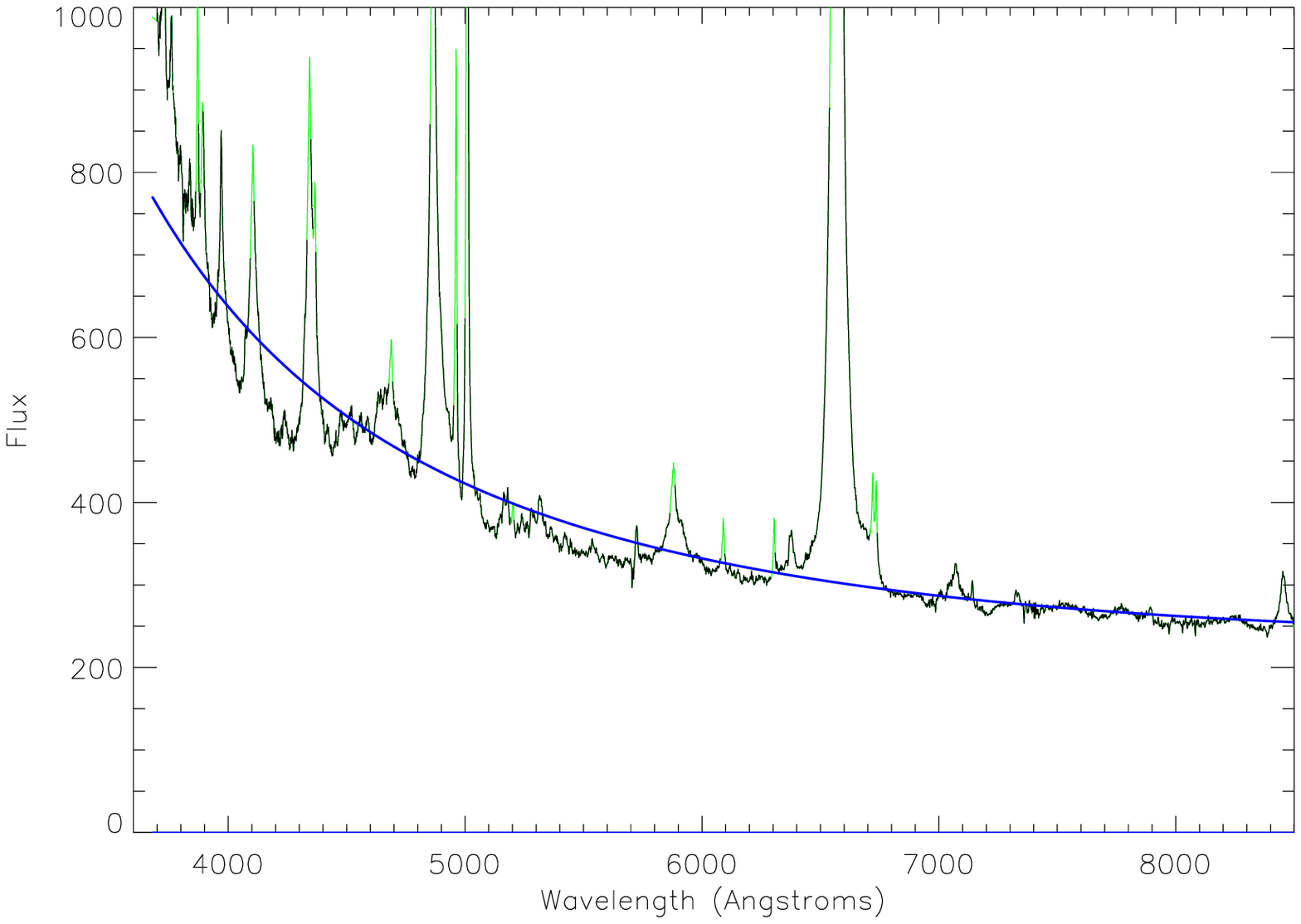}

\caption{Plotted are the best-fit individual components (power law and stellar components, modulated by reddening) fitted to the spectra shown in Figure~\ref{fig-sloansubtr}.  The flux is shown in units of $10^{-17}$\,\flux\,\AA$^{-1}$.  The sources shown include NGC 4992 (top left), Mrk 417 (top right), Mkn 1018 (bottom left), and MCG +04--22--042 (bottom right).  The combined fit is shown in red, while individual stellar components and the power law are each shown in blue.  Masked regions are shown in green.  The first three sources have strong galaxy contributions, each dominated by a contribution from an old population.  The final source is best-fit with a pure reddened power law model.
\label{fig-sloandecomp}}
\end{figure}

In Figures~\ref{fig-sloansubtr} and \ref{fig-sloandecomp}, we show examples of the continuum results.  Both the original and continuum subtracted spectra are plotted in black with the continuum plotted in blue.  For the majority of sources, we find acceptable fits with the stellar + power law continuum models.  Particularly, good fits are obtained for the narrow line sources.  
For the broad line sources, the presence of broad Balmer lines makes it particularly hard to obtain a good fit to the spectrum below $\approx 4500$\AA~(see for example the spectrum of MCG +04-22-042).  

\begin{figure}
\hspace{-1.2cm}
\includegraphics[height=6cm]{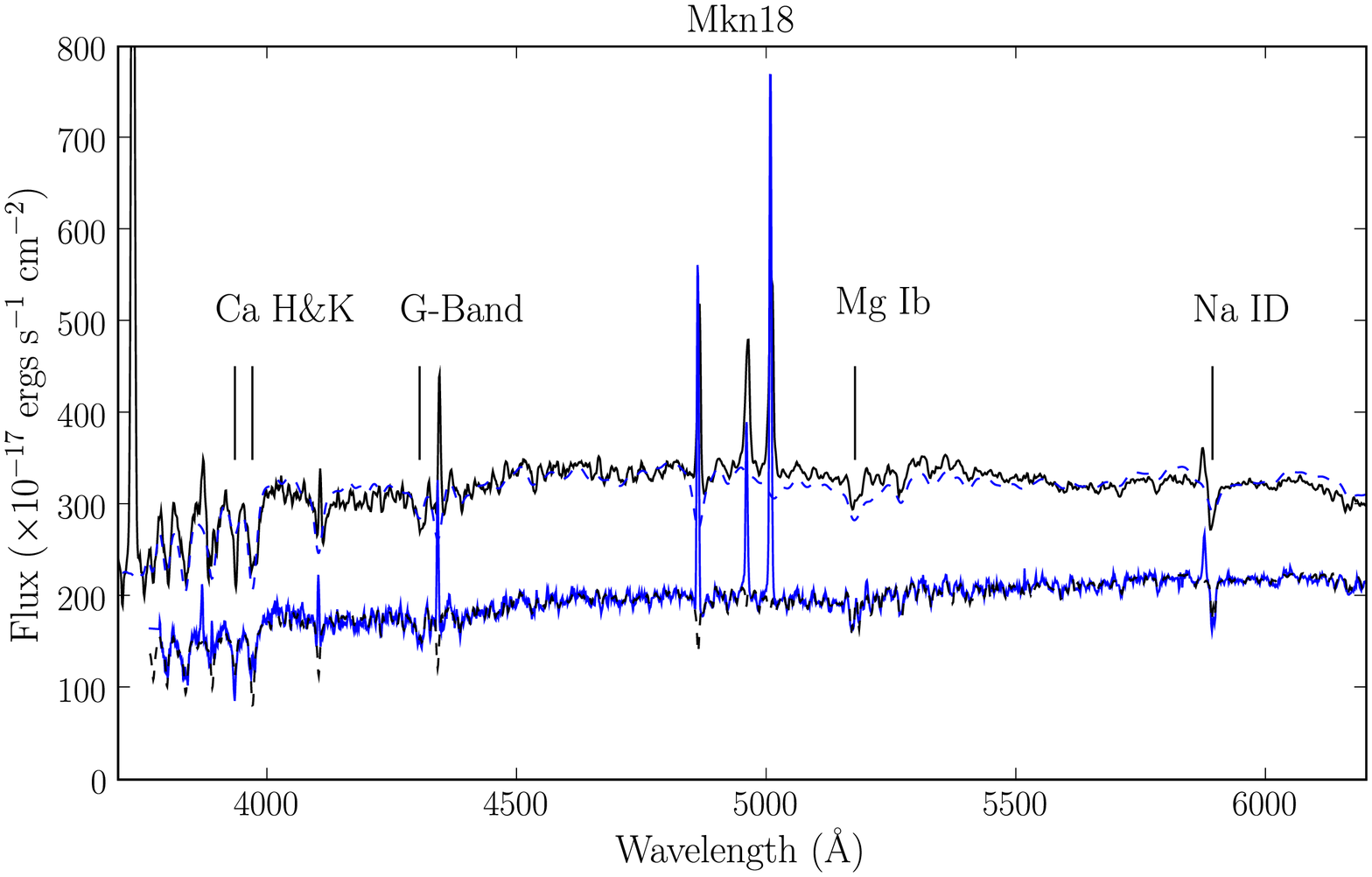}
\hspace{-1.2cm}
\includegraphics[height=6cm]{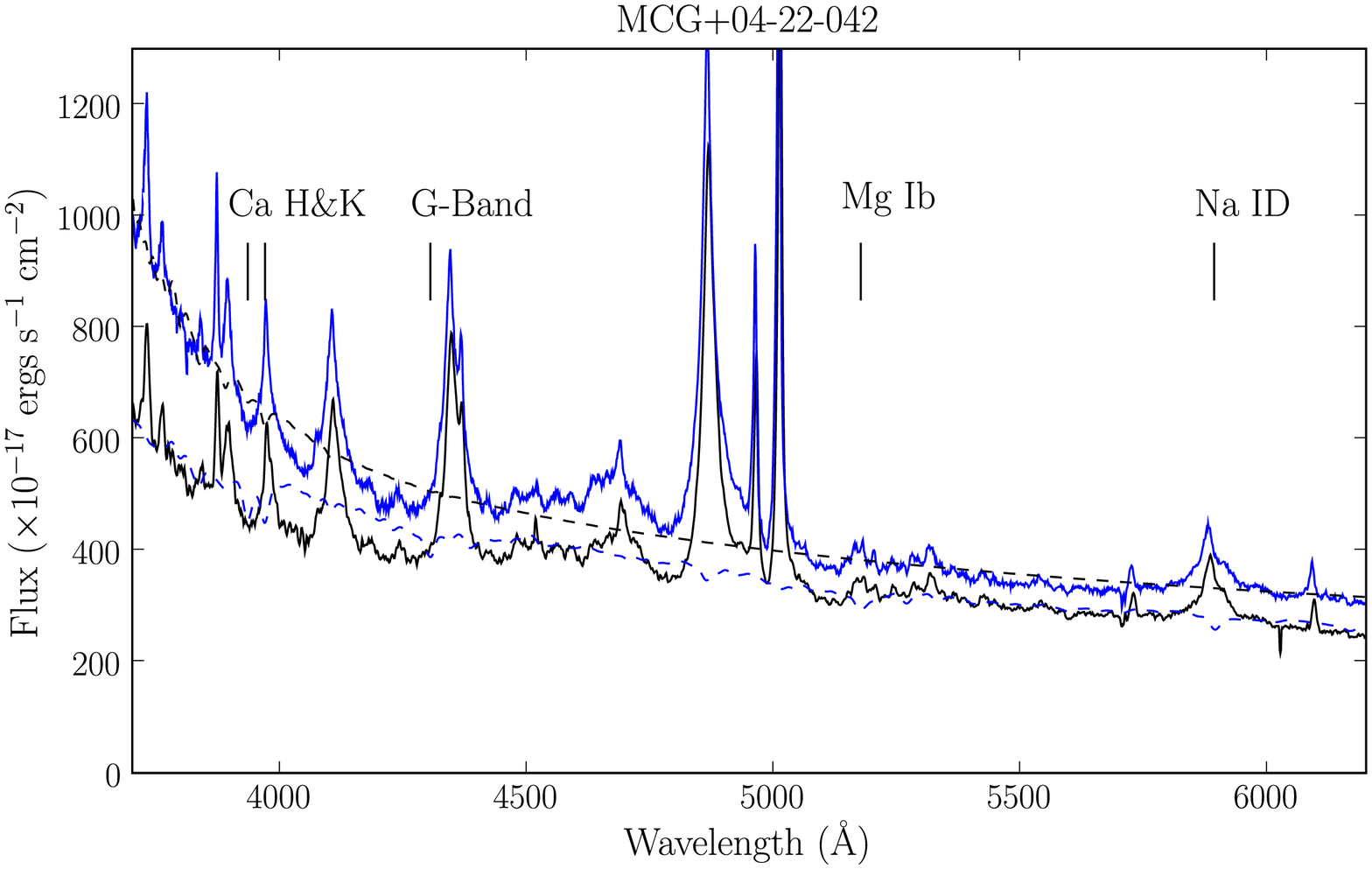}\\

\hspace{-1.2cm}
\includegraphics[height=6cm]{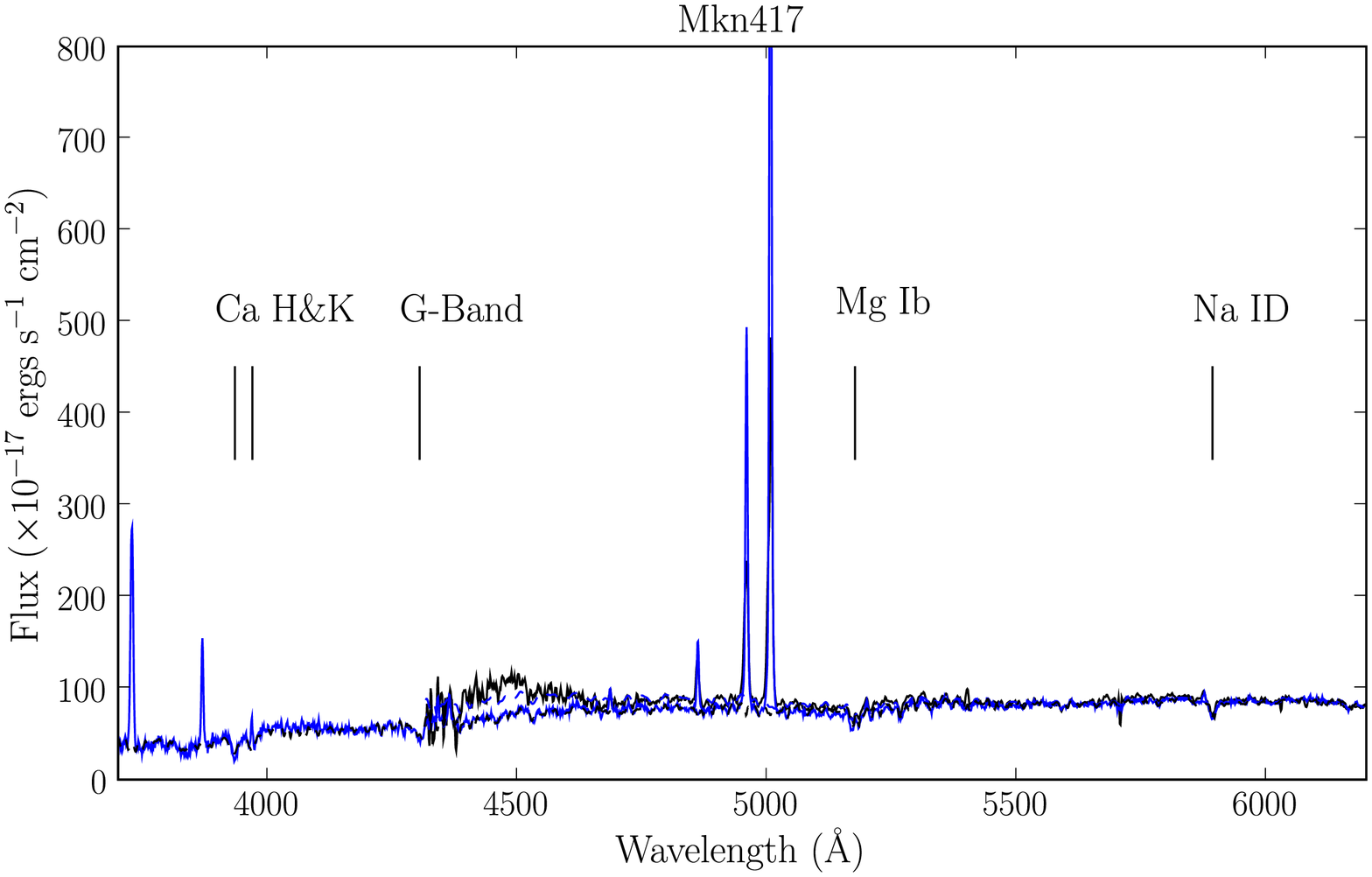}
\hspace{-1.2cm}
\includegraphics[height=6cm]{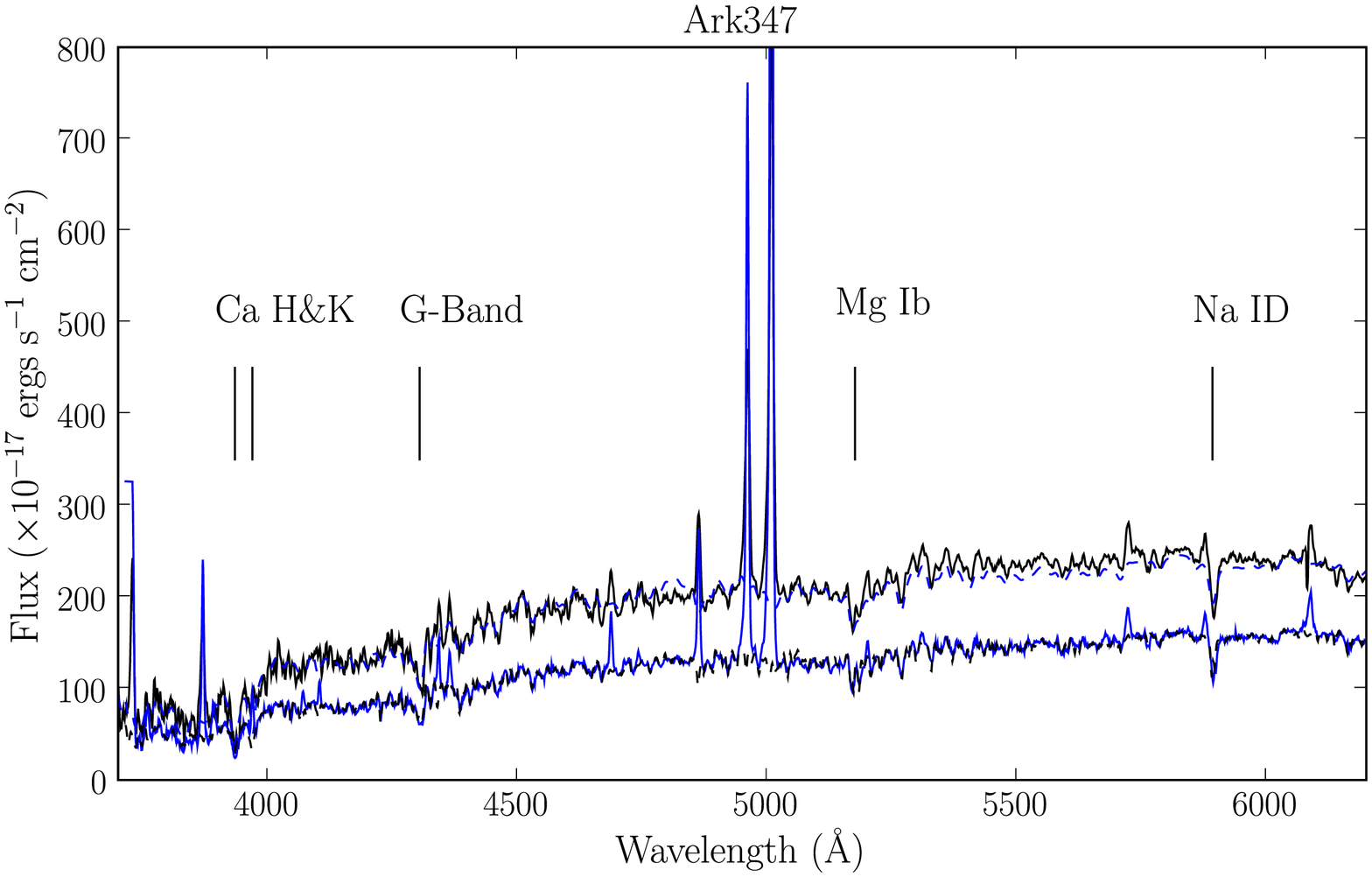}\\

\caption{Plotted are the SDSS spectra (blue) and KPNO spectra (black) of the 4 AGN sources with spectra from both sources, focusing on a region (3700--6200 \AA) which shows both emission (i.e. H$\beta$ and [\ion{O}{3}]) and intrinsic absorption features.  The fitted continuum for each spectrum is shown with the dotted lines.  Comparison of the two sets of spectra for each source show good agreement between the flux measurements and spectral shape.  
\label{fig-sloankpnosubtr}}
\end{figure}

To show how the spectra and continuum models for spectra taken at KPNO compare to the SDSS spectra, we plot the KPNO and SDSS spectra + continuum fits for the four sources with spectra from both in Figure~\ref{fig-sloankpnosubtr}.  We chose to show the region from 3700--6200\AA, a region which includes both prominent emission lines (i.e. H$\beta$ and [\ion{O}{3}]) and intrinsic absorption features (Ca H\&K, the G-Band, \ion{Mg}{1}b, and \ion{Na}{1}D).  Both the SDSS and KPNO spectra of Ark 347 are well fit with a continuum dominated by an old stellar population at solar metallicity.  The KPNO spectrum of Mkn 417 is found to be dominated by a power law, while the SDSS continuum is dominated by a solar metallicity old stellar population.  For the broad line source MCG +04-22-042, neither the KPNO or SDSS spectra are fit well at the blue end of the spectrum (due to the hydrogen Balmer lines), making it unsurprising that the models do not match.

Finally, for Mkn 18, different metallicities (low in the SDSS spectrum and high in the KPNO spectrum) and galaxy contributions are found.  However, as our test models showed, the metallicities are not well-determined with the continuum models.  The young stellar population contributions are similar for both the KPNO and SDSS spectra, leaving the discrepancy in the intermediate and old contributions as a likely effect of the degeneracy we found in our test models between the intermediate and old populations.  The difference in the continuum flux between the KPNO and SDSS spectra of Mkn 18 is an extreme case, likely due to the fact that Mkn 18 is highly elliptical and inclined along the E-W direction of the slit in the KPNO observation (15\arcsec), while the circular fiber of the SDSS (3\arcsec) misses out on this flux.

\subsection{Emission Line Fitting}
\label{emission}
To measure the properties of the emission lines in the KPNO and SDSS spectra (including the  FWHM and flux of each line), we adopted two separate methods for the narrow line and broad line spectra.  For the narrow line spectra, we first measured the prominent lines in two distinct regions, the regions surrounding H$\beta$ and H$\alpha$.  At the blue end of the spectrum, we fixed the positions of the H$\gamma$, H$\beta$, and [\ion{O}{3}] lines ($\lambda 4959$ and $\lambda 5007$), requiring that the velocity offset and FWHM of the lines remain the same for all of the lines measured, and fit for the flux and equivalent width.  For spectra whose wavelength range includes [\ion{O}{2}] $\lambda 3727$, an important diagnostic for distinguishing low-ionization narrow emission-line regions (LINERs)  \citep{1980AA....87..152H}, we include this line in the fits to the blue end of the spectrum.  Additionally, we followed the same procedure to fit the prominent emission lines surrounding and including H$\alpha$, [\ion{O}{1}] $\lambda 6300$, [\ion{N}{2}] $\lambda 6548$, [\ion{N}{2}] $\lambda 6584$, [\ion{S}{2}] $\lambda 6716$, and [\ion{S}{2}] $\lambda 6731$.  The intensities of the [\ion{N}{2}] lines are fixed such that the $\lambda 6548$ line is at a 1:2.98 ratio with the $\lambda 6584$ line, as dictated by atomic physics.  For all of the narrow line fits, the FWHM was corrected for the instrumental resolution (200\,km\,s$^{-1}$ at 5007\AA~ for the KPNO spectra and 75\,km\,s$^{-1}$ for the SDSS spectra) and we placed the restriction that the FWHM values have a lower limit of 50\,km\,s$^{-1}$and an upper limit of 1000\,km\,s$^{-1}$. The results are recorded in Table~\ref{tbl-stronghbeta}. In Table~\ref{tbl-weakblue}, we include the intensity ratios for additional weaker lines (i.e. H$\delta$, [\ion{N}{1}], \ion{He}{1}) measured in the spectra.

For the broad line sources, two complications arise which prevent us from performing the same analysis as for the narrow line sources.  Firstly, greater uncertainties exist in the continuum measurements.  Secondly, the lines can not be fit by simple Gaussians with the same widths.  
While the hydrogen Balmer lines of many of the broad line sources show asymmetries, we chose to fit both H$\alpha$ and H$\beta$ with a combination of narrow and broad Gaussians.  To ensure the uniform measurements of the lines in our spectra, we used an automated process which focused on fitting lines in a narrow region surrounding both the H$\beta$ and H$\alpha$ lines, separately.  

In the H$\beta$ region, defined as the region from 4600--5200\AA, we fit a combination of three narrow Gaussians to [\ion{O}{3}] 5007\AA.  The use of three Gaussians allowed us to reproduce the shape more robustly, since this line often shows extended wings.  The narrow line shape, particularly the widths of these lines, were applied to the narrow \ion{He}{2} 4686\AA, H$\beta$ 4861\AA, and [\ion{O}{3}] 4959\AA~lines.  Both the flux and velocity offset of each line were allowed to vary.  The continuum was fit with a linear function in a region unaffected by the prominent lines.  Finally, these fitted narrow components were combined with a broad  H$\beta$ line, which was modeled with a single broad Gaussian component, and re-fit.  The use of essentially a narrow and broad Gaussian allows us to estimate the flux and width of each component, important in estimating the black hole mass (based on the FWHM in the broad component) and emission line ratios (which depend on the ratio of the narrow lines).  Results of these fits are included in Table~\ref{tbl-bluebroad}, including the measured continuum flux at 5100\AA.  The recorded values of FWHM for the narrow component apply to the strongest narrow line component of the three Gaussians used to fit the [\ion{O}{3}] 5007\AA~line. 

\begin{figure}
\hspace{-1.2cm}
\includegraphics[height=9cm,angle=90]{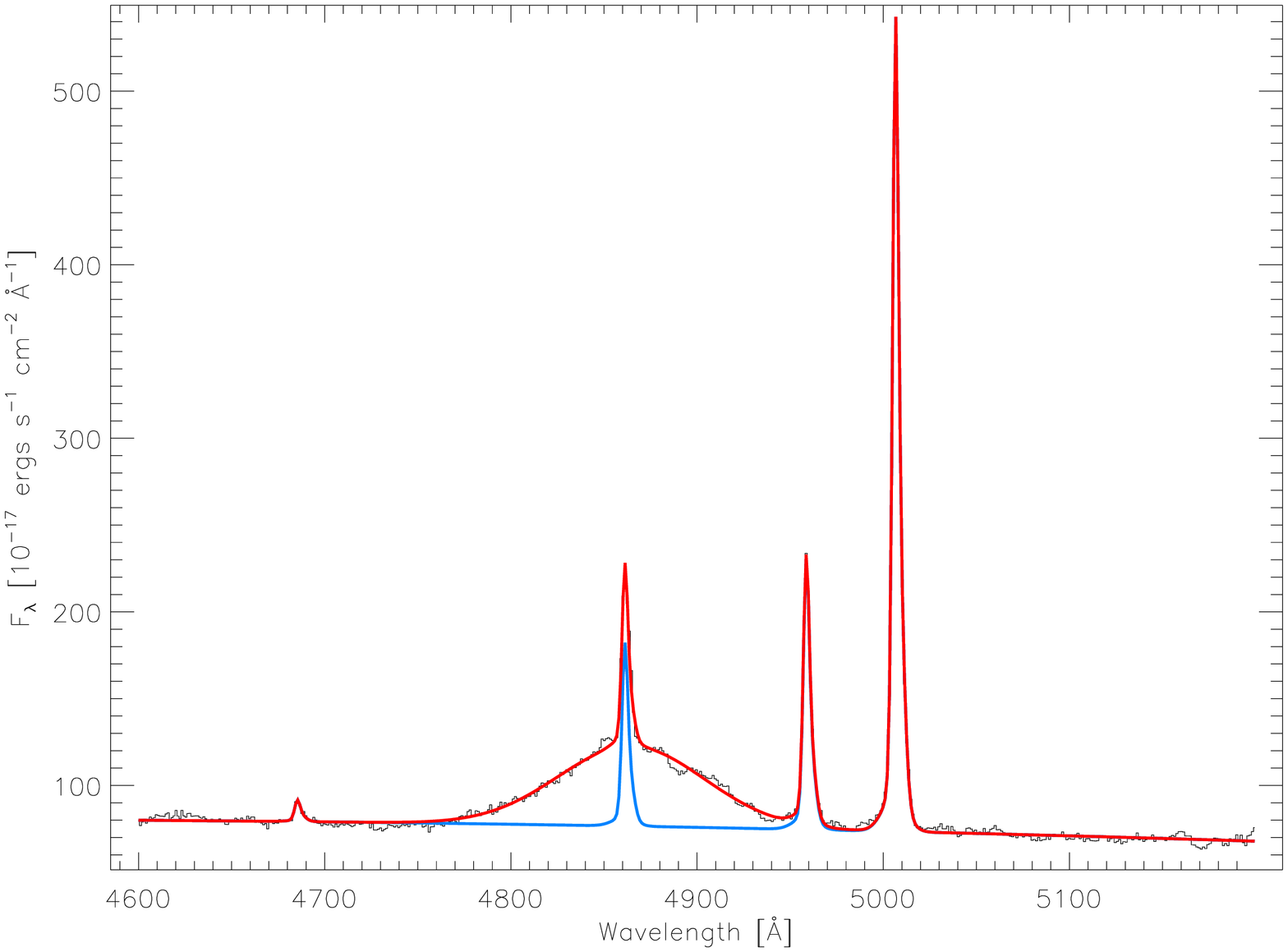}
\includegraphics[height=9cm,angle=90]{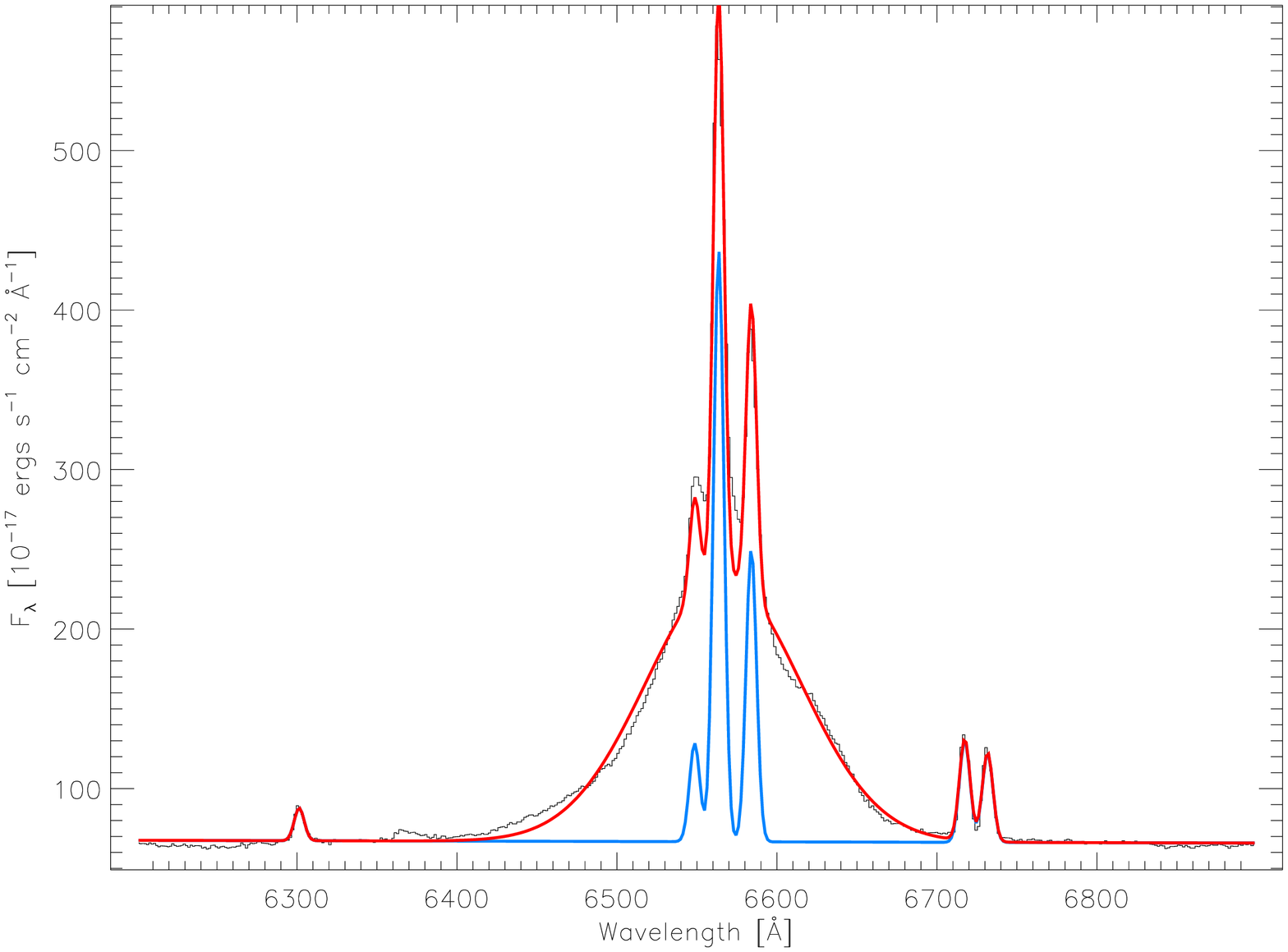}\\
\vspace{-0.8cm}

\hspace{-1.2cm}
\includegraphics[height=9cm,angle=90]{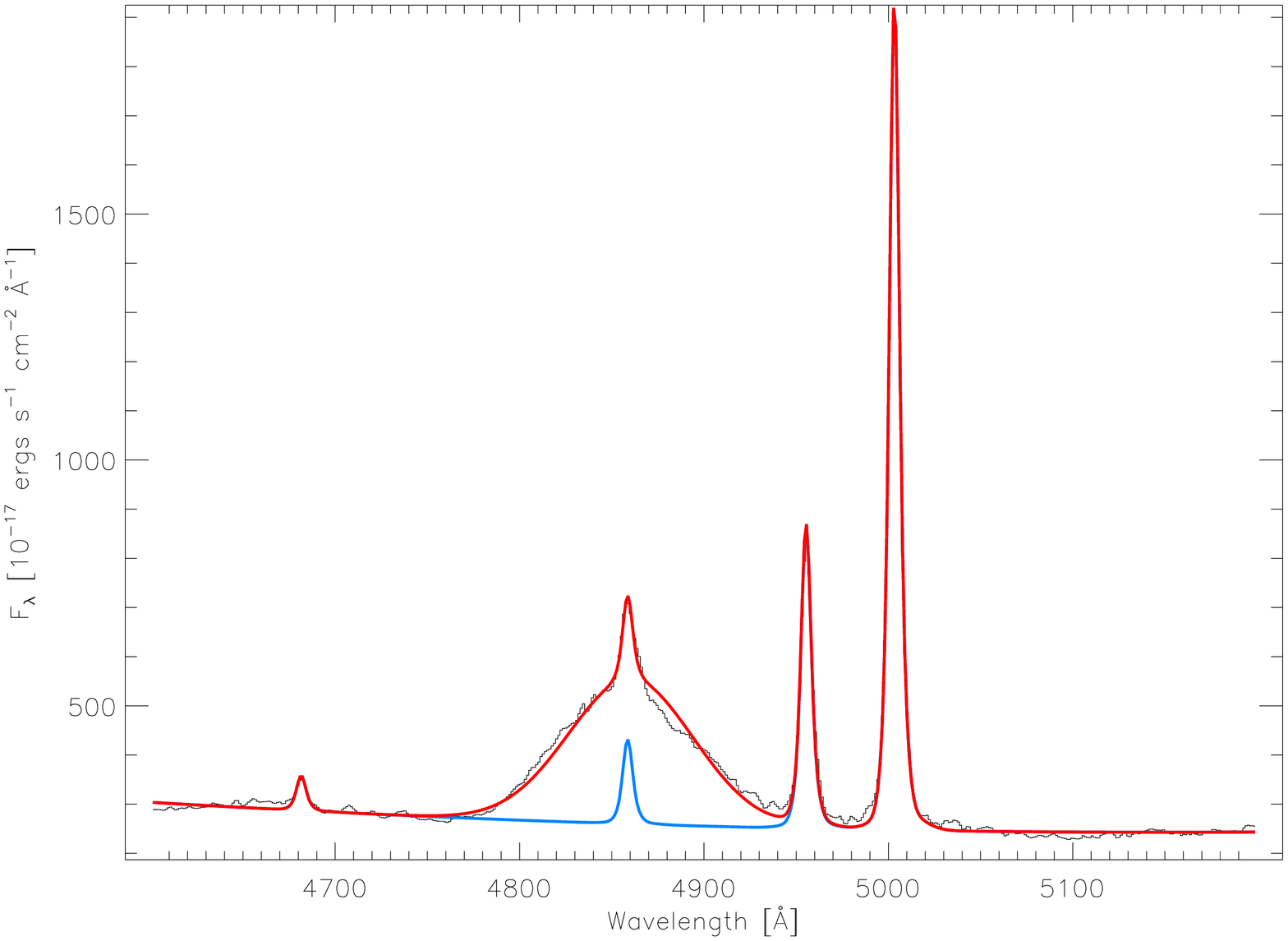}
\includegraphics[height=9cm,angle=90]{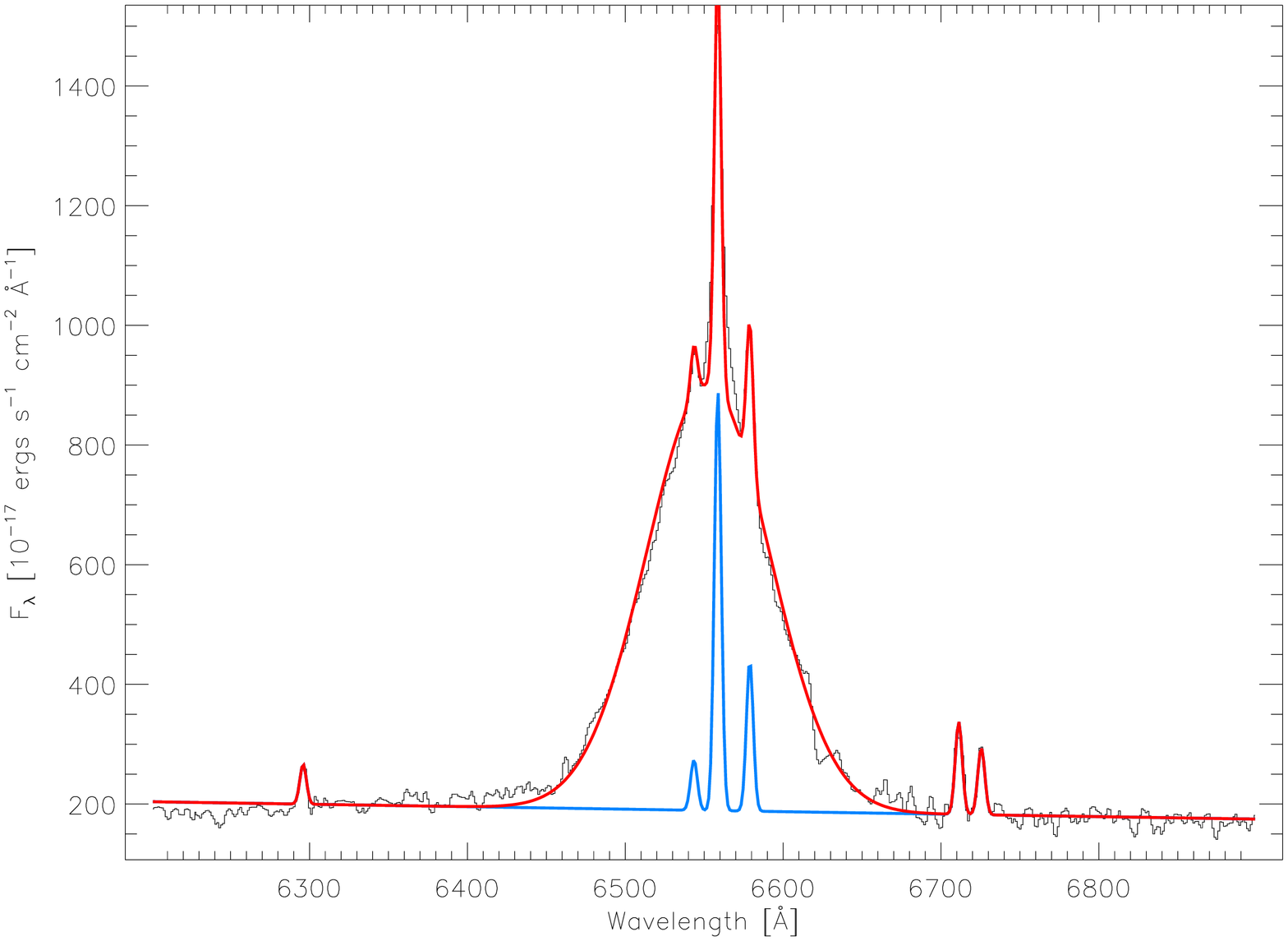}\\

\vspace{-0.8cm}
\hspace{-1.2cm}
\includegraphics[height=9cm,angle=90]{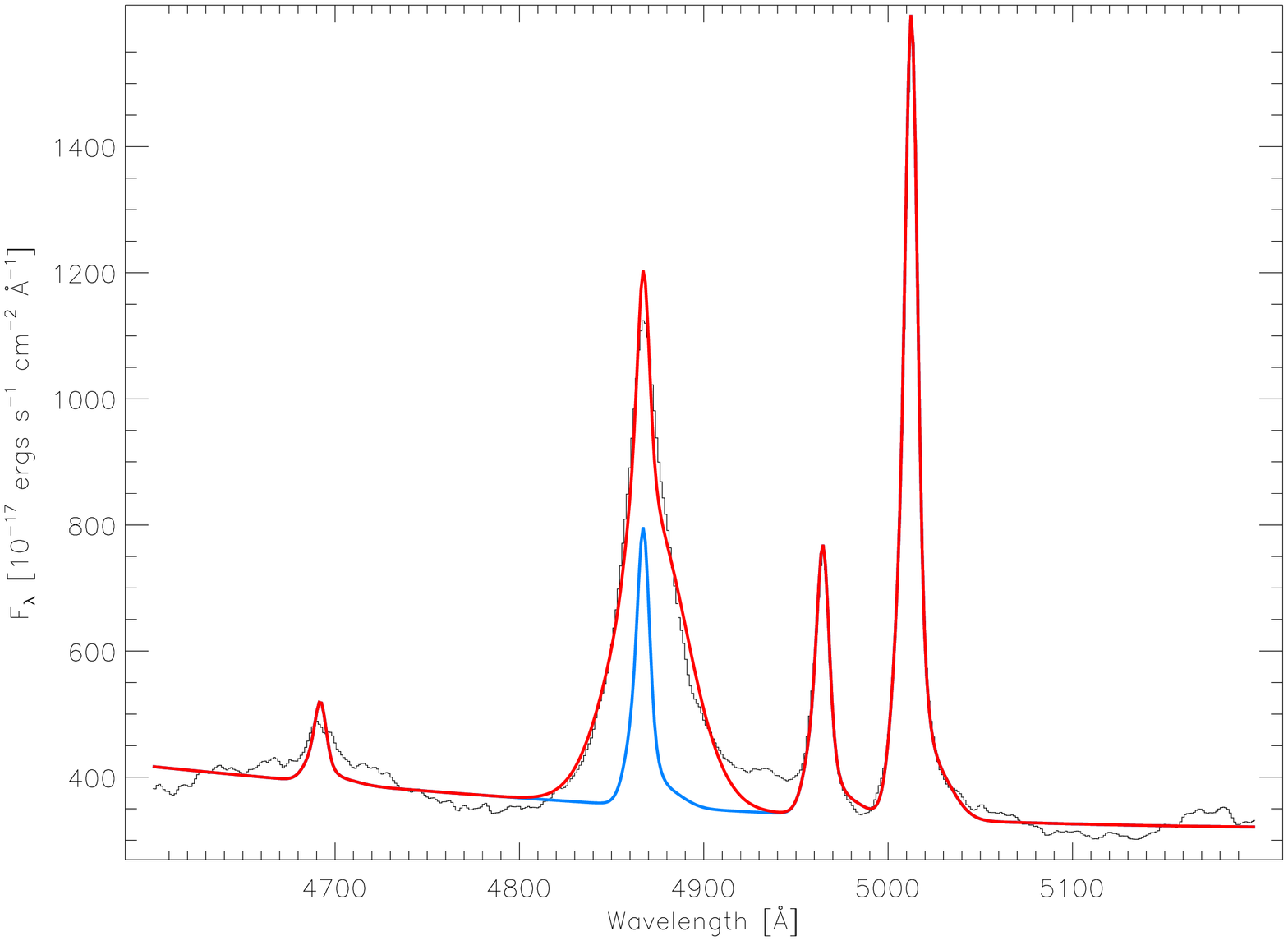}
\includegraphics[height=9cm,angle=90]{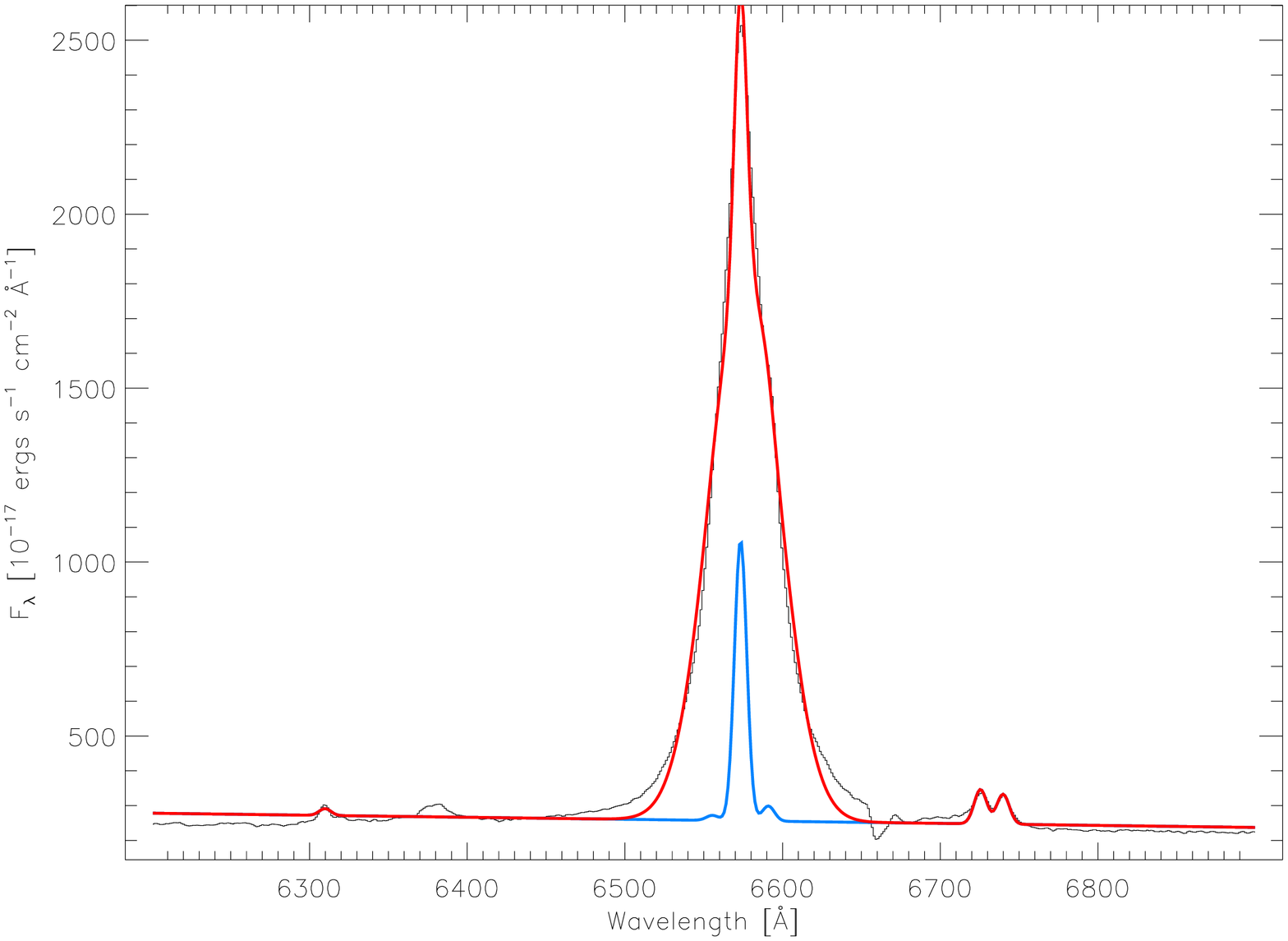}\\

\caption{Plotted are fits to the H$\beta$ and H$\alpha$ regions of the broad line spectra of SDSS J090432.19+553830.1 (top), Mkn 841 (middle), and MCG +04-22-042 (bottom).  The narrow components are shown in blue, while the total fit (broad + narrow lines) is shown in red.
\label{fig-broadfits}}
\end{figure}

In the H$\alpha$ region, defined as the region from 6200--6900\AA, we used the narrow [\ion{O}{1}] 6300\AA~ line to define the initial guess for the velocity offset of the measured lines and the set FWHM of a single Gaussian component.  The offset velocities and fluxes of the remaining narrow H$\alpha$ line, [\ion{N}{2}] lines, and [\ion{S}{2}] lines were allowed to vary.  However, the intensities of the [\ion{N}{2}] lines are fixed such that the $\lambda 6548$ line is at a 1:2.98 ratio with the $\lambda 6584$ line.  A linear continuum was fit in a region unaffected by the emission lines.  The narrow lines were added to a single broad Gaussian for broad H$\alpha$ and re-fit.  Results from these fits are recorded in Table~\ref{tbl-redbroad}.  Examples of fits to both the H$\beta$ and H$\alpha$ regions are shown in Figure~\ref{fig-broadfits}.  The largest uncertainties involved in these fits are associated with the measurements of H$\alpha$ and the two [\ion{N}{2}] lines, which are blended in our broad line spectra, particularly for a source such as MCG +04-22-042.

Additionally, weaker lines that are also present in the spectra were measured by manually selecting a continuum region surrounding the selected emission feature.  The flux of each of these measured lines are included in Table~\ref{tbl-weakbluebroad}.  Where broad lines were present and clearly separable from a narrow component, the indicated flux is for the narrow component.

\subsection{Stellar Absorption Features}\label{gal-absorption}
As an alternate method of determining ages of the host galaxies from the stellar continuum fits, we measure the strength of stellar absorption features directly from the non-galaxy continuum subtracted (both with and without subtraction of the AGN non-thermal component) spectra of our sources.  This method is analogous to the work measuring Lick-indices by \citet{1997ApJS..111..377W}.  However, instead of broadening our spectra to the velocity dispersion of the Lick/IDS spectral library (9\AA), we follow the procedure outlined in \citet{2003MNRAS.341...33K} for SDSS spectra, which instead compares the measured indices to the \citet{2003MNRAS.344.1000B} stellar models.  For further details on the SDSS analysis, along with a comparison of the measured indices with additional high resolution stellar libraries, see the discussion in 
\citet{2003MNRAS.341...33K}.

Two particularly important indicators of the age of a stellar population were used extensively in galaxy studies using SDSS spectra  \citep{2003MNRAS.346.1055K,2003MNRAS.341...33K,2005MNRAS.362...41G,2006MNRAS.372..961K}.  These are the 4000\AA~break (measured with $D_n(4000)$) and the equivalent width of H$\delta$ absorption (measured with H$\delta_A$).  Among these, the 4000\AA~break, or \ion{Ca}{2} break, is observed as a discontinuity in the optical spectrum, caused mainly by the presence of absorption features from metals below 4000\AA.  Since the opacity of metals in young, hot stars is low, this feature is weak in young stellar populations and strong in old populations.  As a measurement of the \ion{Ca}{2} break, we use the definition of
\citet{1999ApJ...527...54B} to compute:
\begin{equation} D_n(4000) = \frac{\int^{4000}_{4100} f_{\lambda} d\lambda}{\int^{3850}_{3950} f_{\lambda} d\lambda}
\end{equation}

While strong \ion{Ca}{2} breaks indicate old populations, strong equivalent widths of H$\delta$ absorption indicate a recent burst of star formation within 0.1--1\,Gyr \citep{1997ApJS..111..377W}.  Therefore, we measure 
\begin{equation} H\delta_A = (4083.50 - 4122.25)(1 - (F_I/F_C)),
\end{equation} where $F_I$ is the flux of the line within the bandpass of the feature (4122.25 -- 4083.50) and $F_C$ is the flux in a pseudo-continuum.  The pseudo-continuum is defined as the line drawn through the average of the flux in the continuum immediately blueward ($\lambda\lambda$ 4041.60 -- 4079.75\AA) and redward ($\lambda\lambda$ 4128.50 -- 4161.00\AA) of the H$\delta$ absorption feature.

\begin{figure}

\includegraphics[height=4.2cm]{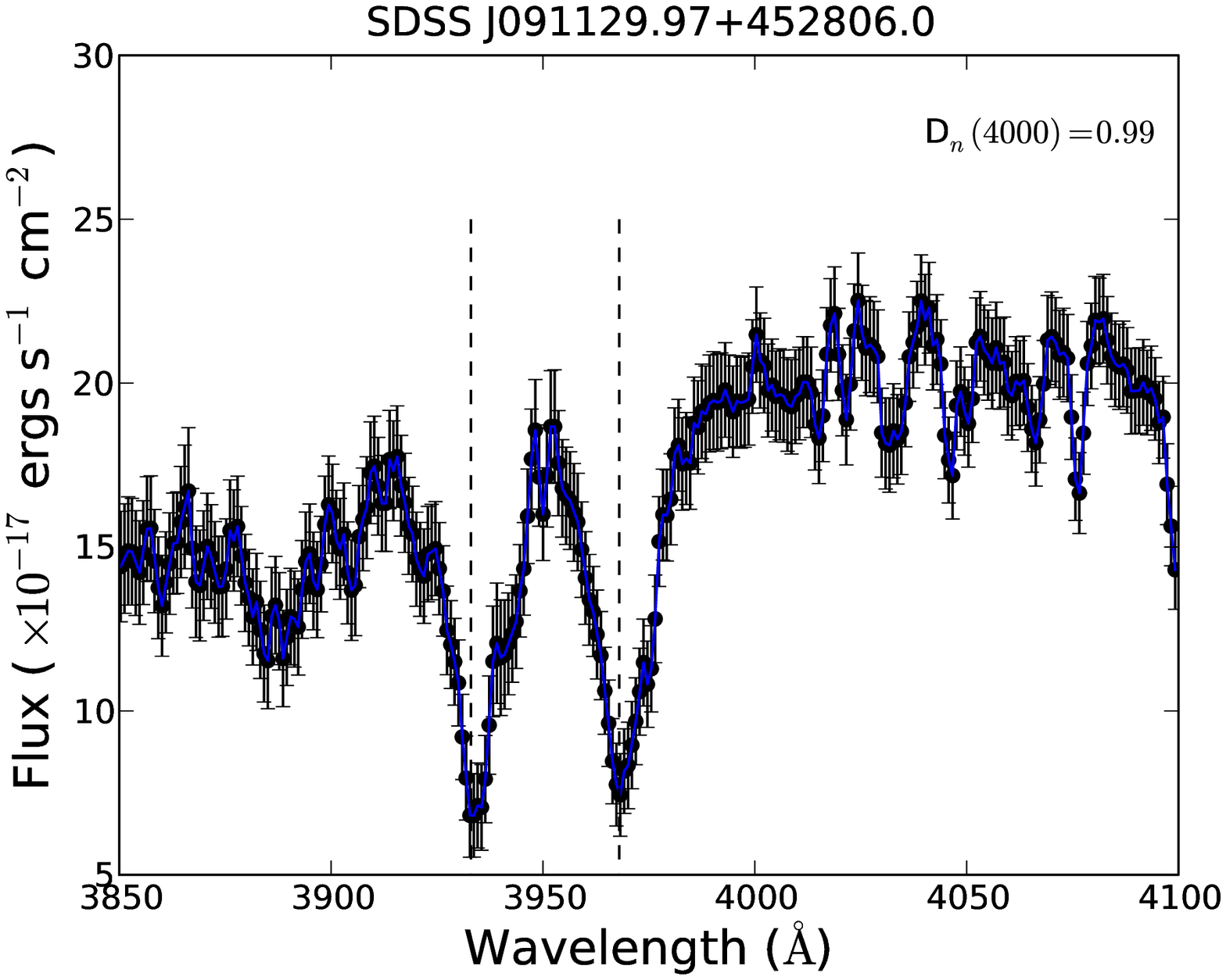}
\hspace{-0.5cm}
\includegraphics[height=4.2cm]{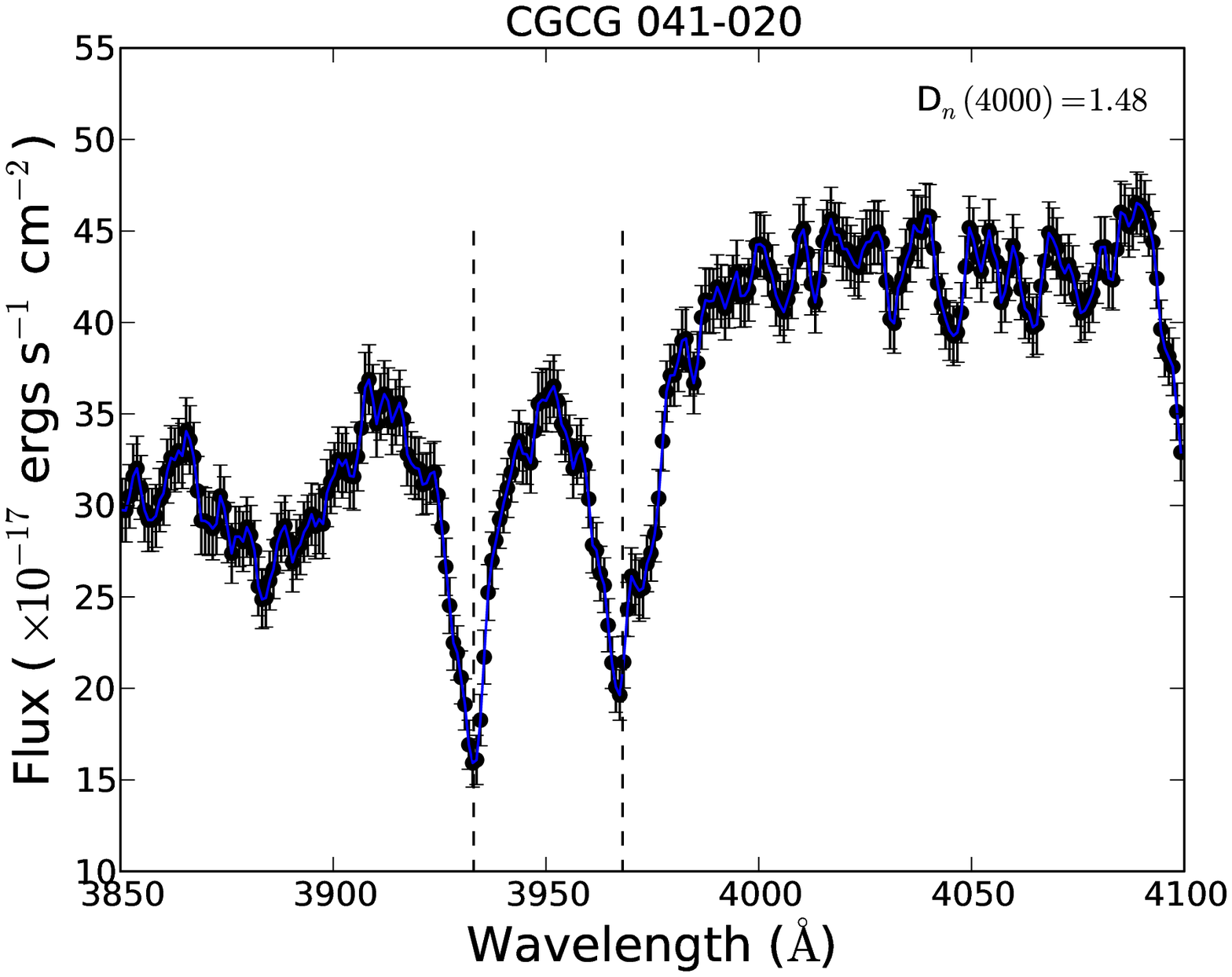}
\hspace{-0.5cm}
\includegraphics[height=4.2cm]{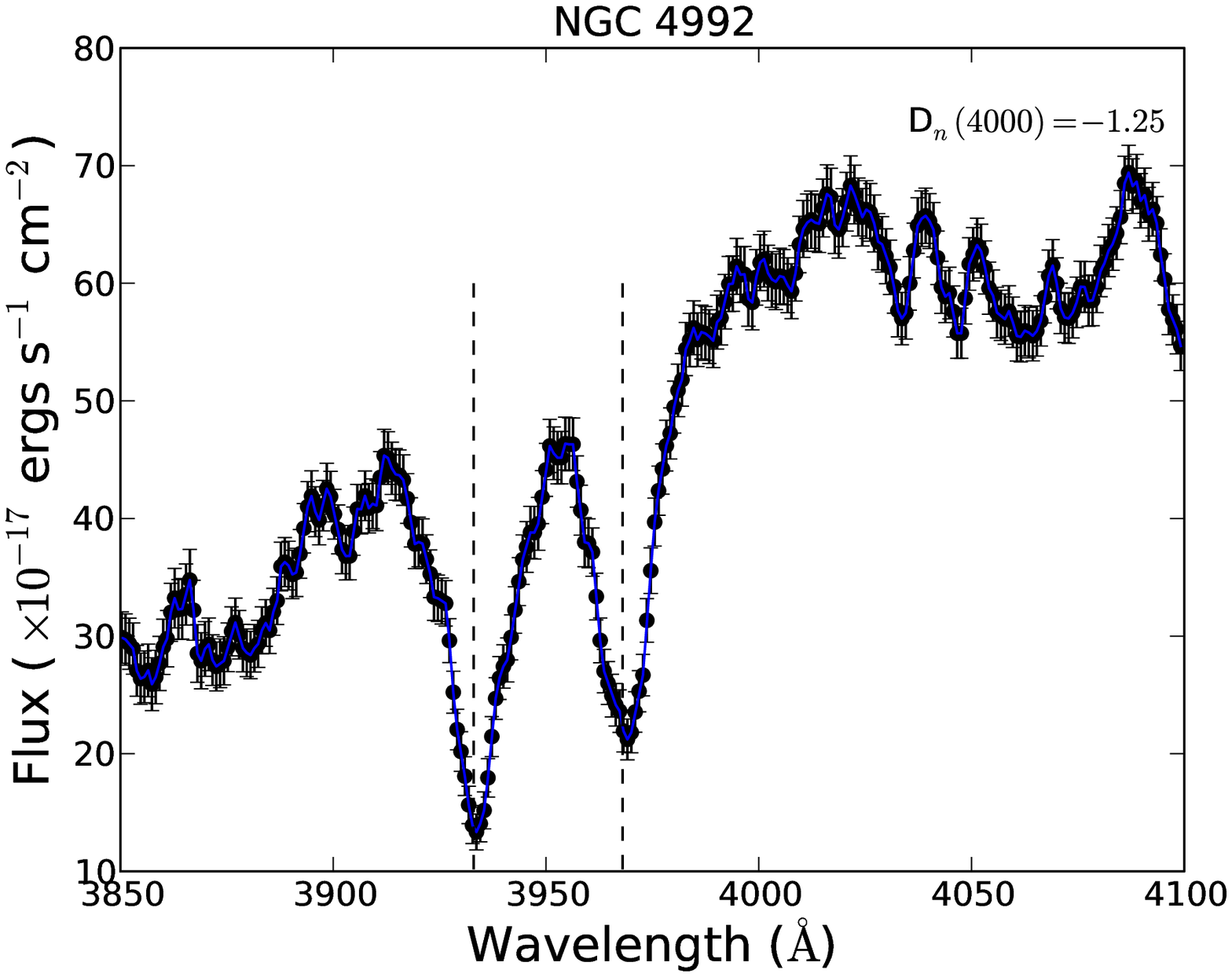}\\

\includegraphics[height=4.2cm]{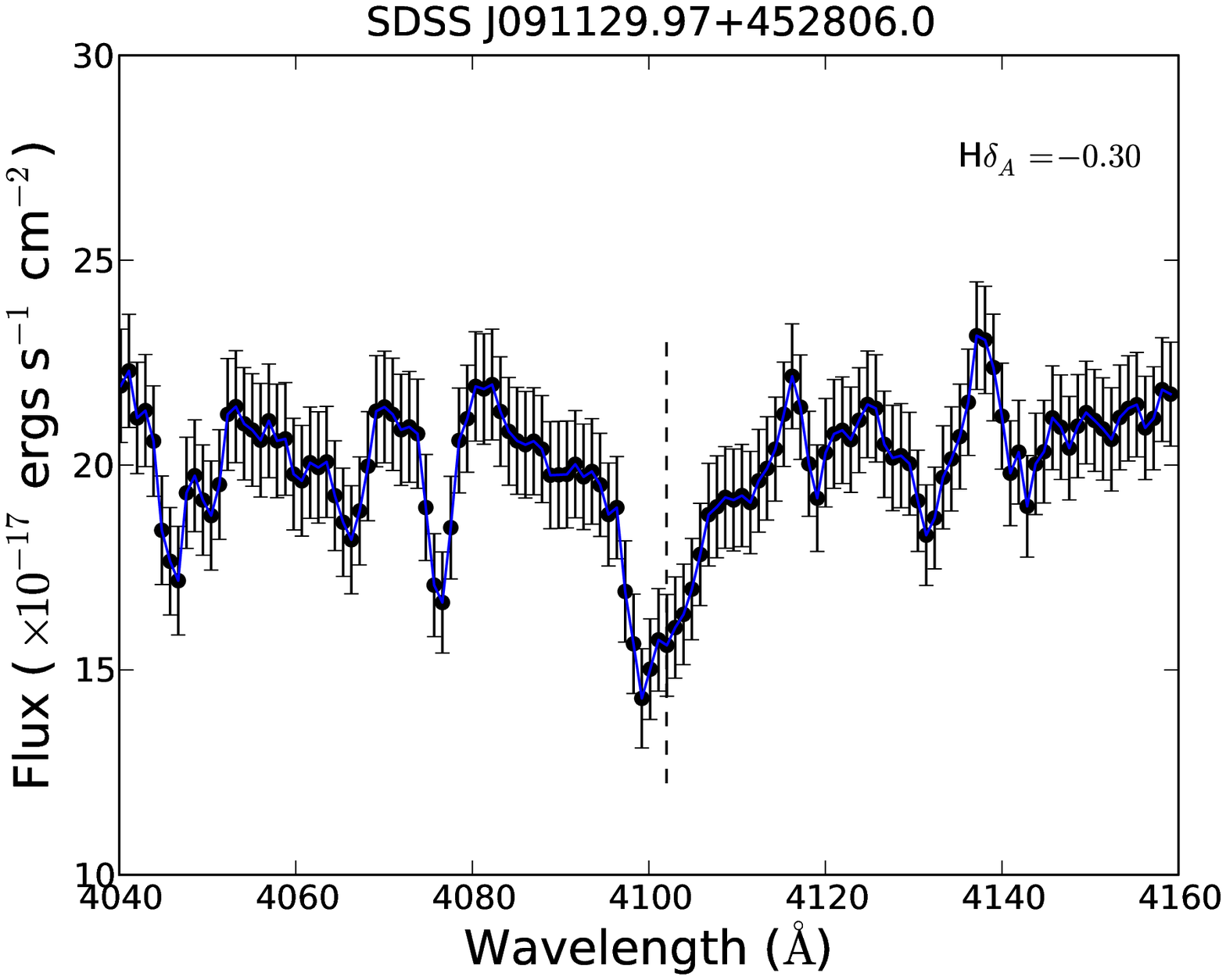}
\hspace{-0.5cm}
\includegraphics[height=4.2cm]{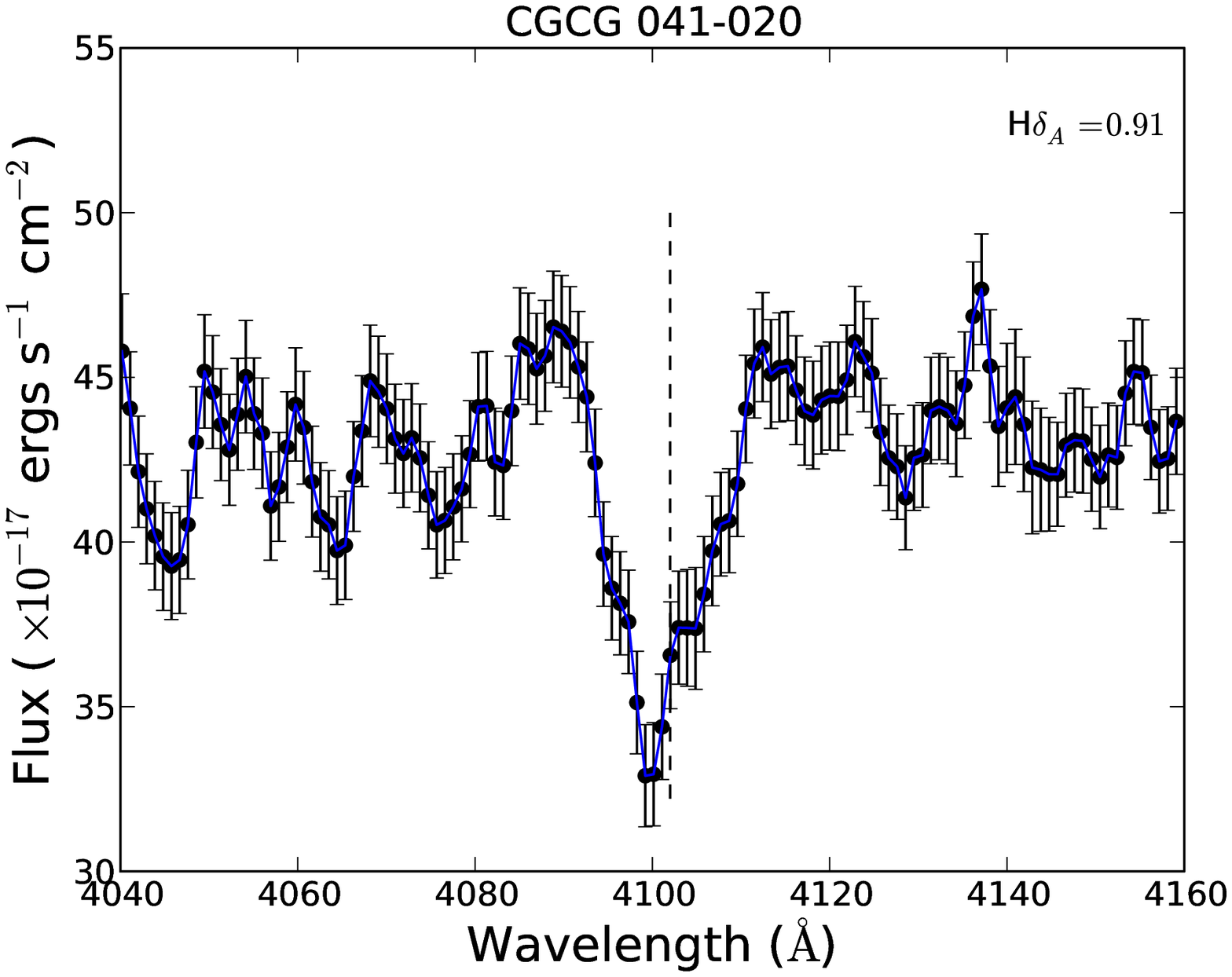}
\hspace{-0.5cm}
\includegraphics[height=4.2cm]{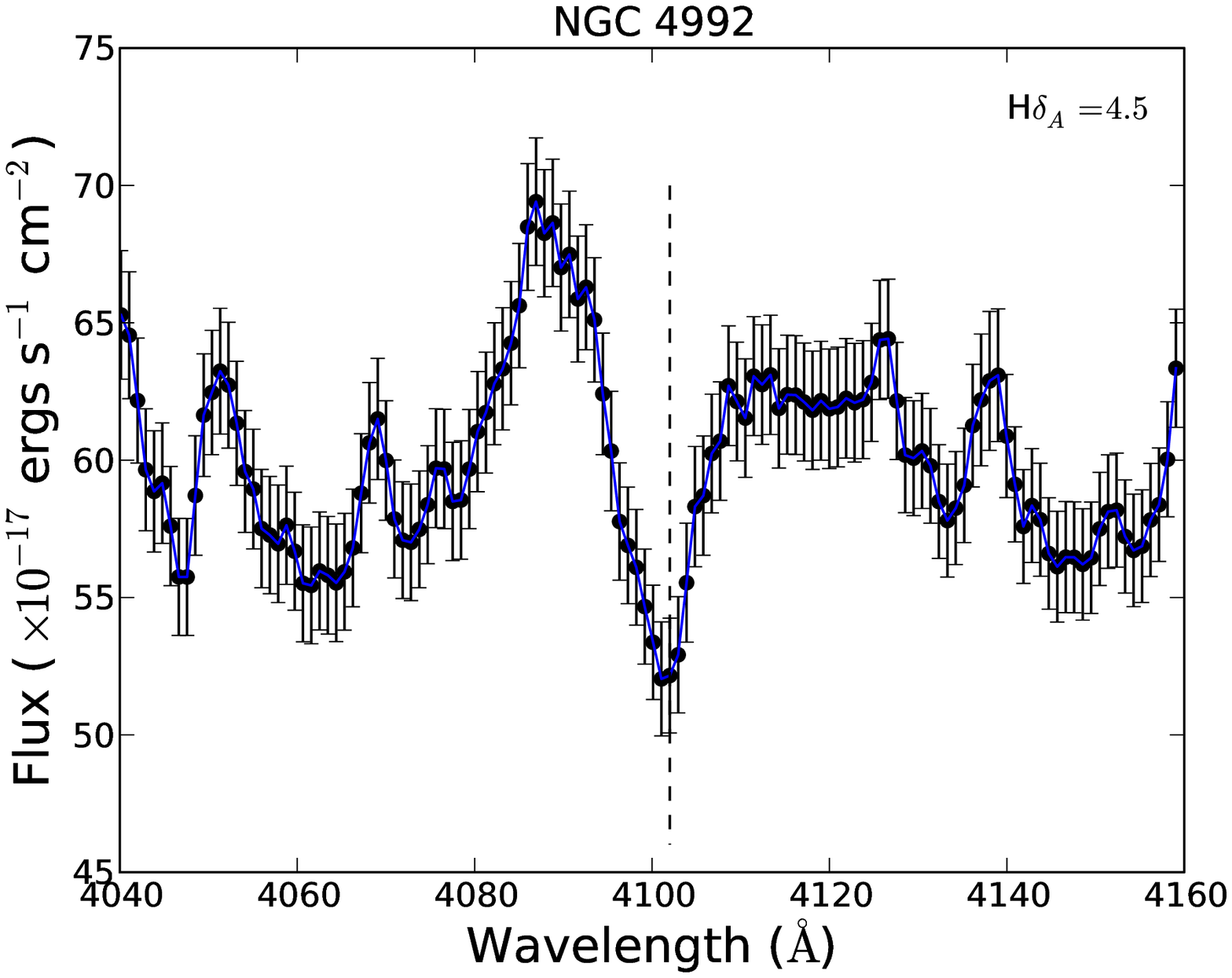}\\

\includegraphics[height=4.2cm]{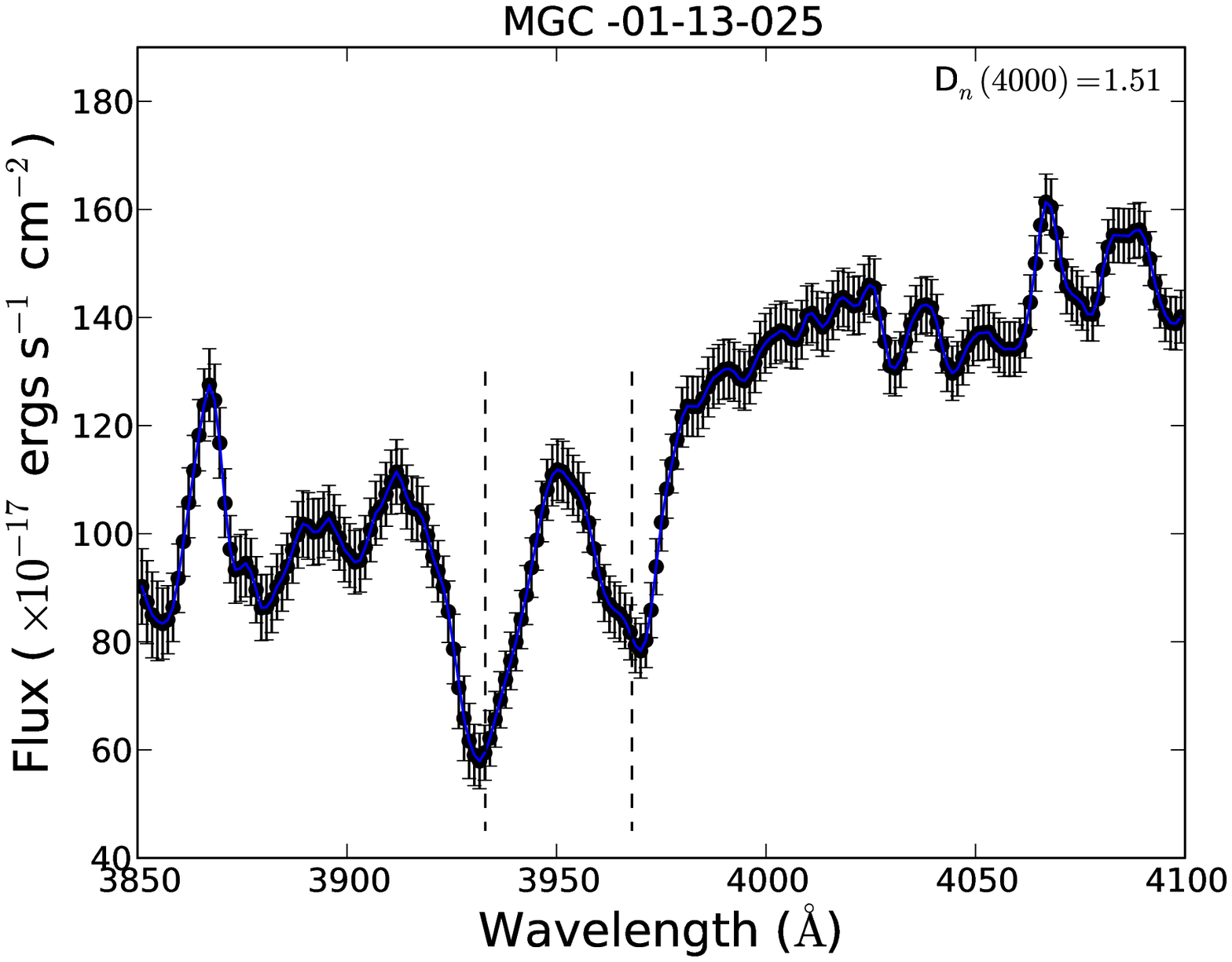}
\hspace{-0.5cm}
\includegraphics[height=4.2cm]{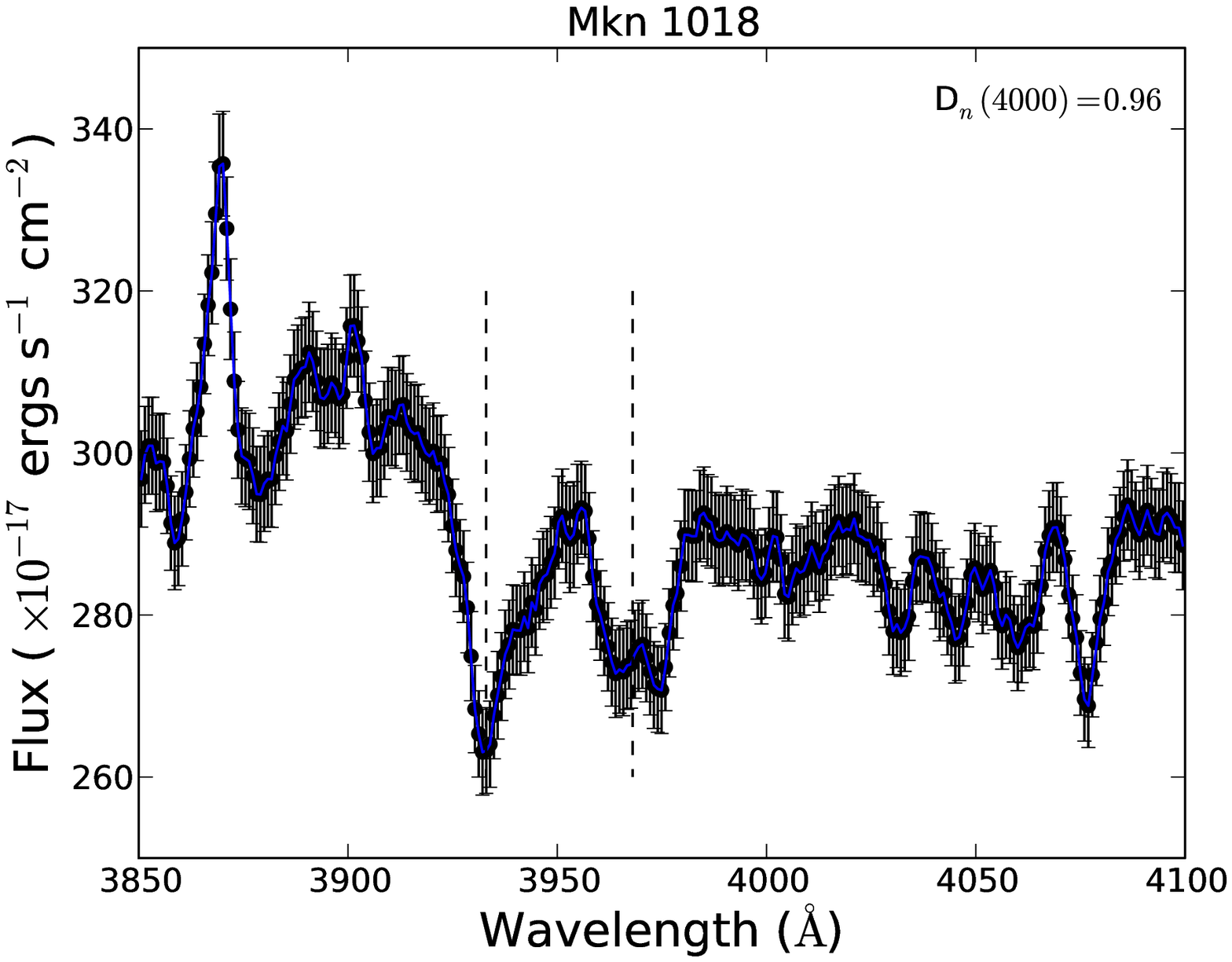}
\hspace{-0.5cm}
\includegraphics[height=4.2cm]{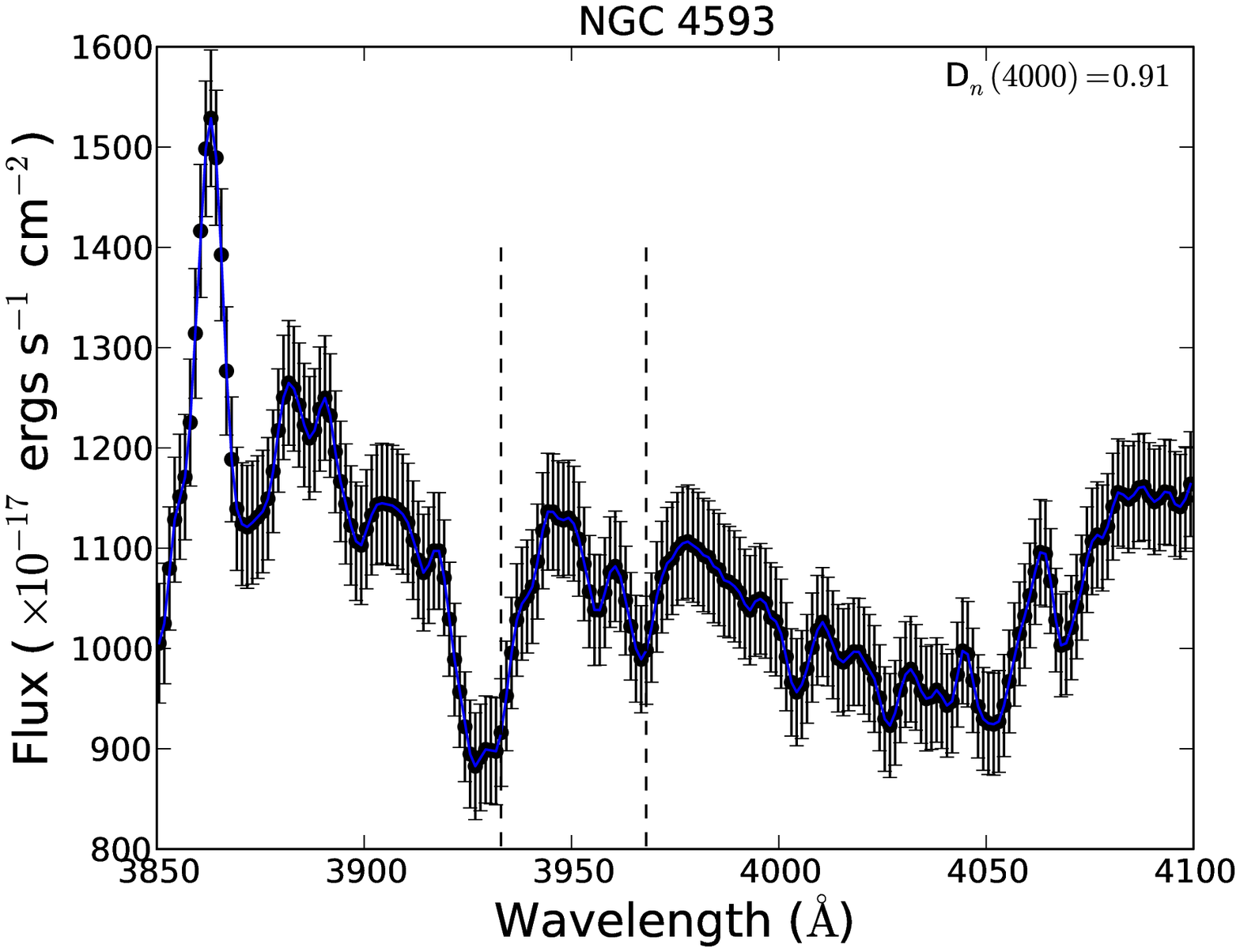}\\

\includegraphics[height=4.2cm]{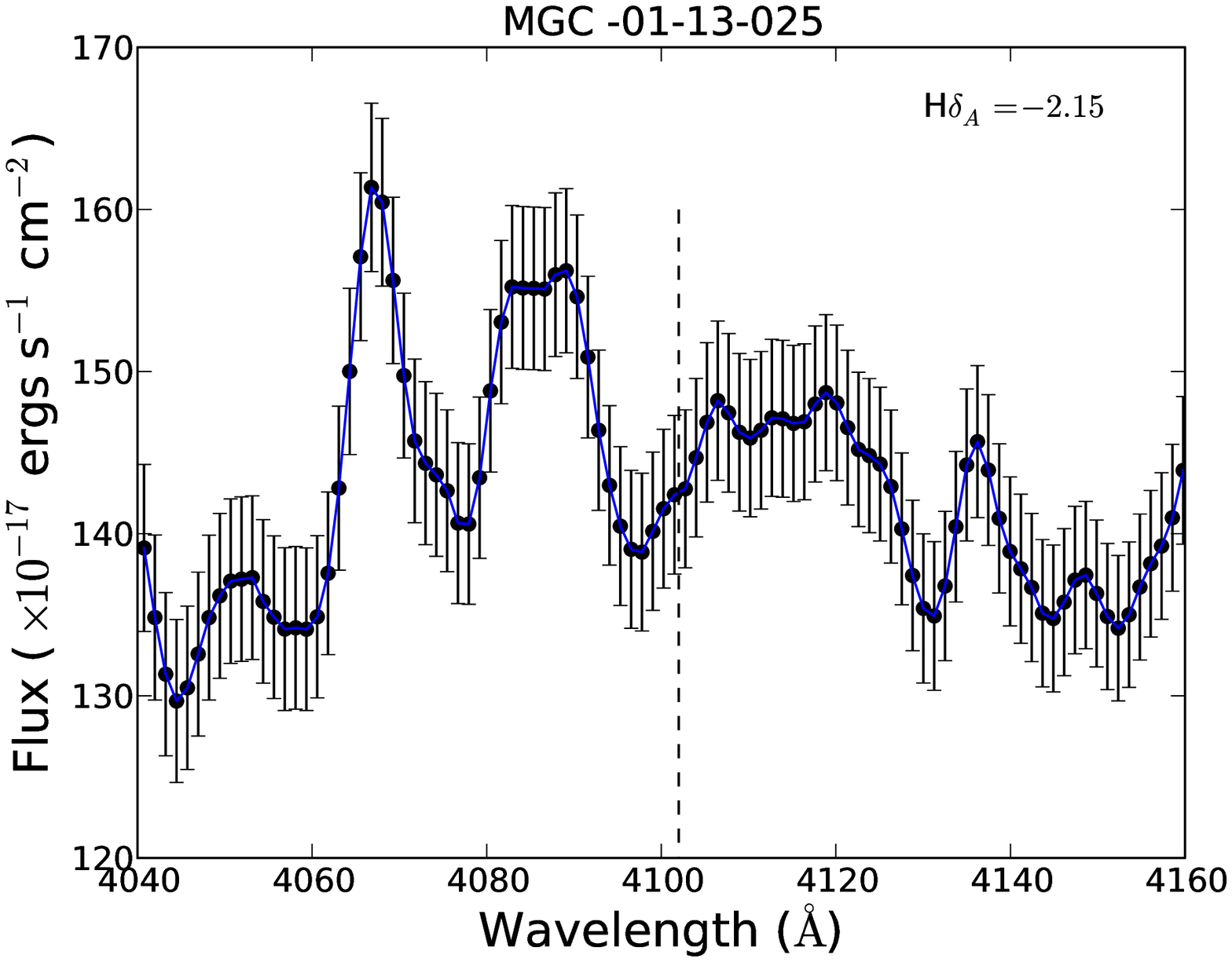}
\hspace{-0.5cm}
\includegraphics[height=4.2cm]{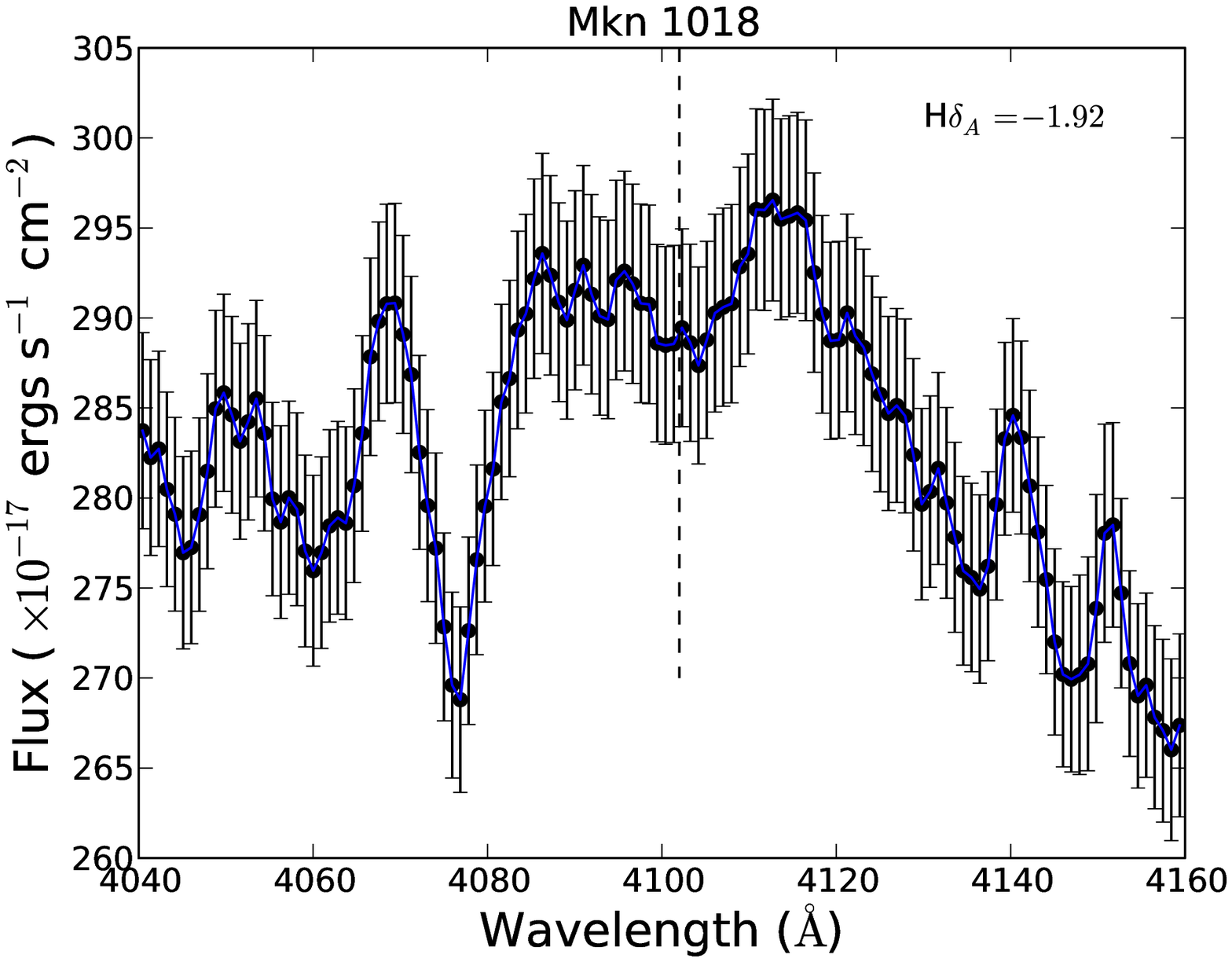}
\hspace{-0.5cm}
\includegraphics[height=4.2cm]{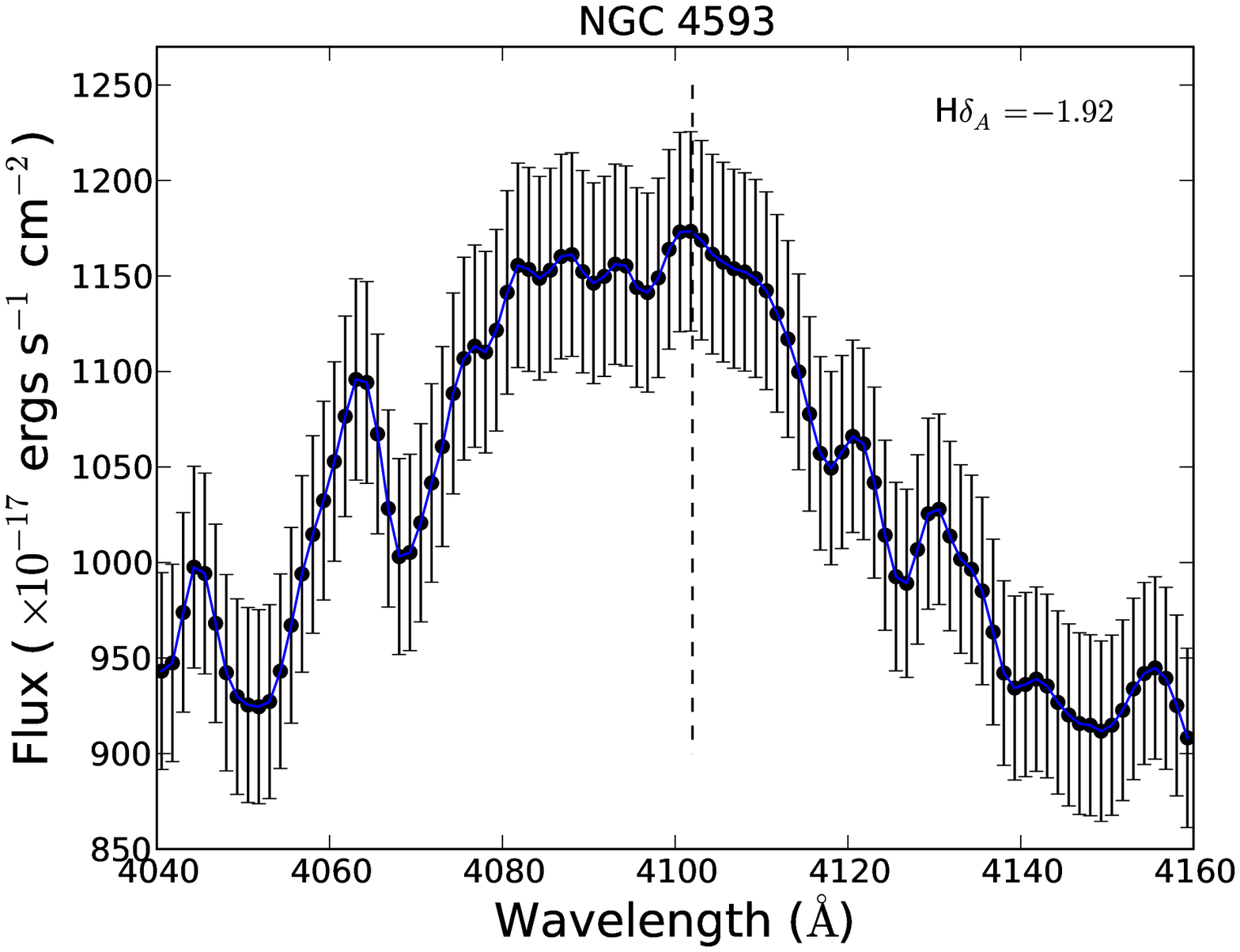}\\

\caption{Plotted are the spectra, with errors, in the regions where $D_n(4000)$ and H$\delta_A$ are measured for representative narrow line (top two panels) and broad line (bottom two panels) sources.  Absorption lines of \ion{Ca}{2} H\&K and H$\delta$ are indicated with dashed lines.  For the narrow line sources, emission lines in the spectra were subtracted.
\label{fig-exlick}}
\end{figure}

Half of the spectra show an H$\delta$ emission line (10 narrow line sources and 16 broad line sources), while emission from [\ion{Ne}{3}] 3869\AA~is often present in the pseudo-continuum from which $D_n(4000)$ is measured.  For the narrow line sources in our sample, we subtracted the measured narrow lines before calculating these age indicators.  Such a calculation is not straight forward for the broad line sources, where broad emission features are often present in the region containing H$\delta_A$ (with H$\delta$ emission) and D$_n(4000)$ (including [\ion{Ne}{3}] + H7 $\lambda 3968$\AA~ and H$\delta$).  In Figure~\ref{fig-exlick}, we plot examples of spectra for both narrow and broad line sources, where stellar absorption features are seen.

\begin{figure}
\centering
\includegraphics[height=8cm]{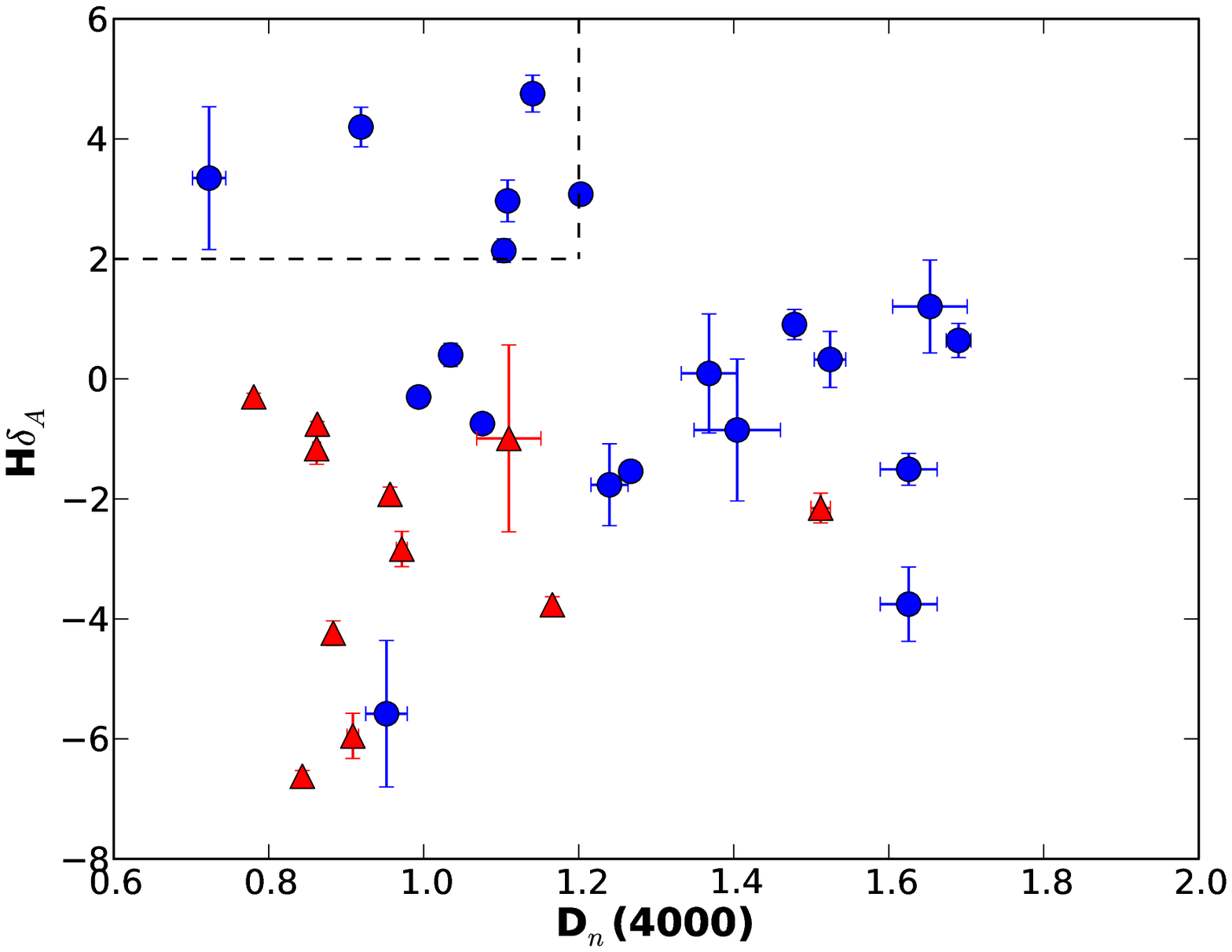}
\includegraphics[height=8cm]{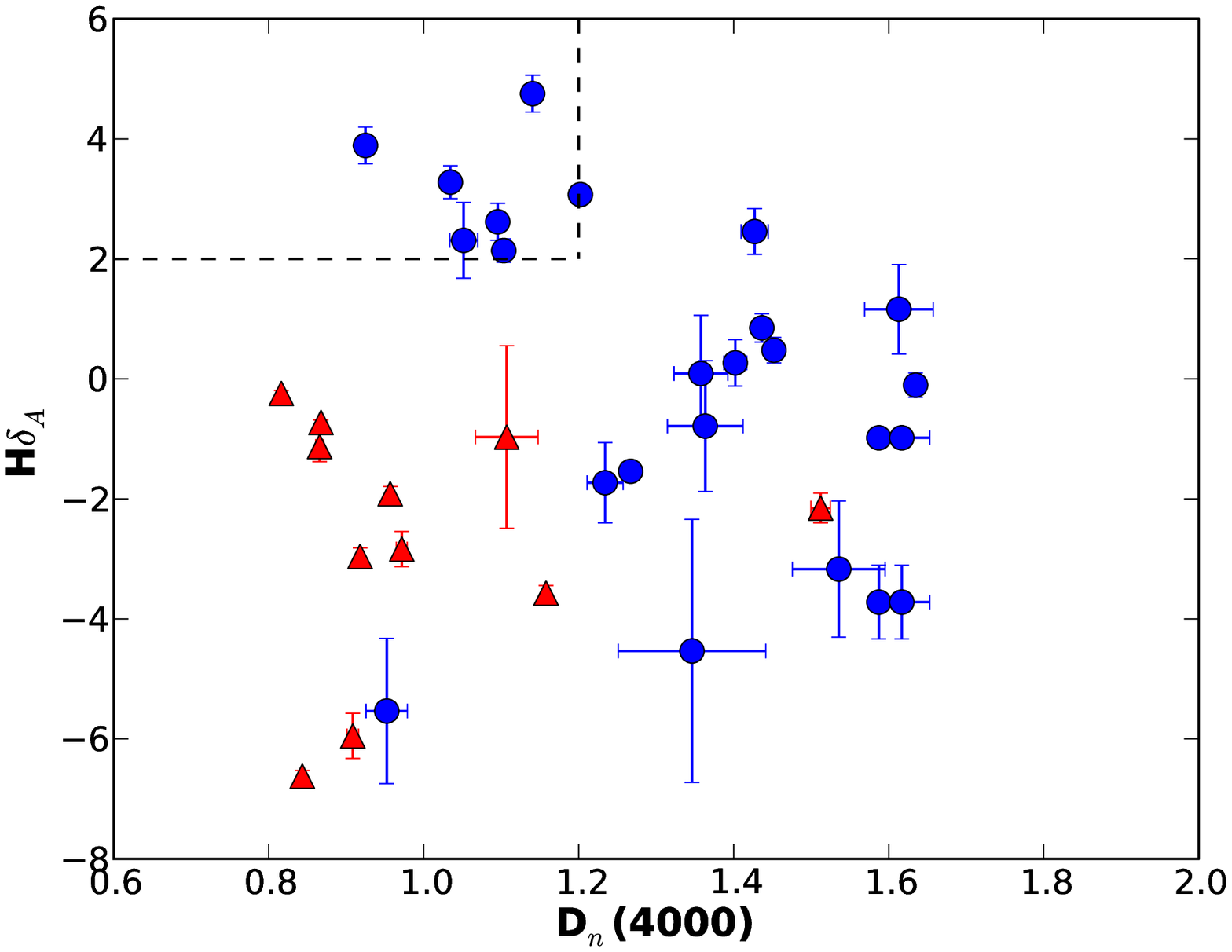}

\caption{Plotted are two age indicators, H$\delta_A$ which measures recent bursts of star formation and D$_n(4000)$ which measures the \ion{Ca}{2} break and is sensitive to old stellar populations.  The circles represent narrow line sources and the triangles represent broad line sources.  In the top plot, we show the values measured after subtracting out the power law components (Table~\ref{tbl-continuum}).  The bottom plot shows the values measured directly from the spectra.  In both plots, the box in the upper left hand corner shows the area where young stellar populations had significant ($\ga 30$\%) contributions in our test galaxy spectra (see \S~\ref{testspectra}, Figure~\ref{fig-hdeltatest}).
\label{fig-agelick}}
\end{figure}

In Figure~\ref{fig-agelick}, we plot H$\delta_A$ versus D$_n(4000)$ for our sources, excluding broad line sources with prominent H$\delta$ emission.  We plot the values measured both after subtracting the power law continuum (Table~\ref{tbl-continuum}; top plot) and from the original dereddened spectrum (bottom plot).  From each of these measurements, the D$_n(4000)$ break does not change appreciably whether or not the power law component is subtracted, with a median value of 1.26 for narrow line sources and 0.91 for broad line sources when the power law is subtracted and 1.41 (narrow) and 0.92 (broad) without the subtraction.  The H$\delta_A$ values are affected, however, for the narrow line sources with median values of 0.81 (narrow) and -2.15 (broad) with the power law subtracted and 1.73 (narrow) and -2.15 (broad) without the subtraction.  To test whether any aperture effects influenced our measurements, we plotted each of these diagnostic measurements against redshift.  With no correlation in either H$\delta_A$ or D$_n(4000)$ with $z$, we conclude that there are no obvious aperture effects to be accounted for in our measurements. 

The majority of the narrow line sources occupy the area expected from our stellar population model tests, discussed in \S~\ref{testspectra} and plotted in Figure~\ref{fig-hdeltatest}.  The broad line sources, however, occupy a region with considerably lower values of H$\delta_A$.  This is true even for sources where an H$\delta$ emission line is not seen in the spectrum (as for the sources plotted).  From visual inspection of the H$\delta$ region of our sources, we find that unlike the narrow line sources, we can not clearly identify an H$\delta$ absorption feature in any of the broad line sources.  In most cases, we see emission features that are often broad.  The low values of H$\delta_A$ measurements for broad line sources are therefore a likely effect of emission in this region.


In addition to these stellar age diagnostics, we measured additional absorption indices for common stellar absorption features.  These values were measured using the same method as used for the H$\delta_A$ index, first subtracting the emission line spectra for the narrow line sources and subtracting the power law component for all of the sources.  Bandpasses and continuum ranges are defined in \citet{1994ApJS...94..687W} and \citet{1997ApJS..111..377W}.  In Table~\ref{tbl-lick}, we present the stellar age indicators (D$_n(4000)$ and H$\delta_A$) along with 6 metallicity indicators, chosen to sample indices sensitive to several different elements (i.e. C, N, Ca, Mg, Fe).  Two of these indices are combinations of other indices, defined in \citet{1993PhDT.......172G}: 
\begin{equation}[{\rm MgFe}] = \sqrt{\rm Mgb <Fe>}{~~~\rm and ~~~} 
<{\rm Fe}> = \frac{1}{2} ({\rm Fe 5270 + Fe 5335}).
\end{equation}  We use the modified form of [MgFe]$'$, defined by \citet{2003MNRAS.343..279T}
as:
\begin{equation}[{\rm MgFe}]' = \sqrt{\rm Mgb (0.72~Fe5270 + 0.28~Fe5335)}. \end{equation}

\begin{figure}
\plotone{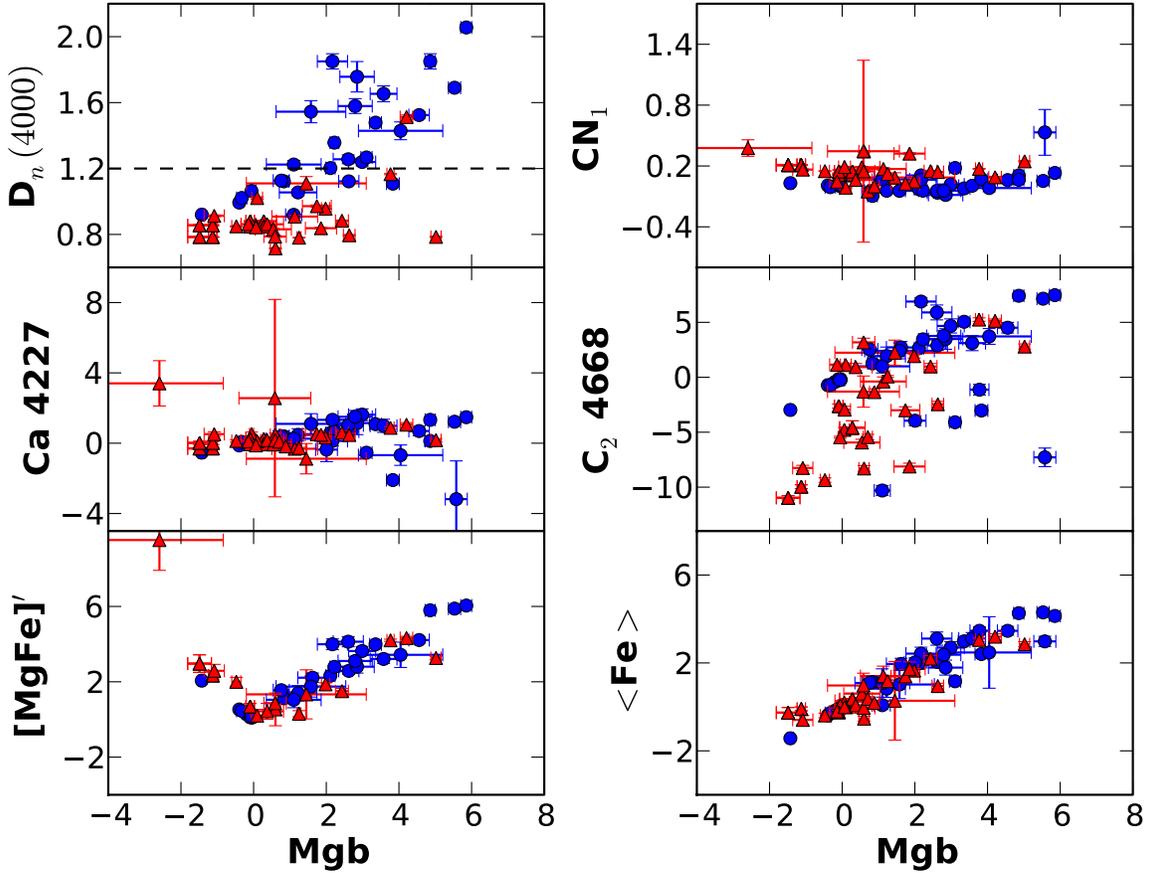} \\
\caption{Plotted are a selection of stellar absorption indices indicating stellar age (D$_n(4000)$) or metallicity of the populations vs. the metallicity indicator Mgb. The circles represent narrow line sources and the triangles represent broad line sources.  In the top left plot, we show a line representing the division in D$_n(4000)$ between populations with a significant contribution from young stars ($\ga 30$\%), as determined in Figure~\ref{fig-hdeltatest}.
\label{fig-metallicitylick}}
\end{figure}

Additionally, to better understand our results, we also measured these stellar absorption indices for a sample of test spectra created from the stellar population models used for the continuum fits.  We discuss these results, where we used different combinations of stellar ages and metallicities, in \S~\ref{testspectra}.   Of the additional stellar absorption indices recorded in Table~\ref{tbl-lick}, H$\delta$ emission could affect the value measured for CN$_1$.  Additionally, \ion{He}{2} 4686~\AA~ is within the range of C$_2$ 4668 and [\ion{N}{1}] 5199~\AA~ is within the range of Mgb.  Since [\ion{N}{1}] 5199~\AA~ is weak in our broad line sources, we expect little error in our Mgb measurements.  In Figure~\ref{fig-metallicitylick}, we plot various metallicity indicators and the age indicator D$_n(4000)$ versus the metallicity indicator Mgb for our target sources.  Comparing with our results from the test spectra, it appears that C$_2$ 4668 is the most affected by ``contaminating'' emission features.  The narrow line sources should be unaffected, however, since we have subtracted the emission components from their spectra.

Based on a comparison of the plots in Figure~\ref{fig-metallicitylick} with the test spectra values, 
we find that the [MgFe]$'$ vs Mgb and $<$Fe$>$ vs Mgb plots are the best indicators of the metallicity of the stellar populations.  However, the only clear result is that we do not find old, high-metallicity (2.5 Z$_{\sun}$) populations within our sample (all of the old population test spectra have Mgb $\la 2$, as determined from the D$_n$(4000) vs Mgb plot).  Since there is little difference in the parameter space occupied by solar and low metallicity populations, we can not discern anything more from our measured stellar absorption indices.

\section{Emission Line Classification}\label{class}
Emission line diagnostic plots, utilizing the optical line ratios of [\ion{O}{3}] $\lambda 5007$/H$\beta$, [\ion{N}{2}] $\lambda 6583$/H$\alpha$, [\ion{S}{2}] $\lambda\lambda 6716, 6731$/H$\alpha$, [\ion{O}{3}] $\lambda 5007$/[\ion{O}{2}] $\lambda 3727$ and [\ion{O}{1}] $\lambda 6300$/H$\alpha$, are an empirical method of separating Seyferts, LINERs, and star-forming galaxies \citep{1981PASP...93....5B,1987ApJS...63..295V}.  The chosen line ratios (1) have small wavelength separations, so that the effects of reddening are minimal, and (2) distinguish between photo-ionization from O and B stars (\ion{H}{2} objects) and a non-thermal/power law continuum (AGNs). In order to construct these diagnostic diagrams for our Swift BAT AGNs, we first corrected the line ratios for reddening. 

To correct our line ratios for extinction, we use the line ratio of the strongest narrow Balmer lines (H$\alpha$/H$\beta$) along with the \citet{1989ApJ...345..245C} reddening curve.  The effect of reddening is represented as 
\begin{equation} \frac{I(H\alpha)}{I(H\beta)} = \frac{F(H\alpha)}{F(H\beta)} 10^{c [f(H\alpha) - f(H\beta)]}
\end{equation}
where $I(\lambda)$ is the intrinsic flux, $F(\lambda)$ is the observed flux, and $f(\lambda)$ is from the reddening curve. We assume an intrinsic H$\alpha$/H$\beta$ ratio ($\frac{I(H\alpha)}{I(H\beta)}$) of 3.1 for our sources, assuming that they are dominated by the underlying AGN.  Additionally, we assume that $R_V = 3.1$ and therefore $E({\rm B} - {\rm V}) = (2.5/3.1)c$.  For 11 of the spectra from KPNO or SDSS, we find that the
ratio of H$\alpha$/H$\beta$ is less than the assumed intrinsic value, for which we do not apply a reddening correction [$E({\rm B} - {\rm V}) = 0$].  The corrected line ratios, along with values found in the literature for an additional 13 sources, are shown in Table~\ref{tbl-literature}.

\begin{figure}
\includegraphics[height=8cm]{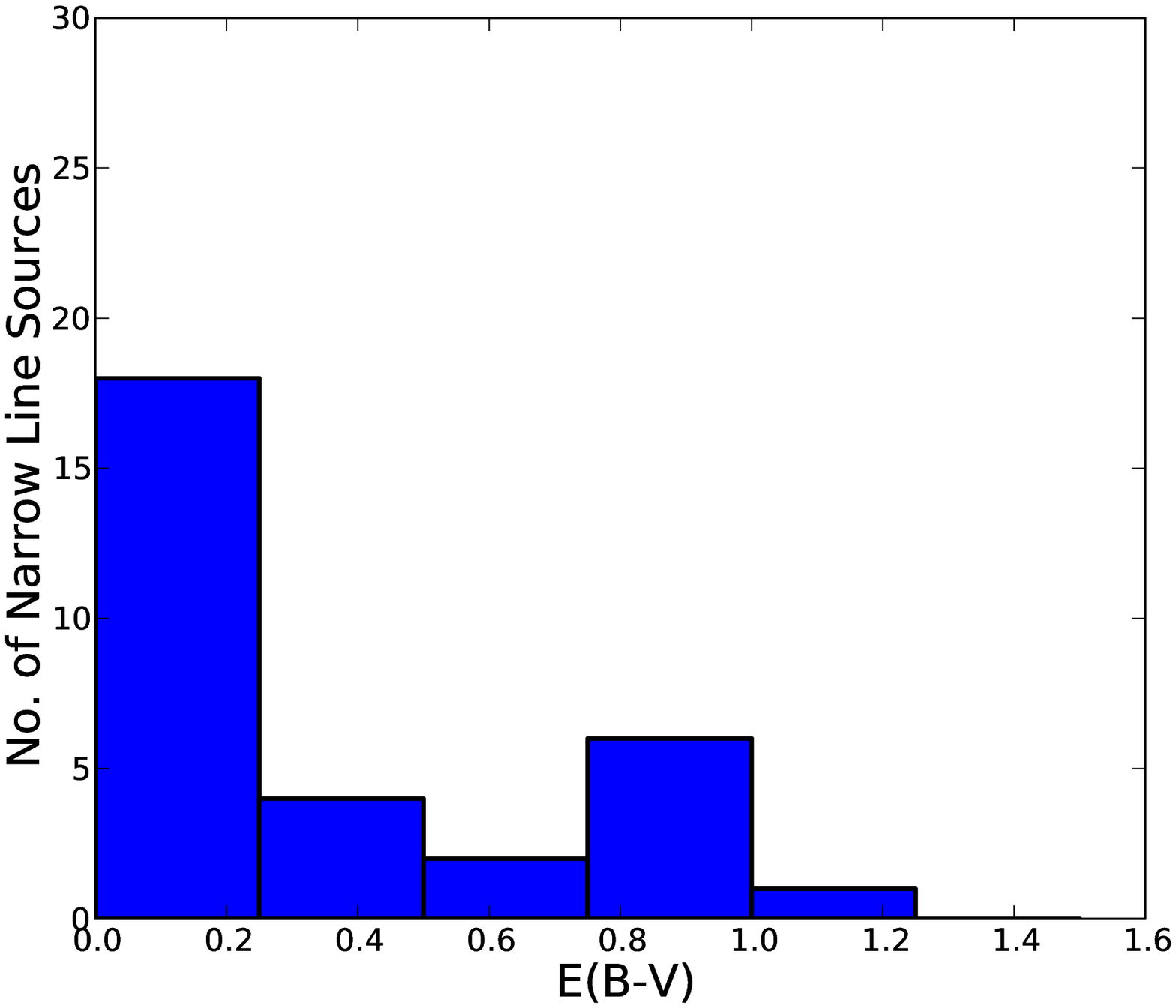}
\hspace{-1cm}
\includegraphics[height=8cm]{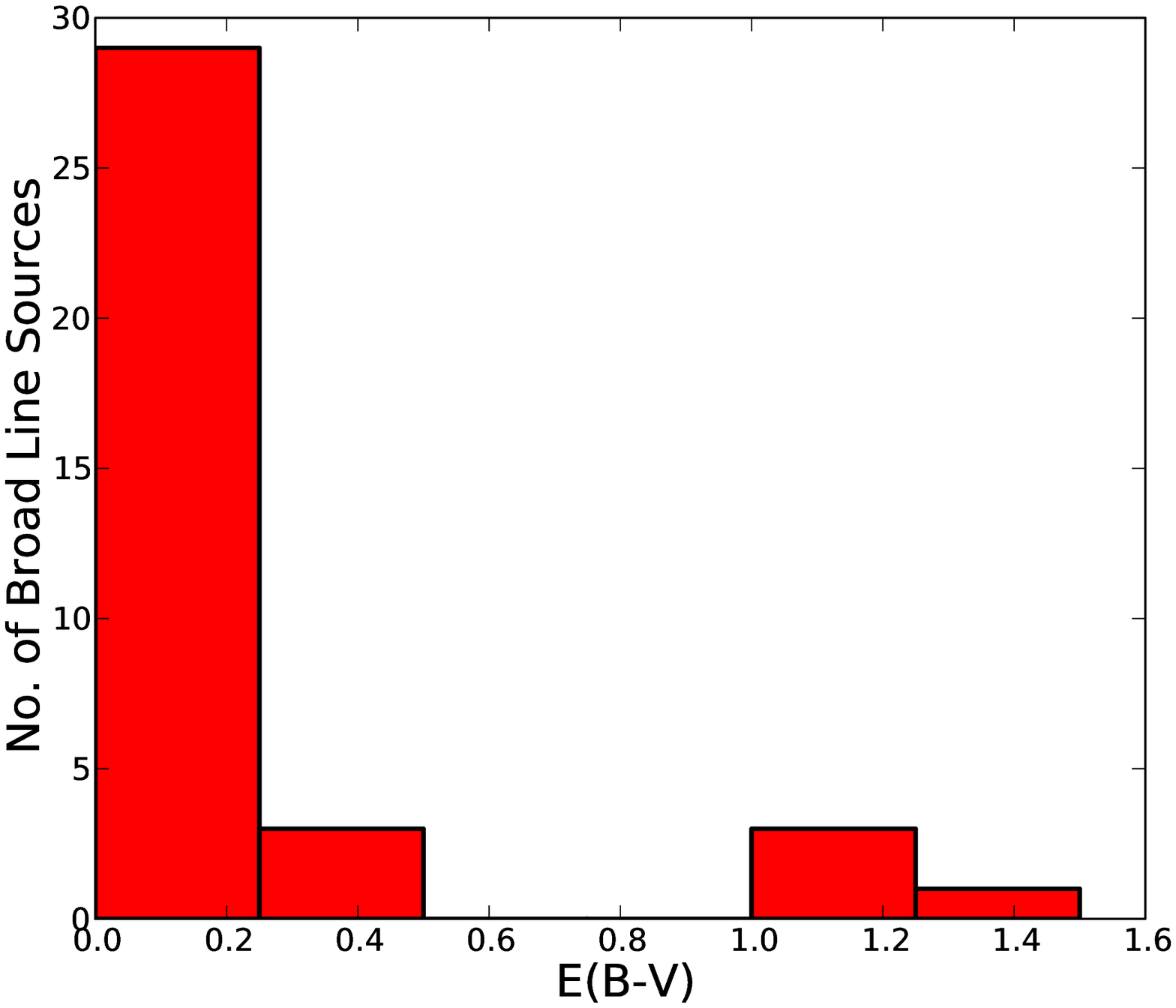}
\caption{Plotted are the distributions of E(B-V) for the narrow line (left) and broad line (right) sources.  The average value and standard deviation are much smaller for the broad line sources, as expected from the unified model.
\label{fig-ebv}}
\end{figure}

In Figure~\ref{fig-ebv}, we plot the distribution of $E({\rm B} - {\rm V})$ for the narrow and broad line sources.  Excluding the few outlying observations with measured values of $E({\rm B} - {\rm V}) > 1.0$, we find that the broad line sources have a lower average value than the narrow line sources and a smaller range of values.  We find the average value of $E({\rm B} - {\rm V}) = 0.08$ with a standard deviation of 0.11 for broad line sources and an average value of $E({\rm B} - {\rm V}) = 0.29$ with a standard deviation of 0.33 for narrow line sources.  The results of a Kolmogorov-Smirnov comparison test show that it is unlikely that the values are drawn from the same distribution with the maximum difference between the cumulative distributions ($D$) of 0.375 and a corresponding probability of 0.016.  This probability is less, but still low, when the outlying points are included ($D = 0.301$ and $P=0.067$).  Thus, the narrow lines in type 2 objects are more extincted.

\begin{figure}

\includegraphics[height=7cm]{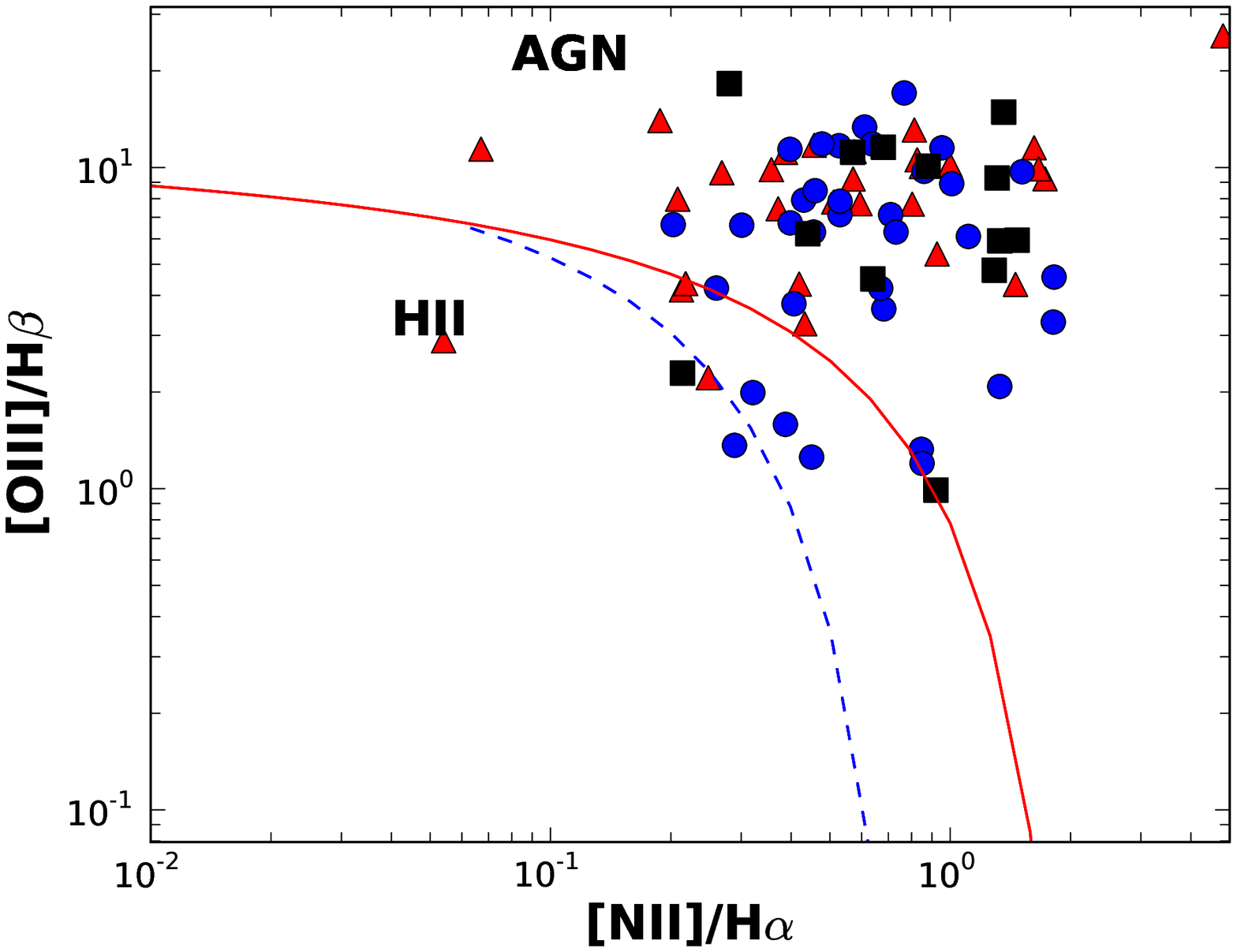}
\includegraphics[height=7cm]{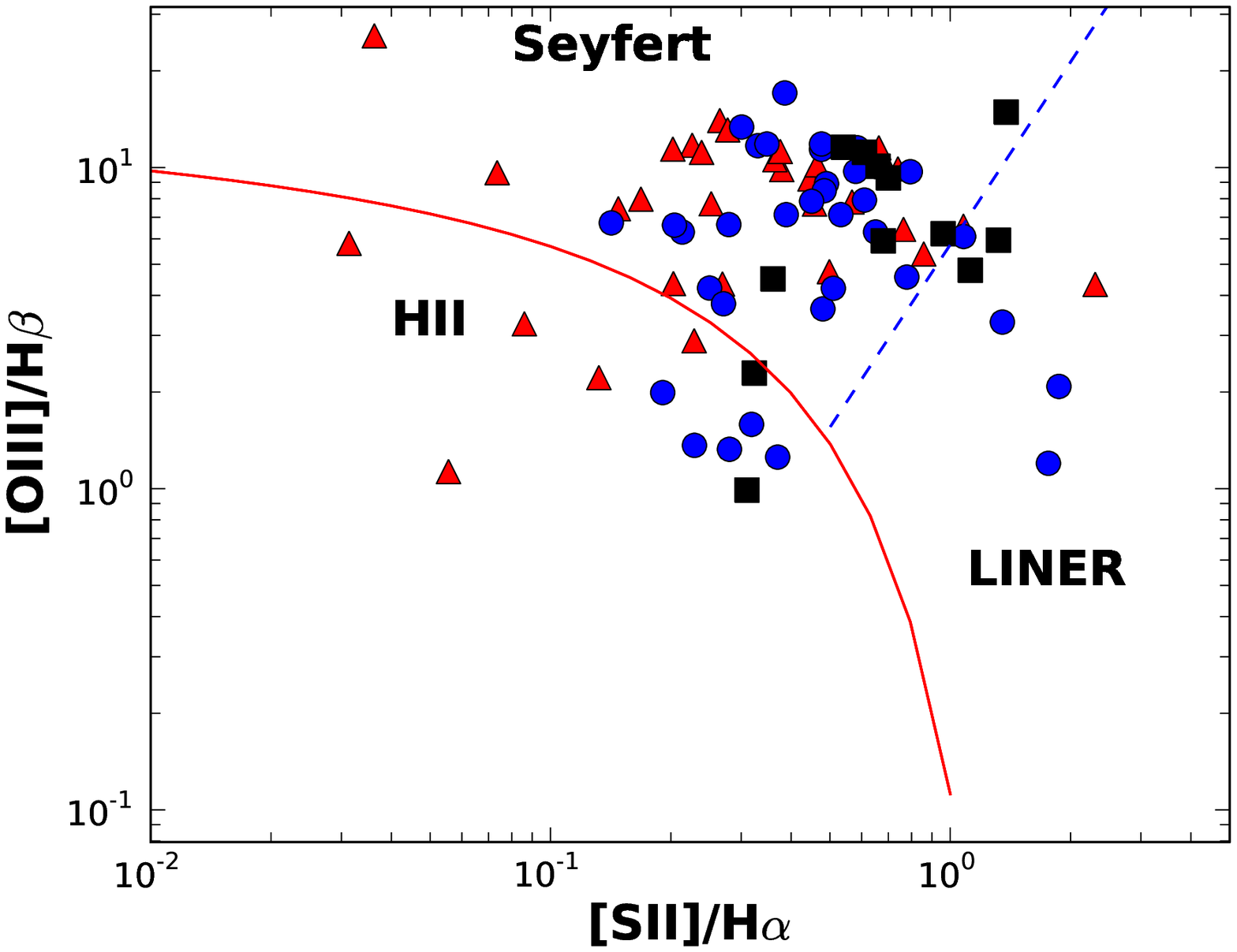}\\

\includegraphics[height=7cm]{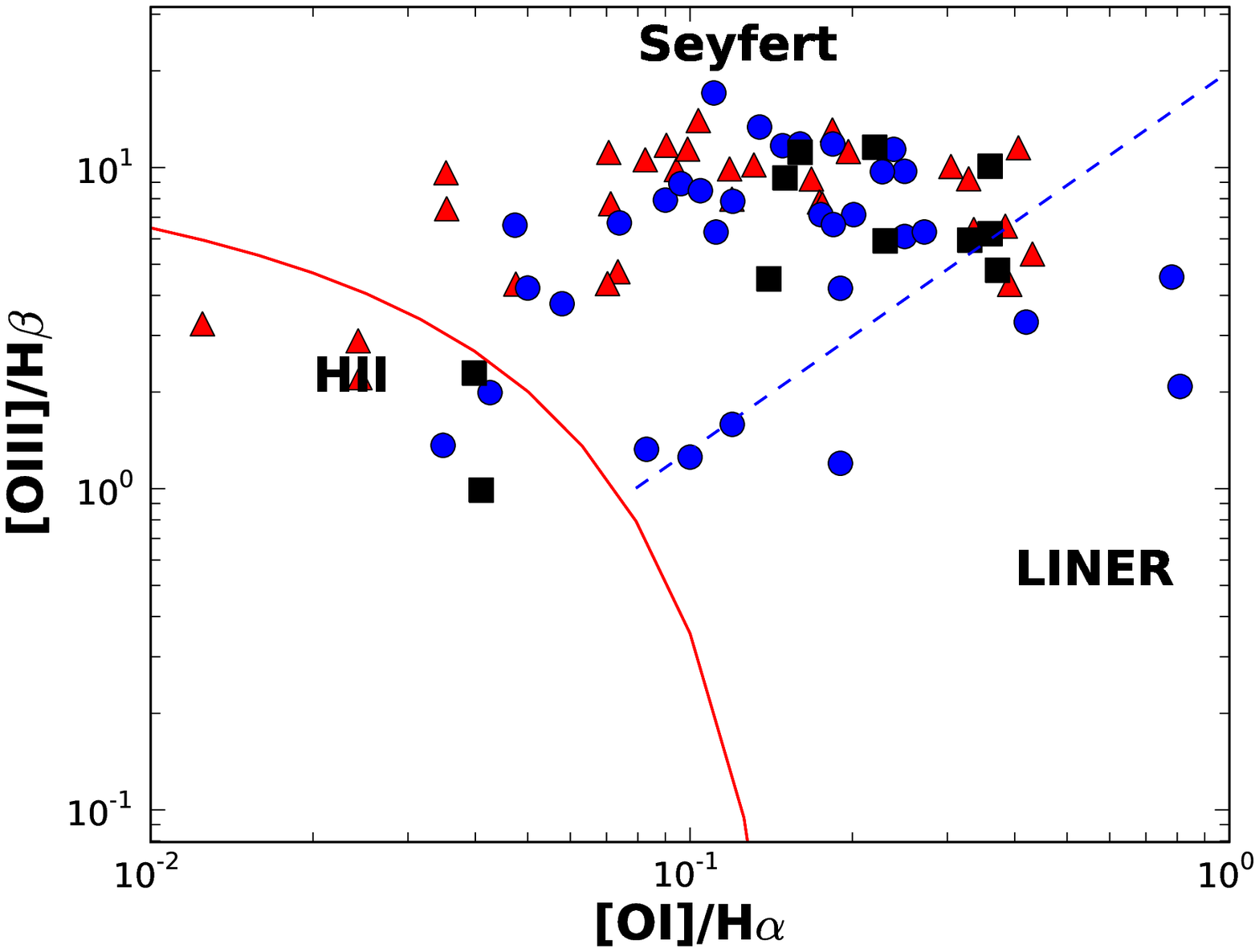}
\includegraphics[height=7cm]{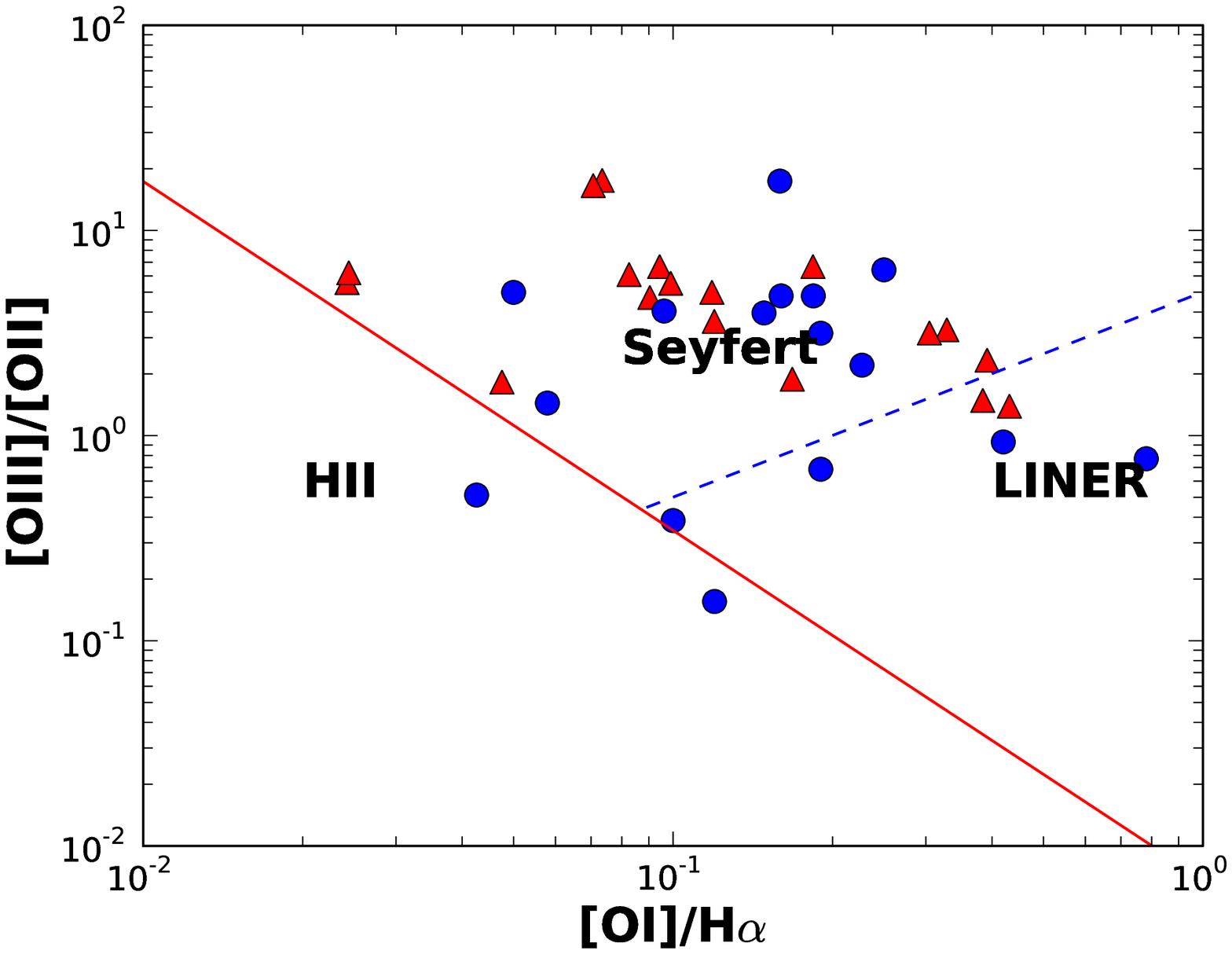}\\

\caption{Plotted are the emission line diagrams showing narrow line (circles) and broad line (triangles) sources from the KPNO or SDSS spectra that we have analyzed, as well as values extracted from the literature (square).  The diagnostic lines separating \ion{H}{2} galaxies from AGNs are shown in red \citep{2001ApJS..132...37K}.  In the [\ion{O}{3}]/H$\beta$ versus [\ion{N}{2}]/H$\alpha$ diagnostic plot, the dashed blue line represents the division between \ion{H}{2} galaxies and composites from \citet{2003MNRAS.346.1055K}.  The blue dashed lines on the remaining plots represent the division between Seyferts and LINERs from \citet{2006MNRAS.372..961K}.
\label{fig-emlinediagrams}}
\end{figure}

We classify our sources as \ion{H}{2} galaxies, composites (COMPs), Seyferts, or LINERs using the classification criteria based on the analysis of the emission line properties of 85224 SDSS galaxies presented in 
\citet{2006MNRAS.372..961K}.  These criteria include a theoretical 'maximum starburst line' from \citet{2001ApJS..132...37K}, shown as a solid line in the diagrams in Figure~\ref{fig-emlinediagrams}, which represents a boundary between \ion{H}{2} galaxies and AGNs.  Additionally, in the [\ion{O}{3}]/H$\beta$ vs. [\ion{N}{2}]/H$\alpha$ diagram, a dashed line shows the empirical division between pure star-forming galaxies and Seyfert-\ion{H}{2} composites from \citet{2003MNRAS.346.1055K}.  Finally, empirically derived divisions between LINERs and Seyferts, from \citet{2006MNRAS.372..961K}, are shown in the [\ion{O}{3}]/H$\beta$ vs. [\ion{S}{2}]/H$\alpha$, [\ion{O}{3}]/H$\beta$ vs. [\ion{O}{1}]/H$\alpha$, and [\ion{O}{3}]/[\ion{O}{2}] vs. [\ion{O}{1}]/H$\alpha$ diagnostic plots.  The emission line diagnostic plots are shown in Figure~\ref{fig-emlinediagrams} and the classifications are shown in Table~\ref{tbl-class}.

Based on these classifications of the narrow line sources (circles and a few squares [values from the literature] in Figure~\ref{fig-emlinediagrams}), 25 spectra are consistent with Seyferts, 1 spectrum corresponds to an \ion{H}{2} object, 5 spectra are consistent with LINERs, 1 is a composite, and 6 are ambiguous.  Among these, we classify the Ark 347 KPNO spectrum as a Seyfert and NGC 4992 as a LINER.  For each of these sources, the \citet{1987ApJS...63..295V} diagram including the [\ion{S}{2}]/H$\alpha$ ratio is the only diagram with a classification inconsistent with the other classification plots.  Errors in this measurement ([\ion{S}{2}]/H$\alpha$) could easily place the spectra within the Seyfert or LINER classification, respectively.  

The LINER sources include NGC 788,   NGC 2110, NGC 4992,  MCG+04-48-002, and NGC 7319.  Of these, NGC 4992 is classified as a possible X-ray bright optically normal galaxy (XBONG) by \citet{2006AA...459...21M}, and MCG+04-48-002 was previously classified as a starburst/\ion{H}{2} galaxy with a hidden Sy 2 nucleus \citep{2006AA...459...21M} (in their spectrum the [\ion{O}{1}] $\lambda 6300$ line was not detected).  All but one of the classified LINERs (NGC 4992) have ratios of H$\alpha$/H$\beta < 3.1$.  

The spectra classified as starburst/\ion{H}{2} galaxy and composite, respectively, are the SDSS spectrum of Mkn 18 and UGC 11871.
Finally, the 6 ambiguous sources include: 2 spectra with COMP/LINER properties (the KPNO spectrum of Mkn 18 and NGC 6240, a luminous infrared galaxy known to show contributions from both the AGN and starbursts \citep{1988ApJ...325...74S}), 2 spectra with Seyfert/\ion{H}{2} (both the KPNO and literature spectra of NGC 4102), and 2 spectra with Seyfert/LINER properties (NGC 1275 (which is in the middle of a strong emission nebulae associated with the cooling flow in the Perseus cluster) and NGC 4138).  In general, there is good agreement between classifications of sources with multiple spectra.  Both Mkn 417 and Ark 347 spectra indicate a Seyfert and the NGC 4102 spectra show an ambiguous source between Seyfert/\ion{H}{2}.  While the Mkn 18 classifications are not the same, they both point to having at least some \ion{H}{2}-like emission line ratios (particularly [\ion{S}{2}]/H$\alpha$).

While it is clear that broad line sources are Seyfert 1s, it is of interest to examine how they would be classified based on their narrow line ratios.  If the predictions of the unified model are true then, if the broad line region is absorbed out, the narrow line ratios should classify these objects as Seyferts also.  We find, much to our surprise, that a significant fraction of the broad line objects have narrow line ratios which lie outside the AGN region based on the \citet{2006MNRAS.372..961K} classifications.  While the majority (75\%) of broad line sources have narrow line ratios consistent with classification as Seyferts (30 spectra), some (in particular NGC 931, 1RXSJ193347.6+325422, UGC 6728, and IGR21247+5058) are not, being classified as composites or \ion{H}{2} sources, though H$\alpha$ and the [\ion{N}{2}] lines were too heavily blended to separate for the latter two.  Additionally, 7 spectra (including the KPNO and SDSS spectra of MCG +04--22--042) have ambiguous classifications.  There is good agreement between classifications of sources with multiple spectra (i.e. MCG +04--22--042, NGC 4151, NGC 3227, NGC 3516).  The source NGC 4051, classified as ambiguous from the KPNO spectrum due to the [\ion{N}{2}]/H$\alpha$ diagram result showing a COMP, should more likely be classified as a Seyfert (as in the spectrum analyzed in the literature).

Therefore, the Swift BAT AGN optical classifications are mostly Seyferts.  There are a total of 29 individual narrow line sources represented, and of these, about 66\% are Seyferts, 16\% LINERs, 13\% ambiguous, 3\% composites, and 3\% \ion{H}{2} galaxies.  Of the 35 broad line sources, about 75\% are Seyferts, 14\% are ambiguous, and 11\% are composites or \ion{H}{2} galaxies.  We find no broad line sources with narrow emission consistent with LINERs.

Since we are studying in this paper the optical properties of a hard X-ray detected sample, it is useful to make a comparison with optically selected samples, in particular the recent results of the SDSS.  In this comparison, we find that the optically selected emission-line sources from the 85224 SDSS galaxy sample of \citet{2006MNRAS.372..961K} consist of very different percentages of the various classification categories than our hard X-ray selected sample.  The SDSS sample consists of 75\% star-forming/\ion{H}{2} galaxies, 3\% Seyferts, 7\% LINERs, 7\% composites, and 8\% ambiguous.  It is no surprise, that the majority of our 14--195\,keV X-ray sample consists of the much more energetic (across multiple bands) Seyferts.  However, comparing the SDSS results solely with our narrow line sources, we are finding far fewer LINERs than we might expect.  In the optically selected SDSS sample, the LINER class contains more than twice the number of sources as Seyferts, while we are finding four-times as many Seyferts as LINERs among the narrow line sources.

There are a few possibilities as to why the hard X-ray sample selects fewer LINERs.  The most obvious reason could be that LINERs are less luminous X-ray sources (we discuss this further in \S~\ref{lum}).  Indeed, \citet{2006MNRAS.372..961K} did find that LINERs had substantially lower reddening corrected [\ion{O}{3}] 5007\AA~luminosities than Seyferts.  If L$_{[OIII]}$ is an indicator of bolometric luminosity and scales with the Swift BAT luminosity, we may simply not be detecting many LINERs with BAT because their X-ray fluxes are below the current detection threshold.  Further, studies such as the Chandra snapshot analysis of \citet{2003ApJ...583..145T} also find LINERs as less luminous than Seyferts in X-rays.  Also, based on the nuclear X-ray luminosities of local LINER sources determined from the Chandra analysis of  \citet{2006ApJ...647..140F} who used the IR-selected LINER sample of \citet{2006ApJ...653L..13S}, the typical local LINER would have a BAT flux  far below the flux detection limit of the Swift survey.

It is also possible that LINERs are more absorbed in the X-rays.  In \citet{2009ApJ...690.1322W}, we have shown that the more X-ray absorbed (i.e. highest neutral hydrogen column density) sources have lower X-ray luminosities, on average.  If this is the case, we would expect to find a higher number of LINERs as Swift BAT detects more heavily absorbed and less luminous sources.  In support of this possibility, the average value of the X-ray derived N$_{\rm H} = 6 \times 10^{23}$\,cm$^{-2}$ of our LINERs is high \citep{2009ApJ...690.1322W}.  This is in contrast, however, to the optical reddening, where we noted that the ratio H$\alpha$/H$\beta$ is below the accepted value for AGN (3.1) and the theoretically expected value for case B recombination (2.85) for most of our LINERs.  \citet{2006MNRAS.372..961K} also found this in 45\% of their LINER sample, which could be the result of a higher nebular temperature 
\citep{oster89} or shocks.  In these cases, it is unclear how to relate the optical Balmer decrement to the X-ray derived column density.

With lower luminosities than typical AGN sources and emission line ratios potentially indicating a shock origin, it is possible that LINERs are typically not powered by accretion processes.  As  \citet{2006ApJ...647..140F} show, many LINERs do not have any detected X-ray emission.  Further, recent work by the SAURON team \citep{2009arXiv0912.0275S}
 and SEAGAL collaboration \citep{2009arXiv0902.0523C}
 indicates that the majority of LINERs are not powered by AGN but instead by evolved stellar populations.  Therefore, we would expect to detect few LINERs in the Swift BAT band.

\section{Additional Diagnostic Lines}
Comparisons of the intensities of multiple emission lines from the same ion provide important diagnostics of the gas in which they are produced.  In the optical range probed by our spectra, the relative population and therefore intensity of [\ion{S}{2}] $\lambda 6716/\lambda 6731$ depends on the density of the gas (with only a slight dependence on temperature of the order $T_e^{1/2}$).  The [\ion{O}{3}] $\lambda 4363$ emission line comes from a different upper energy level than the $\lambda 4959$ and $\lambda 5007$ lines, where the relative rates of excitation to these upper levels is strongly dependent on temperature.  An equation relating the ratio of the [\ion{O}{3}] lines to temperature and density is given in  \citep{oster89} as:
\begin{equation}
\frac{I_{\lambda 4959} + I_{\lambda 5007}}{I_{\lambda 4363}} = \frac{7.73 \exp((3.29 \times 10^4)/T}{1 + 4.5 \times 10^{-4} (N_e/T^{1/2})}.
\end{equation}

\begin{figure}
\centering
\includegraphics[height=8cm]{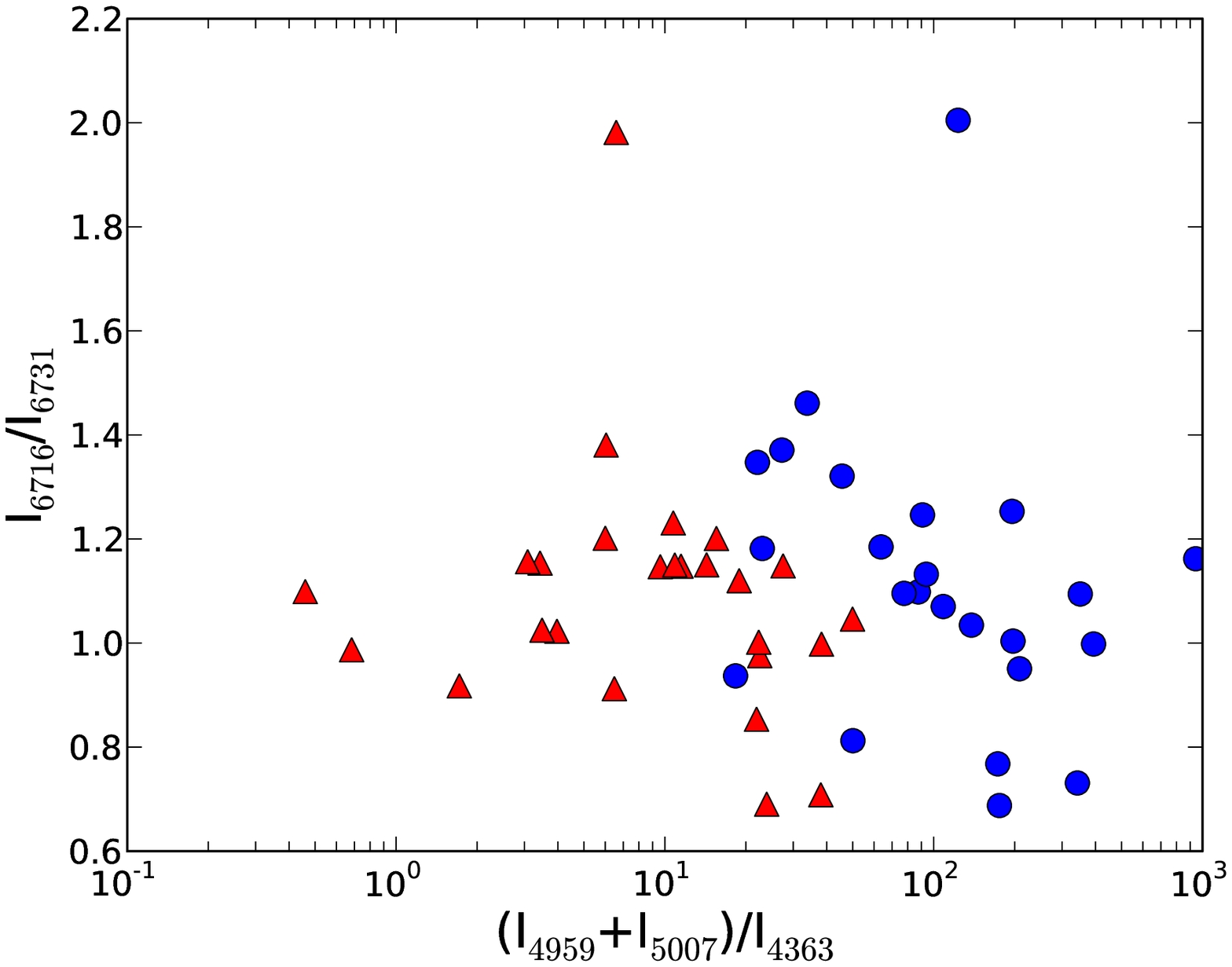}\\
\includegraphics[height=9cm]{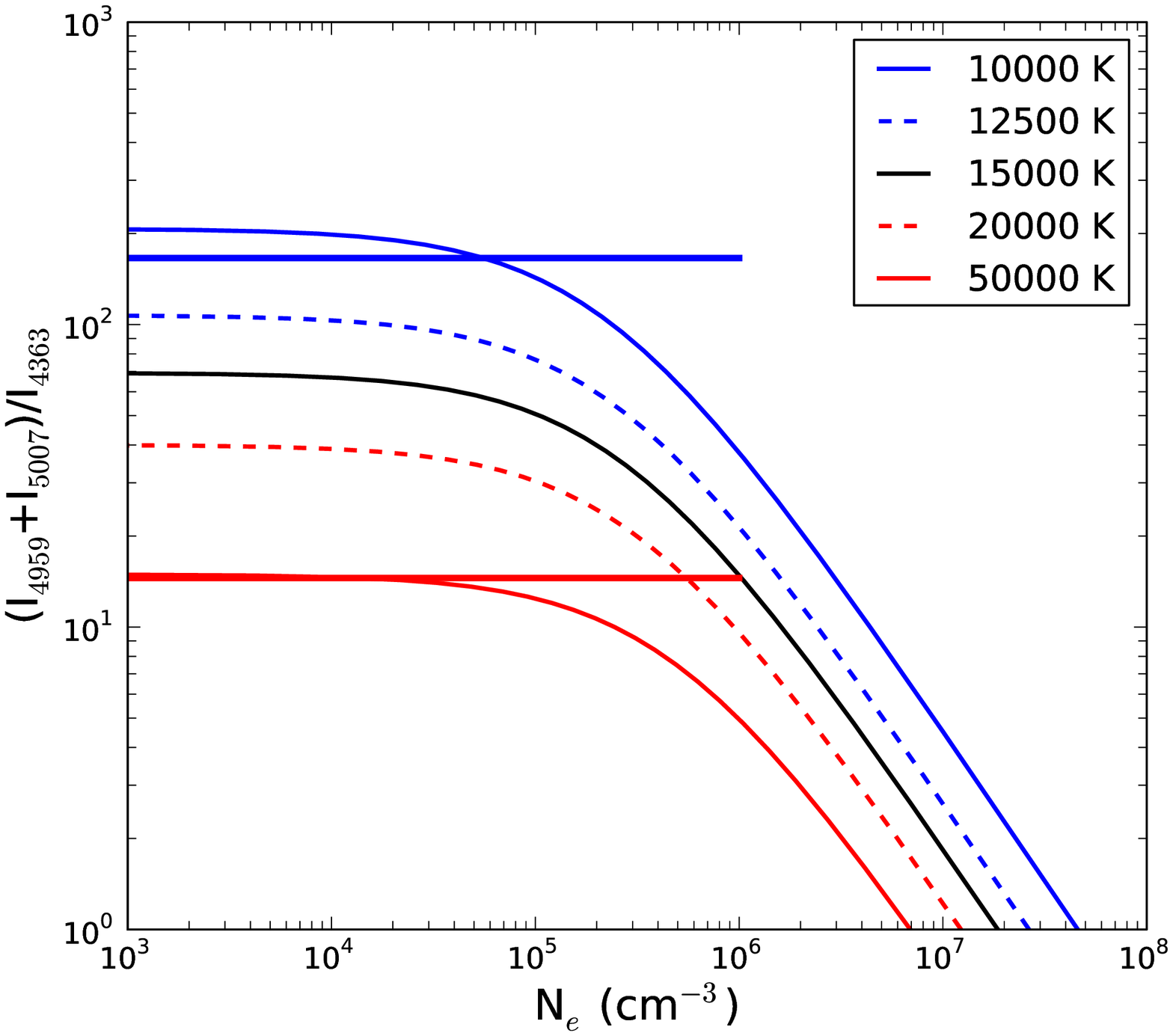}
\caption{Plotted at the top is a comparison of the ratio of intensities of [\ion{S}{2}] $\lambda 6716/\lambda 6731$ versus the ratio of [\ion{O}{3}] $(\lambda 4959 + \lambda 5007)/\lambda 4363$ for the narrow line (circle) and broad line (triangle) sources. A K-S test shows that the ratios of [\ion{S}{2}] (an indicator of density) are likely from the same population, while the [\ion{O}{3}]-temperature diagnostic is not.  In the bottom plot, we show the average diagnostic value for the narrow (horizontal blue line) and broad (horizontal red line) line sources versus density for values of constant temperature.  This diagnostic points to a much higher temperature for the broad line sources, if the densities are low.  If the densities are high in this $\rm{O}^{+2}$ emission region for broad line sources ($10^6  {\rm cm}^{-3} \la N_e \la 10^7 {\rm cm}^{-3}$) and low ($N_e \la 10^4 {\rm cm}^{-3}$) for narrow line sources, the temperatures are similar.
\label{fig-s2o3}}
\end{figure}

In Figure~\ref{fig-s2o3}, we plot the reddening corrected flux ratios for both of these diagnostics ([\ion{S}{2}] and [\ion{O}{3}]).  While both intensity ratios do not necessarily probe the same regions of the narrow line region, this figure is useful in illustrating the range of values measured for our sample.  One of the results of our analysis is that the ratio of [\ion{S}{2}] $\lambda 6716/\lambda 6731$ is similar for both the broad and narrow line sources.  Using a Kolmogorov-Smirnov comparison test, we find that both distributions are likely to be drawn from the same population with $D = 0.22$ and $P = 0.50$.  The average and standard deviations of these values are 1.12 and 0.27 for the narrow line sources and 1.09 and 0.23 for the broad line sources.  These values of the ratio of [\ion{S}{2}] $\lambda 6716/\lambda 6731$ correspond to electron densities of $N_e \approx 10^3$\,cm$^{-3}$ (assuming $T_e = 10^4$\,K as in figure 6.2 of \citet{peterson-97}).  These results are consistent with average narrow line region densities of 2000\,cm$^{-3}$ found by \citet{1978ApJ...223...56K}.  Thus, the hard X-ray detected Swift BAT AGN have the same densities as optically selected AGN in this region (which produces the [\ion{S}{2}] emission), regardless of whether broad lines are present.

The temperature sensitive diagnostic [\ion{O}{3}] $(\lambda 4959 + \lambda 5007)/\lambda 4363$ clearly is not the same for the narrow and broad line sources.  The Kolmogorov-Smirnov comparison test yields a P-value of 0.000.  The average and standard deviation of [\ion{O}{3}] $(\lambda 4959 + \lambda 5007)/\lambda 4363$ is 166.0 and 193.0 for the narrow line sources and 14.53 and 12.71 for the broad line sources.  To better illustrate what these values mean, in Figure~\ref{fig-s2o3} we also plot the relationship of the [\ion{O}{3}] temperature diagnostic versus electron density for fixed temperatures.  The average values of both the narrow and broad line sources are indicated with a horizontal line.  In the low density limit ($N_e < 10^4$cm$^{-3}$), the average temperature of the [\ion{O}{3}] emitting gas is approximately 10000\,K for narrow line sources and 50000\,K for broad line sources.  Typical temperatures for narrow line regions are between 10000--25000\,K, with an average value of 16000\,K reported in \citet{1978ApJ...223...56K}.  

If the temperature of the narrow line region in the type 1s and 2s is different, this would be a violation of the unified model.  However, if the densities are different, this might be due to geometrical effects wherein the dense regions in type 2s are blocked from our view or have very high reddening values.  However, there is uncertainty in the measurement of [\ion{O}{3}] $(\lambda 4959 + \lambda 5007)/\lambda 4363$ associated specifically with the measurement of the faint [\ion{O}{3}]\,$\lambda 4363$\AA~line, which is just 1\% of the bright $\lambda 4959$\AA~and $\lambda 5007$\AA~lines.  We note that it is particularly hard to measure this line in the broad line sources where H$\gamma \lambda 4340$\AA~may be producing a tilted pseudo-continuum.

The result of broad line sources having lower values of [\ion{O}{3}] $(\lambda 4959 + \lambda 5007)/\lambda 4363$ than narrow line sources has been noted before and is attributed to broad line sources having stronger $\lambda 4363$ emission \citep{1978PhyS...17..285O}.  Instead of a higher temperature in the narrow line regions of broad line sources, \citet{1978PhyS...17..285O} suggests densities of $10^6$--$10^7$cm$^{-3}$ in broad line sources and $\la 10^5$\,cm$^{-3}$ in narrow line sources.  To reconcile these high densities with lower densities derived in the   S$^+$ emission region, the narrow line region must consist of a range of densities, among which low densities are found in low-ionization regions.  Under this interpretation, the temperatures of the narrow line region producing O$^{+2}$ are the same for broad and narrow line sources, provided the densities differ in this higher ionization region.     

\section{[\ion{O}{3}] and Hard X-ray Luminosities}\label{lum}

\begin{figure}
\hspace{-1cm}
\includegraphics[height=8cm]{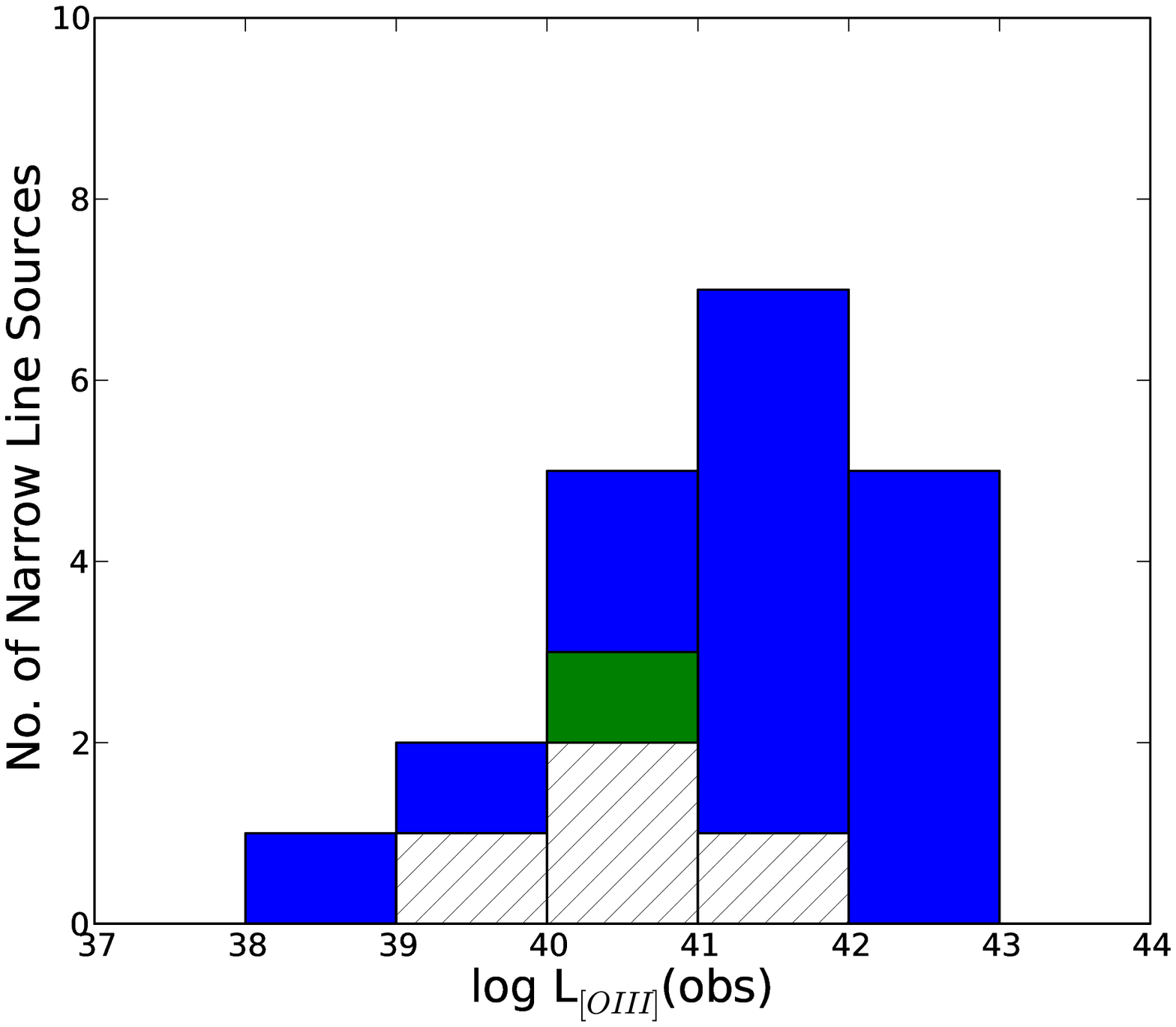}
\hspace{-1cm}
\includegraphics[height=8cm]{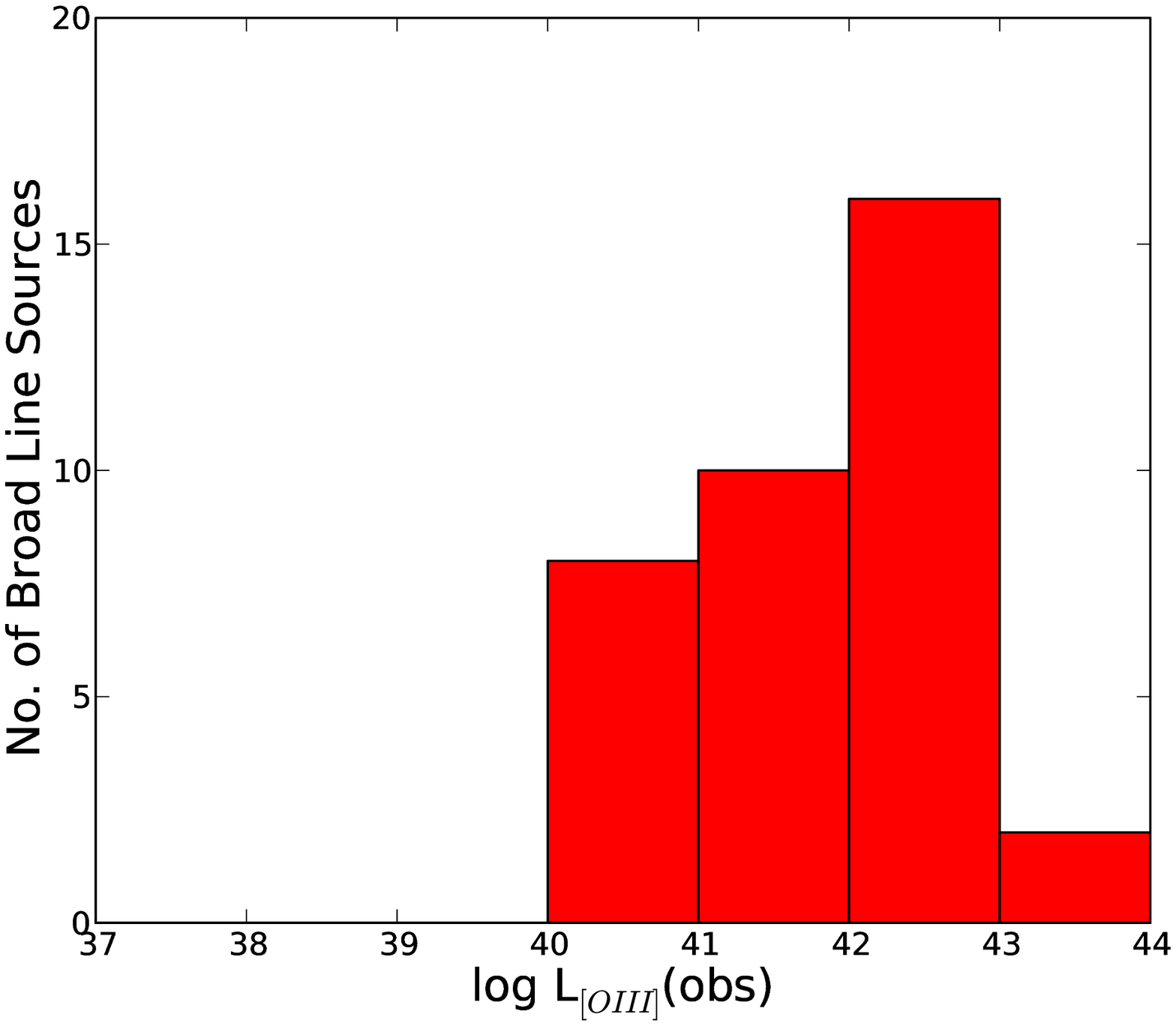} \\
\vspace{-0.5cm}
\hspace{-1cm}
\includegraphics[height=8cm]{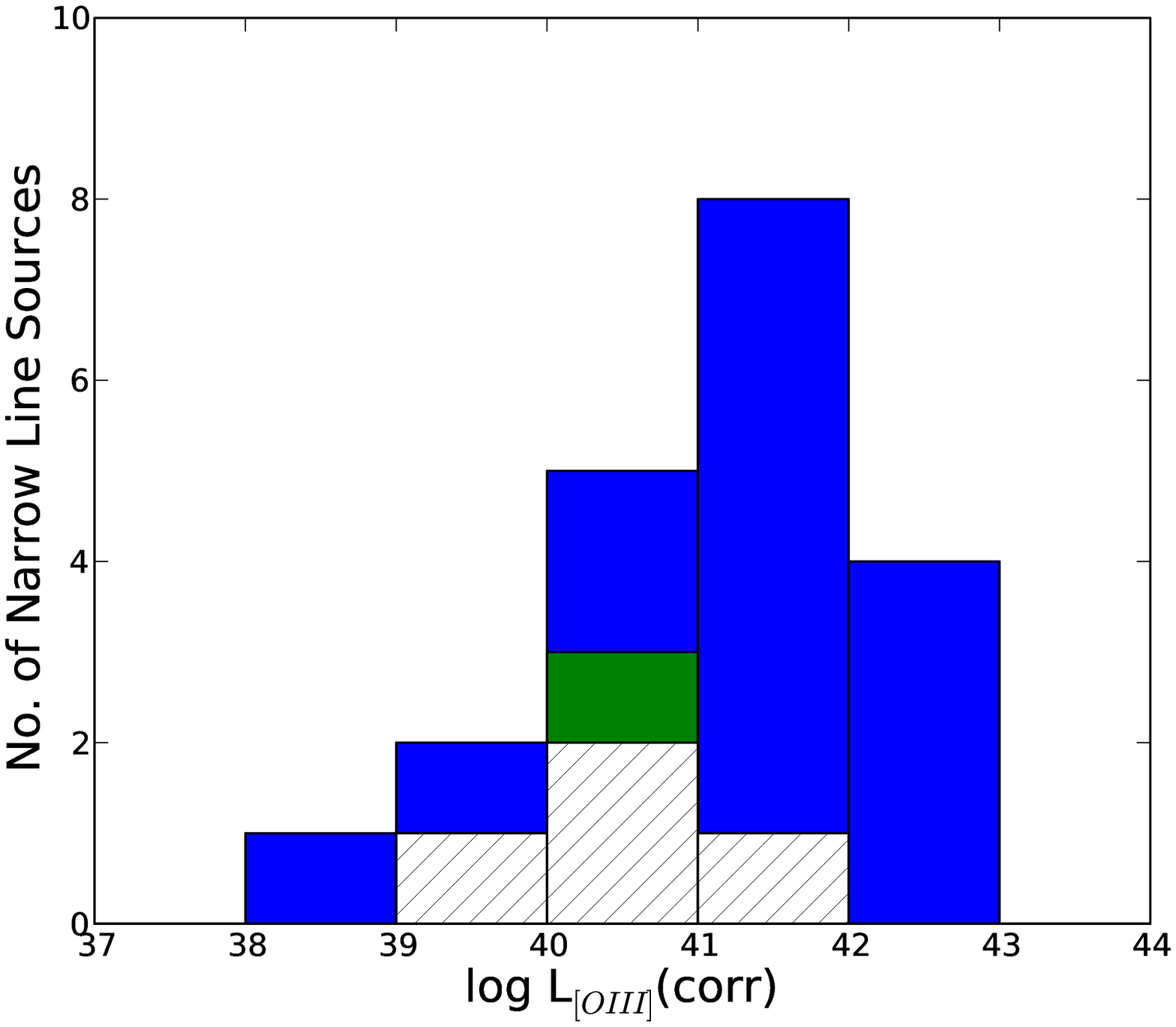}
\hspace{-1cm}
\includegraphics[height=8cm]{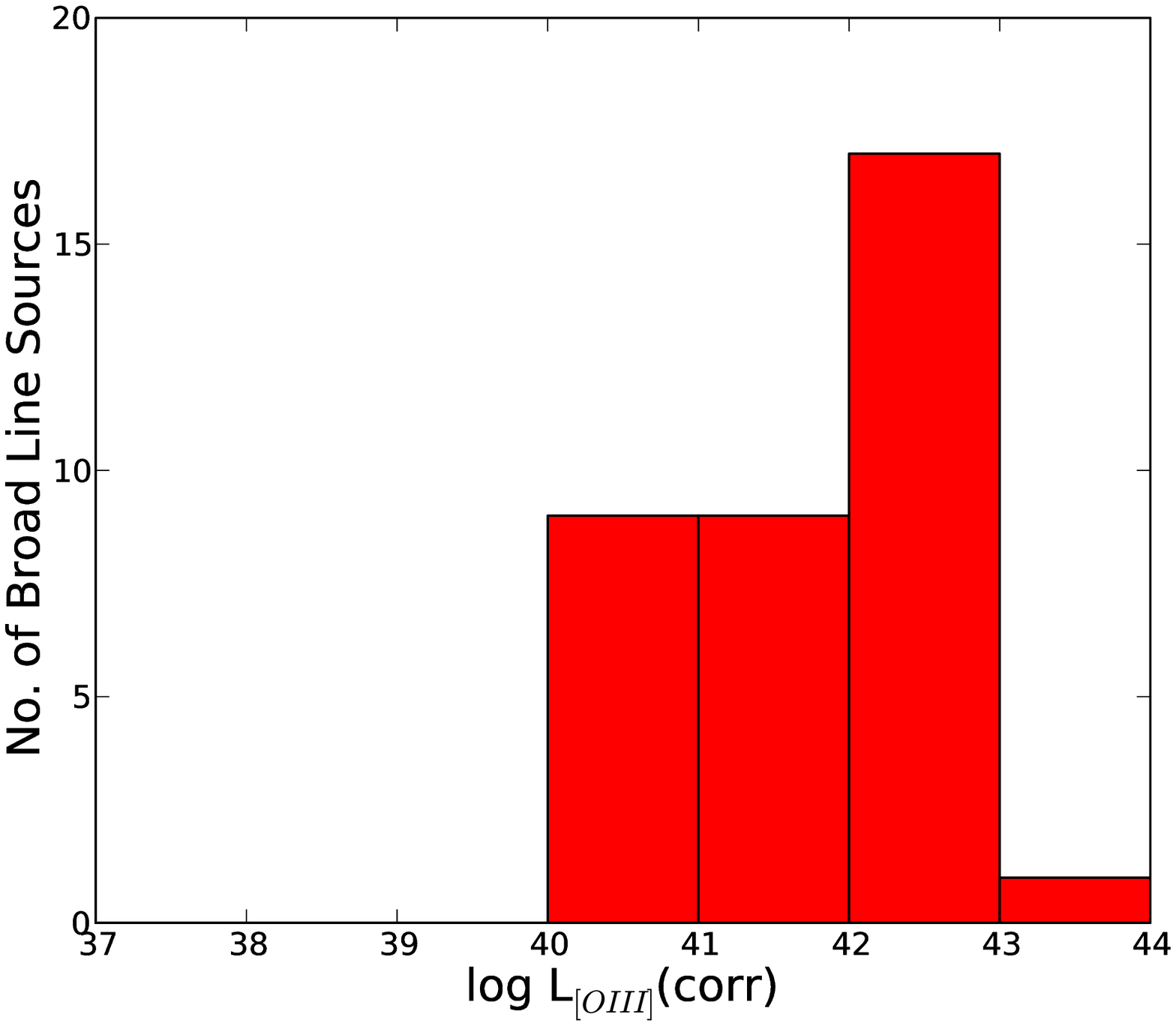}
\caption{Plotted are histograms of the [\ion{O}{3}] 5007\AA~emission line luminosity for the narrow line (Seyferts = blue, LINERs = green, others = hatched) and broad line (red) sources, showing both the observed (obs, top plots) and extiction-corrected (corr, bottom plots) luminosities.  The broad line sources are more luminous on average than the narrow line sources, in both the observed and extinction-corrected luminosities.  The mean value for the extinction-corrected luminosity distribution of broad line sources is $\log {\rm L}_{[OIII]} = 41.79$, while the narrow line sources have a mean value of $\log {\rm L}_{[OIII]} = 40.82$. 
\label{fig-lo3}}
\end{figure}

A fundamental property of an AGN is its power, measured through luminosity.  In Figure~\ref{fig-lo3}, we plot the distributions of both the observed and extinction-corrected [\ion{O}{3}] 5007~\AA~ luminosities for both our narrow line and broad line sources.  For sources with multiple measurements, we averaged the values together to obtain a single measurement of observed and extinction corrected luminosity per source (these values are included in Table~\ref{tbl-masses}).  We find that the extinction corrections do not significantly change the luminosity measurements, with the corrected values being on average 1.1 (broad line sources) and 1.3 (narrow line sources) times larger than the observed luminosities.

The mean value for the distribution of extinction corrected luminosity for the broad line sources is $\log {\rm L}_{[OIII]} = 41.79$ with a standard deviation of 0.90, while the narrow line sources have a mean value of $\log {\rm L}_{[OIII]} = 40.82$ with a standard deviation of 1.16.  The results of a Kolmogrov-Smirnov comparison test suggest that these values are not drawn from a single population ($D = 0.49$ and $P = 0.001$).  Therefore, the broad line sources appear to be more luminous than the narrow line sources, on average.  This is also true of the observed luminosities (the averages and standard deviations are 41.76, 0.79 (broad line sources) and 40.87, 1.08 (narrow line sources)) and therefore not an effect of incorrect reddening corrections.  If the [\ion{O}{3}] 5007\AA~ emission line is indeed an estimator of the AGN power (assuming that the contamination from star formation is not great), these results agree with our X-ray results for the BAT AGNs.  Namely,  \citet{2009ApJ...690.1322W} showed that the unobscured  X-ray sources (presumably optical broad line sources) in the sample were also intrinsically more luminous.  

In \S~\ref{class}, we described that previous optical and X-ray studies find LINERs as less luminous than Seyferts.  Comparing the extinction-corrected [\ion{O}{3}] luminosities for the narrow line sources, we confirm these results with our unbiased hard X-ray detected sample.  We find Seyferts have an average value of $\log {\rm L}_{[OIII]} = 41.55$ with a standard deviation of 0.85, LINERs have 
an average value of $\log {\rm L}_{[OIII]} = 40.73$ with a standard deviation of 0.60, and sources in other categories (including ambiguous classifications, \ion{H}{2} galaxies, and composites) have an average value of $\log {\rm L}_{[OIII]} = 40.33$ with a standard deviation of 0.65.  Of particular importance, we find that the narrow line Seyferts have luminosities consistent with those of broad line sources.

\begin{figure}
\hspace{-1.2cm}
\includegraphics[height=8cm]{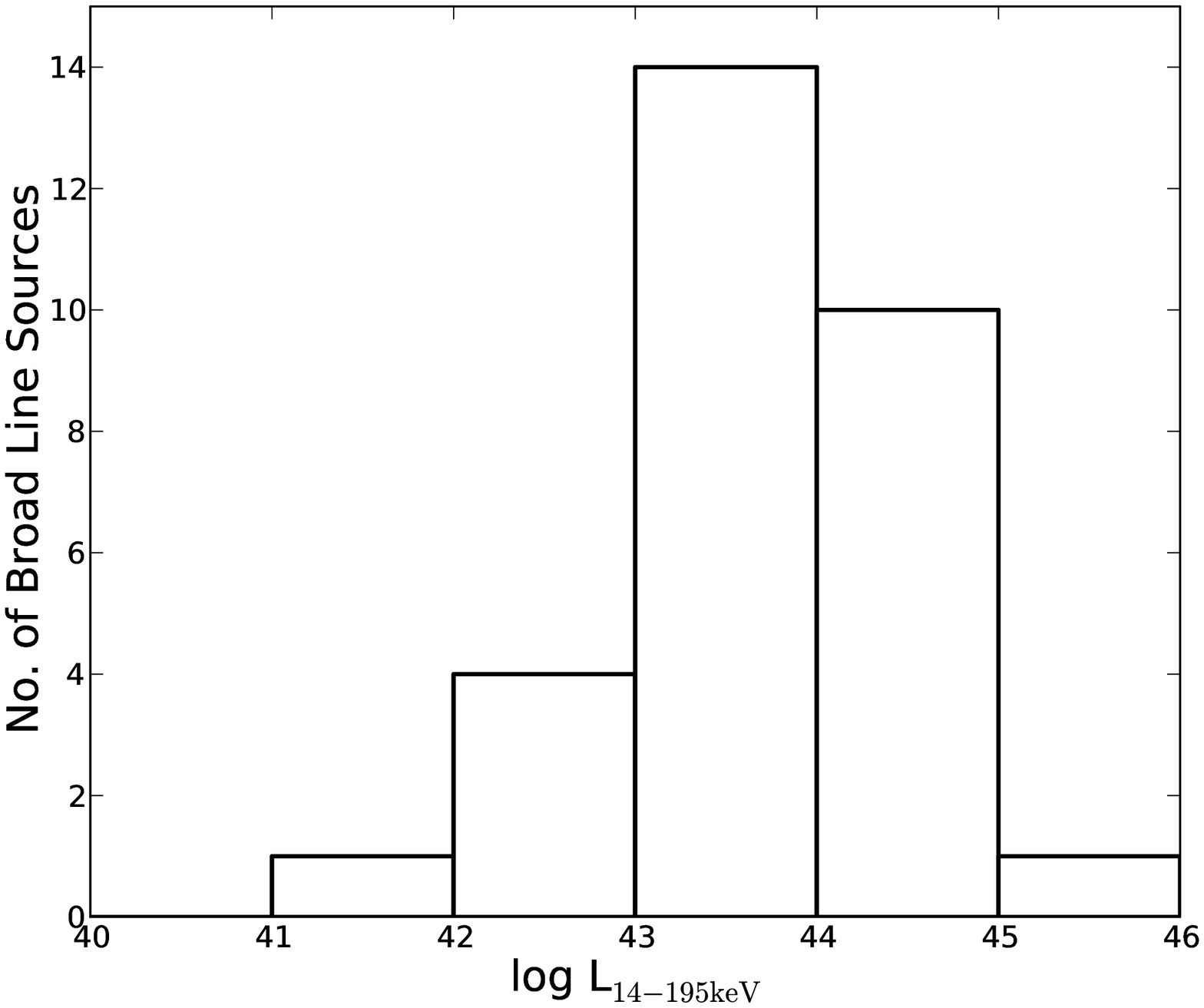}
\hspace{-1.0cm}
\includegraphics[height=8cm]{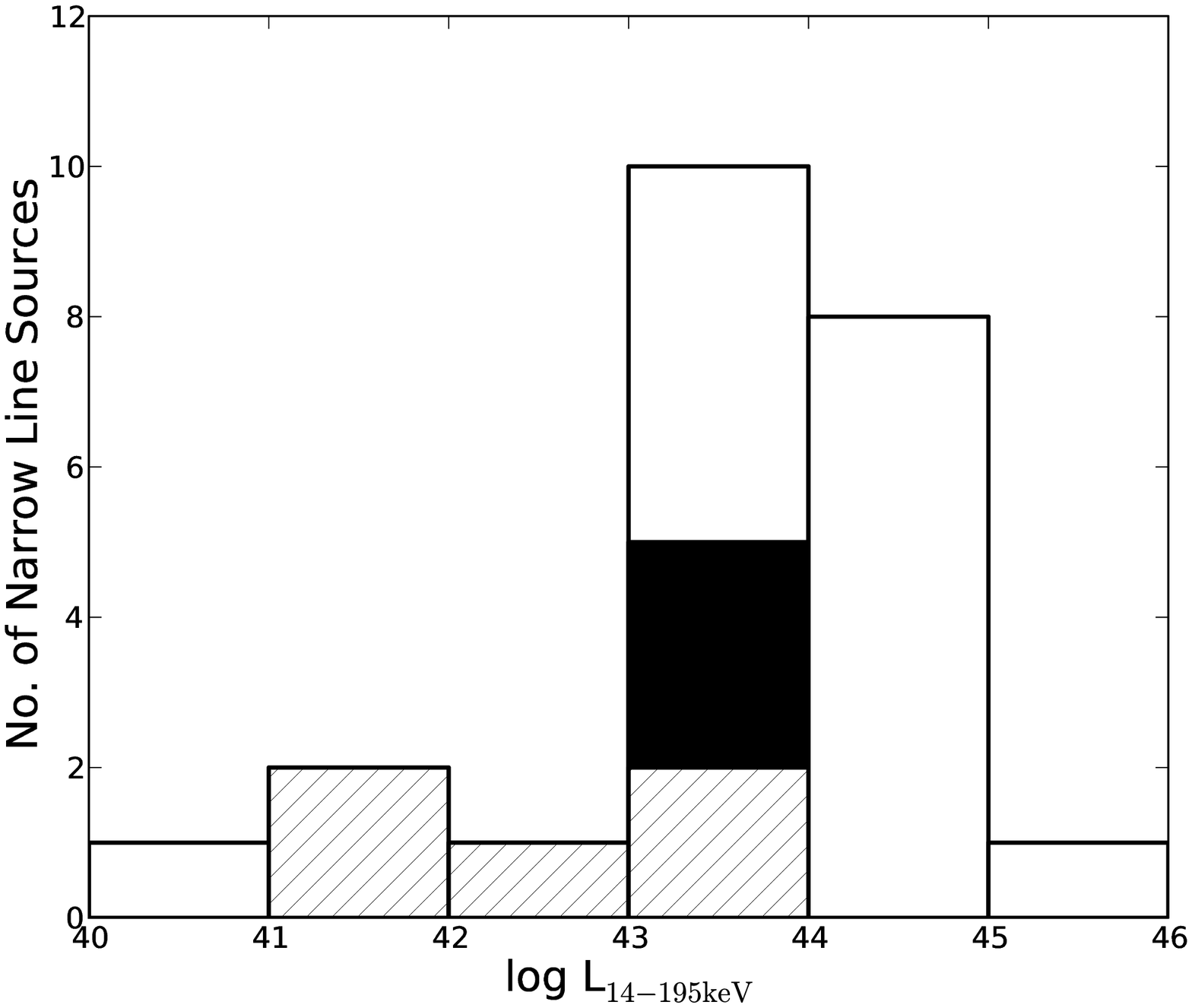}
\caption{Plotted are the distributions of the 14--195\,keV luminosity for broad line (left) and narrow line (right) sources.  The narrow line classifications are represented as Seyferts (white), LINERs (black), and either ambiguous/\ion{H}{2} galaxies/composites (hatched).  The Seyferts are the most luminous of the narrow line sources, both in the hard X-ray band and the optical (indicated by the [OIII] 5007\AA~luminosities).  While the broad line sources have an average X-ray luminosity higher than the narrow line sources, the distribution of luminosities for narrow line Seyferts is the same as for the broad line sources as a whole.
\label{fig-lumxray}}
\end{figure}

Further, we find that the hard X-ray luminosities (in the 14--195\,keV band) show these same trends.  To illustrate these results, we plot the distribution of hard X-ray luminosity for our sources in Figure~\ref{fig-lumxray}.  For the narrow line sources, we find that the Seyferts have an average value of $\log {\rm L}_{14-195 {\rm keV}} = 43.87$ with a standard deviation of 0.94, LINERs have 
an average value of $\log {\rm L}_{14-195 {\rm keV}} = 43.50$ with a standard deviation of 0.16, and sources in other categories have an average value of $\log {\rm L}_{14-195 {\rm keV}} = 42.69$ with a standard deviation of 0.98.  Once again, the \ion{H}{2}/composites/ambiguous sources have the lowest luminosities while the Seyferts are most luminous.  Also, the X-ray luminosities of the narrow line Seyferts are consistent with those of the broad line sources (which have an average value of $\log {\rm L}_{14-195 {\rm keV}} = 43.74$ with a standard deviation of 0.74).

Based on X-ray surveys, several studies had found the fraction of obscured sources to increase at lower 2--10\,keV luminosities, including those by  \citet{2003ApJ...598..886U} and \citet{2003ApJ...596L..23S}, as well as our own study of the Swift sources  \citep{2009ApJ...690.1322W}.  Based on an optically selected sample, \citet{2009ApJ...698..623D} also found differences in the distributions of 2--10\,keV and [\ion{O}{3}] $\lambda 5007$\AA~luminosities for Sy 1s and Sy 2s in the revised Shapley-Ames sample \citep{1987rsac.book.....S}.  A clear explanation for the differences in the X-ray selected samples is that the lowest luminosity X-ray sources, which tend to be absorbed sources, are not optical Seyferts, as found in our current study.  In an optically defined sample, we would expect both the obscured and unobscured Seyferts to have the same luminosity distributions.  However, in this same optically-selected sample \citet{2009ApJ...698..623D} find that the [\ion{O}{4}] $\lambda 25.89$\,$\mu$m line, an indicator of bolometric luminosity \citep{2008ApJ...682...94M}, does not show this difference in distributions between Sy\,1s and Sy\,2s.  It is unclear how to interpret these results.  Since the study of \citet{2009ApJ...698..623D} consists of only sources for which multi-wavelength luminosity measurements are available (18 Seyfert 1s and 71 Seyfert 2s), it is potentially biased (considering that X-ray surveys find equal numbers of absorbed and unabsorbed sources) compared to the Swift sample.  However, the \citet{2009ApJ...698..623D} sample also includes a high percentage of Compton-thick sources (20\%), which the Swift sample is currently not finding (due to the low X-ray flux in the BAT band of Compton thick sources).

Since the hard X-ray luminosities are at high enough energies to cut through much of the gas and dust around the AGN, they are a good estimate of the bolometric luminosity.  In lieu of these measurements, the optical [\ion{O}{3}] luminosities are often used as a measurement of the AGN total power.  Further, in support of using the [\ion{O}{3}] luminosities as a proxy for the bolometric luminosity, \citet{2005ApJ...634..161H} found a relationship between the observed hard X-ray (3-20\,keV) and observed [\ion{O}{3}] luminosities for a sample of AGN in the RXTE slew survey.
However, the results from a sample of Swift BAT AGN dispute the claim that [\ion{O}{3}] luminosities are good estimates of bolometric luminosity. \citet{2008ApJ...682...94M} found that [\ion{O}{3}] was not well-correlated with the hard X-ray (14--195\,keV).  

\begin{figure}
\hspace{-1.2cm}
\includegraphics[height=8cm]{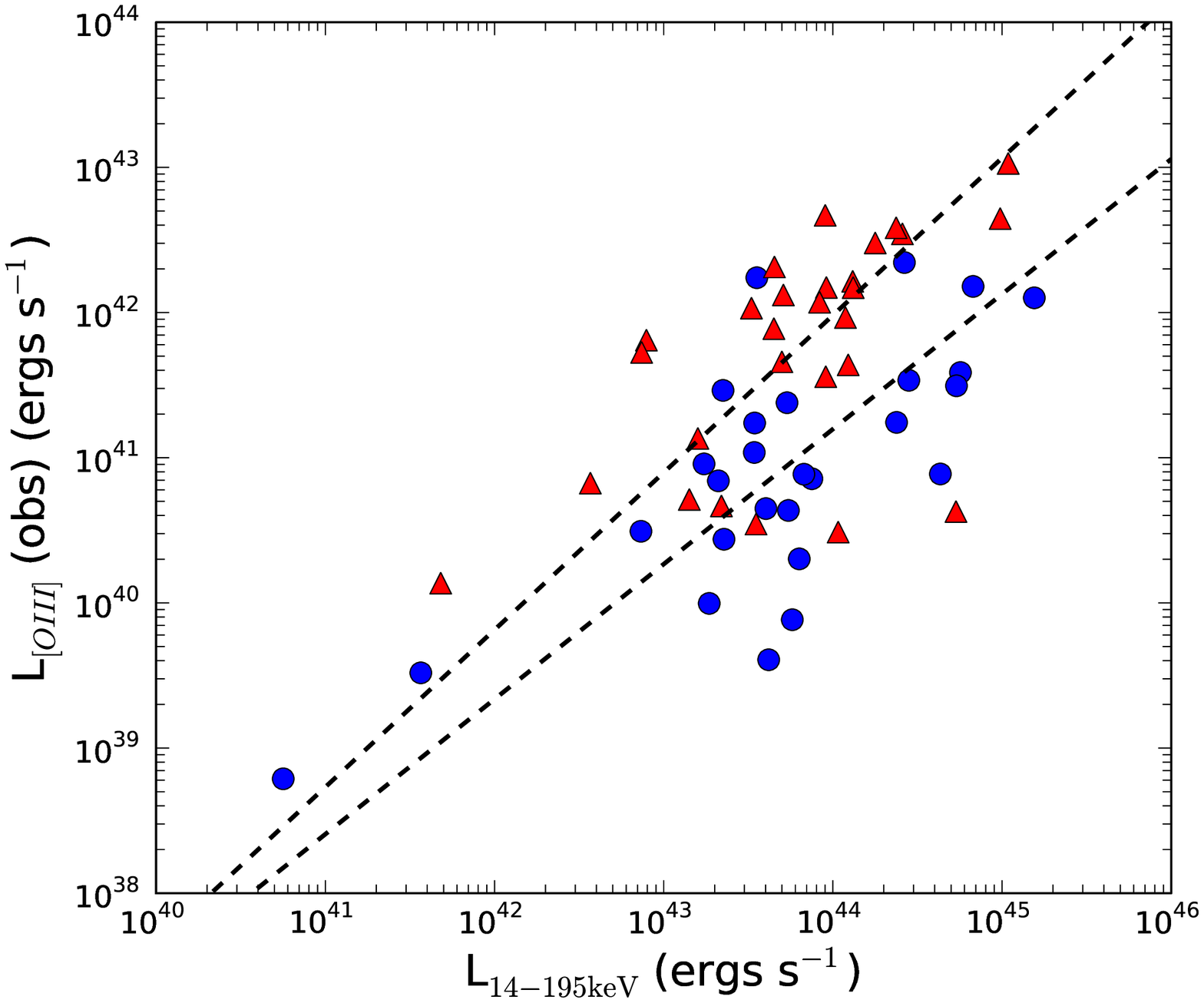}
\hspace{-1cm}
\includegraphics[height=8cm]{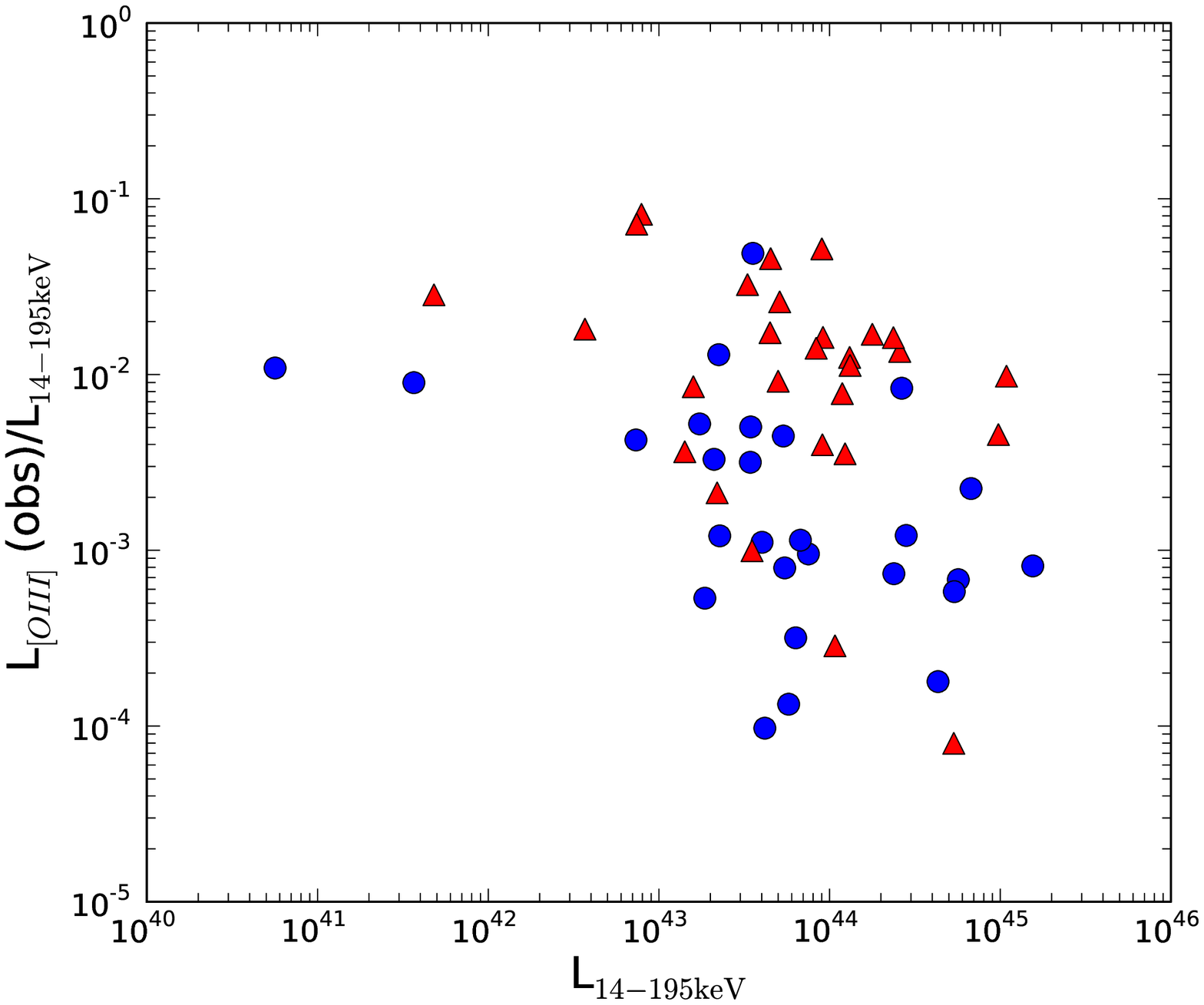}\\

\hspace{-1.2cm}
\includegraphics[height=8cm]{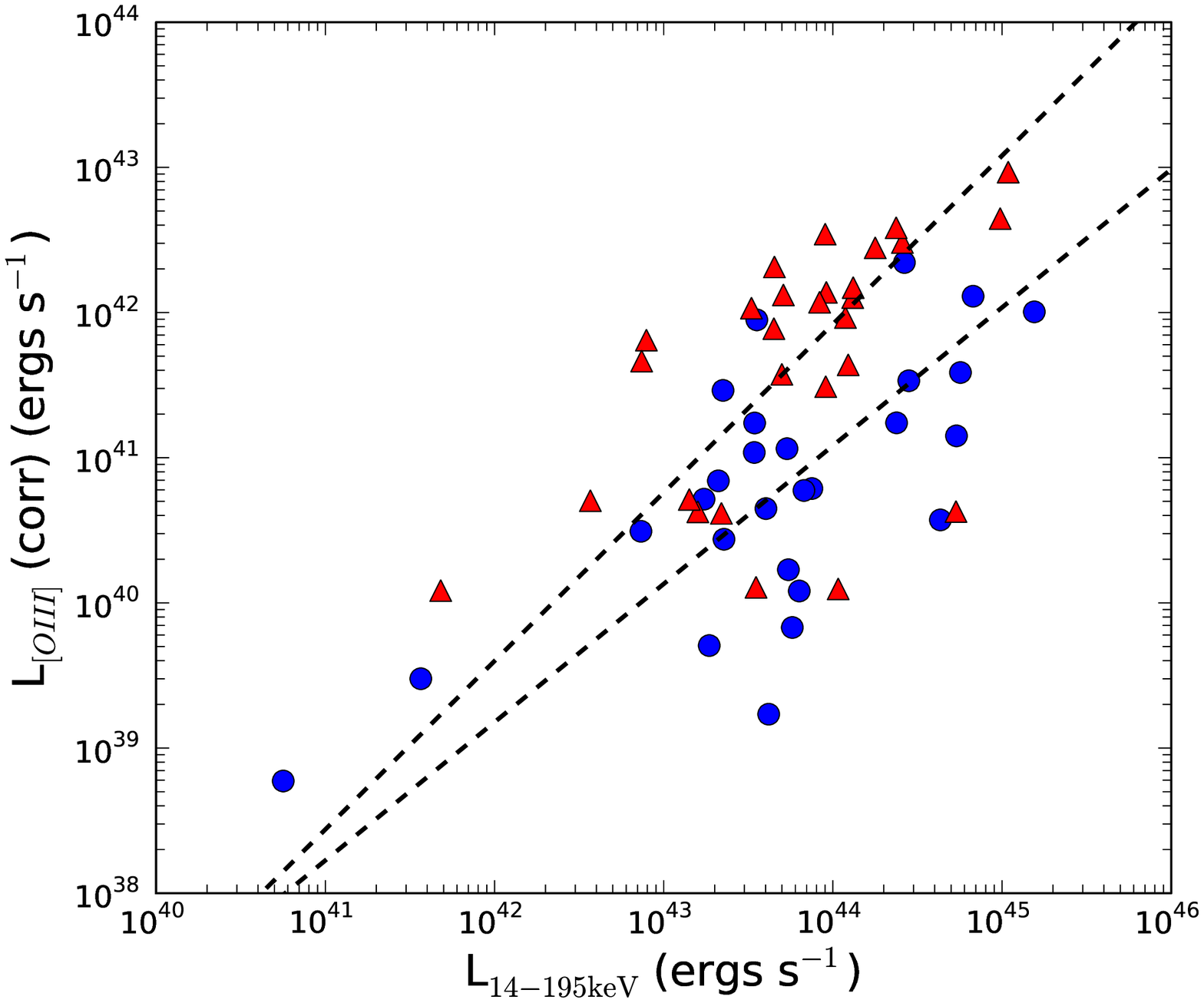}
\hspace{-1cm}
\includegraphics[height=8cm]{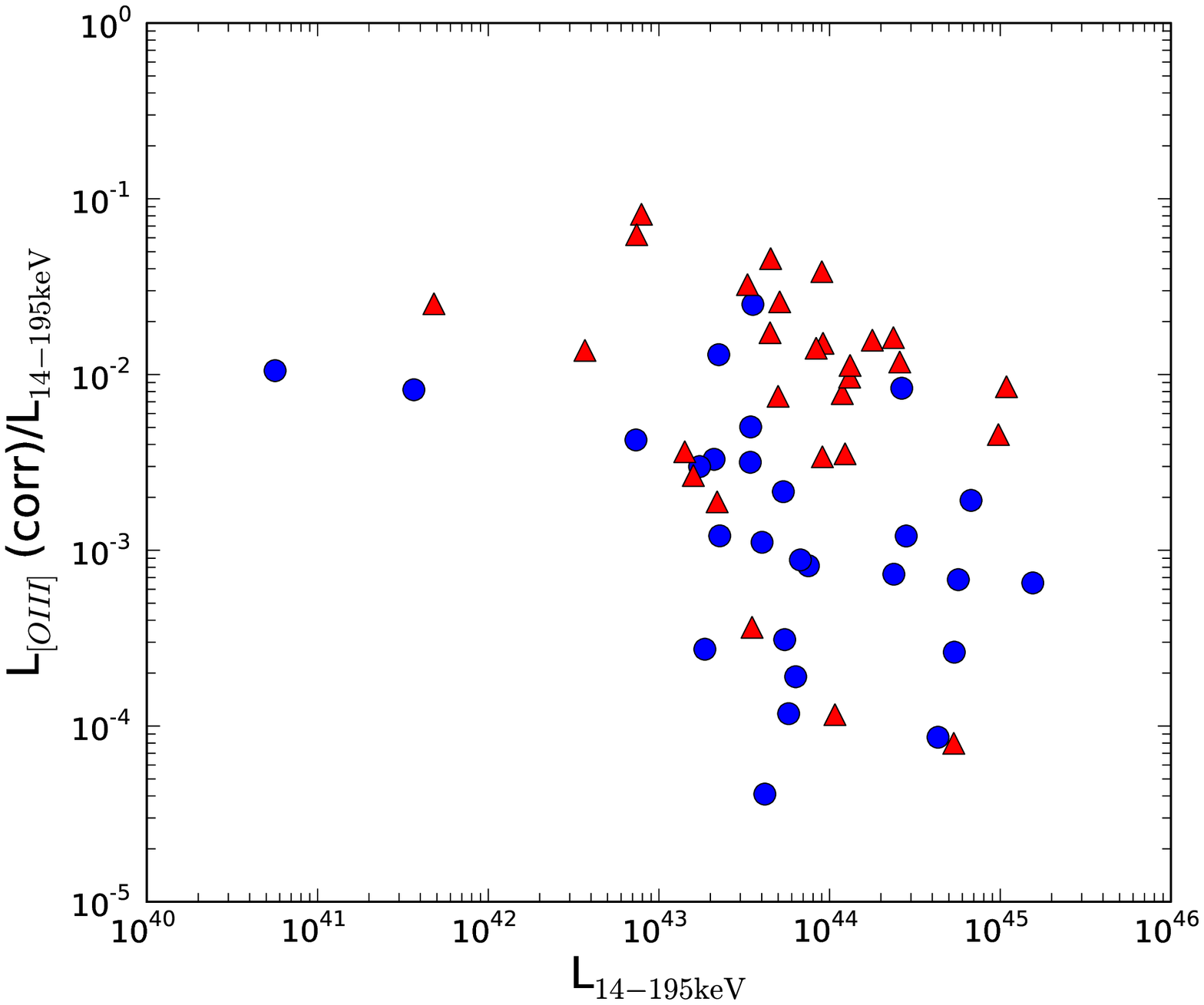}
\caption{Plotted is the relationship between observed (top) and reddening corrected (bottom) [OIII] 5007\AA~luminosities and the 14--195\,keV Swift BAT luminosities (left) and the ratio of these luminosities versus the Swift BAT luminosity (right).  In the plots, broad line sources are indicated with red triangles, while the narrow line sources are indicated with blue circles.  As the left plots show, L$_{[OIII]}$ is not well correlated with the hard X-ray luminosity.  The lines indicate the weak correlations seen for the Seyfert 1s ((corrected) $R^2 = 0.35$) and narrow line sources ((corrected) $R^2 = 0.38$).  In the right-hand plots, it is clear that there is a great deal of scatter in the optical/X-ray ratio for a given X-ray luminosity.
\label{fig-xrayopt}}
\end{figure}

With our larger and more uniformly measured extinction-corrected [\ion{O}{3}] sample than in the \citet{2008ApJ...682...94M} sample (drawn from the 3-month Swift catalog \citep{2005ApJ...633L..77M}), we tested for a correlation between the BAT and [\ion{O}{3}] luminosities.  In Figure~\ref{fig-xrayopt}a, we plot the results of our comparison.  We find weak linear correlations between the 14--195\,keV and [\ion{O}{3}] luminosities for the broad and narrow line sources. Using the ordinary least-squares (OLS) bisector method \citep{1986ApJ...306..490I}), we found $L_{[OIII]} ({\rm corr}) \propto L_{14-195 {\rm keV}}^{1.16 \pm 0.13}$ and $R^2 = 0.34$ ($P \approx 0.005$) for the broad line sources and $L_{[OIII]} ({\rm corr}) \propto L_{14-195 {\rm keV}}^{1.16 \pm 0.24}$ and $R^2 = 0.42$ ($P \approx 0.002$) for the narrow line sources.  Here, $R^2$ is the correlation coefficient.  As further illustrated in the ratio of optical/hard X-ray luminosity in Figure~\ref{fig-xrayopt}b, there is a great deal of scatter in these relationships (e.g. more than 2 magnitudes at $\log L_{14-195 {\rm keV}}$).  Our results support those of  \citet{2008ApJ...682...94M}, showing that even the reddening corrected $L_{[OIII]}$ is affected by extinction.  This effect is most pronounced for the narrow line sources, which show the greatest amount of scatter.

As shown in \citet{2009ApJ...700.1878R}, at high levels of absorption the luminosities measured in the Swift BAT band are affected by extinction.  Using models from \citet{2000MNRAS.318..173M}, they show the difference between the emergent and input BAT flux at a variety of column densities.  For column densities less than $10^{23}$\,cm$^{-2}$, this effect is minimal ($\le 4$\%). Since none of our targets are Compton thick ($N_H < 10^{24}$\,cm$^{-2}$ in the Swift sample, see \citet{2009ApJ...701.1644W}
for Suzaku observations of heavily obscured sources confirming their Compton thin nature), the effects on our sample are confined to a factor of $\approx 10 - 20$\% for the highest column density sources (25\% of the narrow line sources).  Even with this level of scatter introduced in the BAT luminosities, clearly the scatter seen in the narrow line sources in Figure~\ref{fig-xrayopt} is not accounted for by a 20\% underestimate in BAT luminosity.

\section{Mass and Accretion Rate Estimates}
For each of the broad line spectra, we were able to derive the mass of the central black hole using the FWHM of the broad component of H$\beta$ and the continuum luminosity at 5100\AA\,.  The continuum luminosity at 5100\AA\, was computed from a power law continuum fit to the H$\beta$ region.  We calculated the H$\beta$ masses using our measurements in Table~\ref{tbl-bluebroad} and equation 5 from \citet{2006ApJ...641..689V}.  The computed values of extinction corrected $\lambda L_{\lambda}$ (5100\AA) and M$_{\rm{H}\beta}$ are included in Table~\ref{tbl-masses}.  At the resolution of our spectra, we found that the H$\beta$ line is often more complicated than a simple combination of narrow and broad Gaussian profiles.  Additional structure or asymmetries are seen in a number of sources, making our measurements an approximation of the broad H$\beta$ line FWHM (see Figure~\ref{fig-broadfits} for example fits).

\begin{figure}
\begin{center}
\includegraphics[height=8cm]{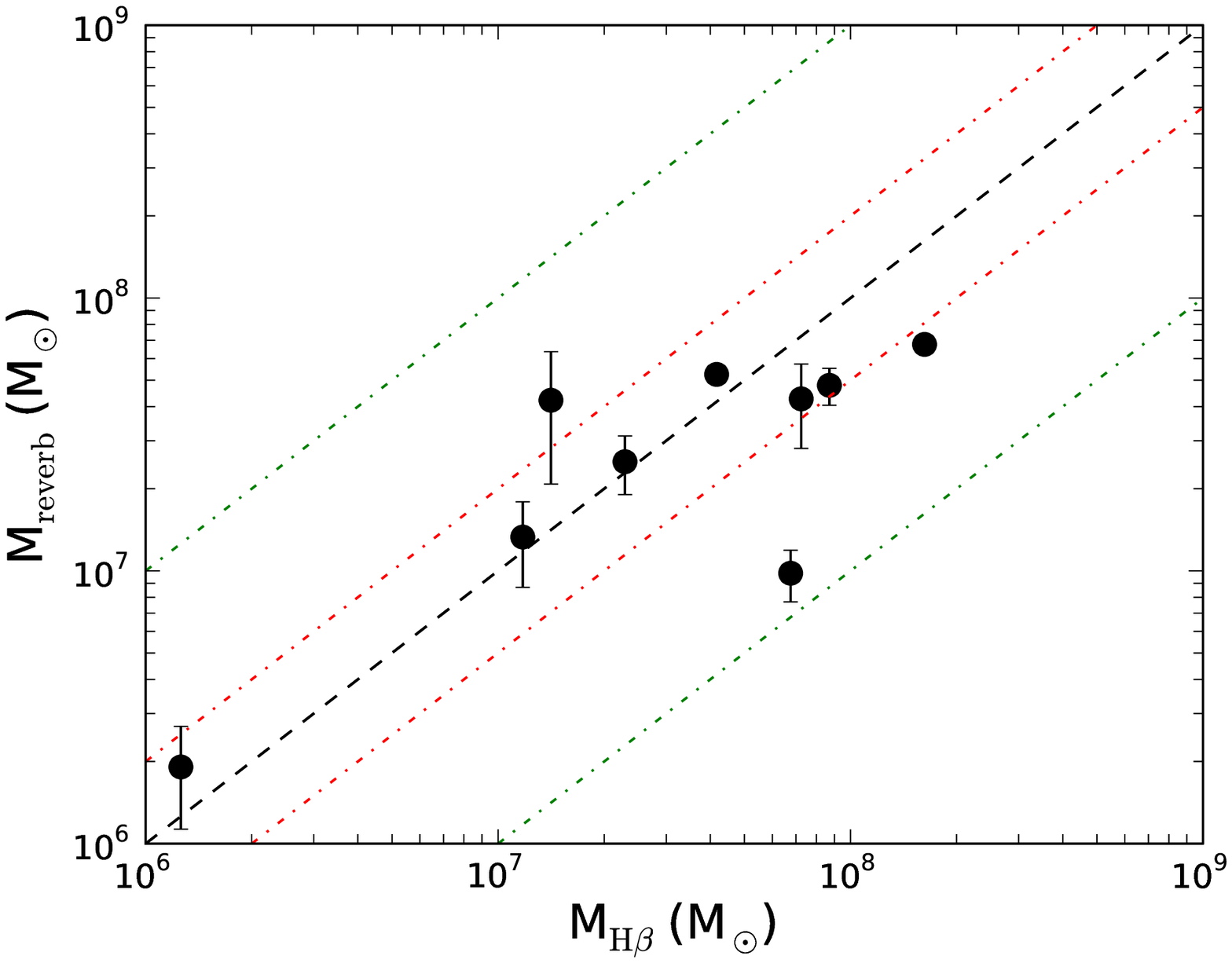}\\
\includegraphics[height=8cm]{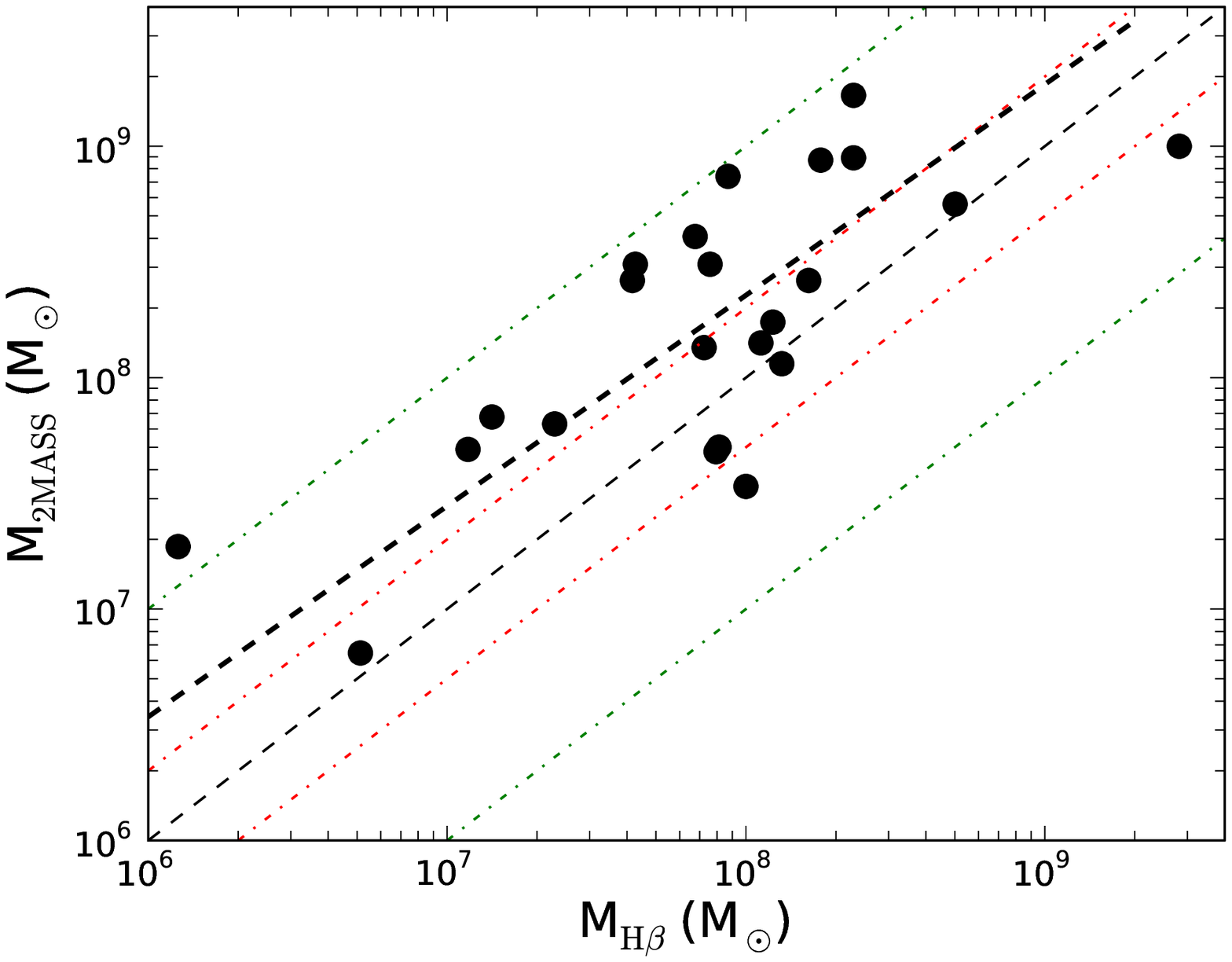}
\end{center}
\caption{Plotted are comparisons of the H$\beta$ derived masses from the FWHM of
the broad component of H$\beta$ and $\lambda L_{\lambda}$(5100\AA) and two additional mass estimates.  The first comparison is with reverberation mapping masses, largely from
 \citet{2004ApJ...613..682P}.  We find good agreement between the H$\beta$ masses and this more direct mass measurement.  The second comparison is with masses derived from the 2MASS K-band stellar magnitudes \citep{2008ApJ...684L..65M,2009ApJ...690.1322W}.  We also find a correlation between these two mass measurements (IR and H$\beta$ derived), indicated by the bolder dashed line. The additional dashed lines on the second plot represent values with (from the top to bottom most line) 10$\times$ difference, 2$\times$ difference, 1:1 correspondence, 1/2 difference, or 1/10 difference.
\label{fig-masses}}
\end{figure}

To test how our values of M$_{\rm{H}\beta}$ compare with other mass estimates, we plot our values versus reverberation mapping masses and masses derived from the stellar K-band light from 2MASS photometry in Figure~\ref{fig-masses}.  The mass estimates from reverberation mapping were obtained for 9 sources from \citet{2004ApJ...613..682P} and are listed in Table~\ref{tbl-masses}.  As shown, our H$\beta$ derived masses are in good agreement with the reverberation mapping results (with the exception of NGC 4593).  There are no systematic offsets between the two methods.  

Not surprisingly, there are larger differences between the IR derived and H$\beta$ derived masses. The 2MASS K$_s$ band derived masses \citep{2008ApJ...684L..65M,2009ApJ...690.1322W,2009arXiv0907.2272V} were calculated by subtracting the central luminosity of a point source (the size of the 2MASS PSF).  This presumed AGN luminosity was subtracted from the integrated luminosity of the galaxy to determine the luminosity of the stellar bulge.  The relation defined by \citet{2006ApJ...637...96N} was then used to convert the bulge luminosity to stellar mass.  Approximately 40\% of the mass estimates from the 2MASS K$_s$-band and H$\beta$ are within a factor of 2 of each other.  A greater majority of the IR masses are higher (typically, by up to a factor of 10). 

Despite the fact that the 2MASS K$_s$ band derived masses are a less accurate mass determination than those using reverberation mapping or the H$\beta$ FWHM method, we find that the 2MASS and H$\beta$ FWHM masses are linearly correlated.  Using the OLS bisector method, we find \begin{equation} \log M_{2MASS} = (0.91 \pm 0.14) \times \log M_{H\beta} + (1.07 \pm 1.13),
\end{equation} with $R^2 = 0.56$.  This is encouraging since the 2MASS derived measurements are the only uniform estimates that we have for the narrow line and broad line sources.  Therefore, we use the 2MASS derived masses to compare estimated masses, and later accretion rates, between all of our sources.  Unlike our comparison of luminosities ($\log L_{[OIII]}$), we find that the 2MASS derived masses show a great probability (from the K-S test) of the narrow and broad line masses being derived from the same population ($D = 0.21$ and $P = 0.71$).  The mean and standard deviations of $\log (M_{IR}/M_{\sun})$ are 8.07 and 0.83 (narrow line sources) and 8.19 and 0.62 (broad line sources).  With the more accurate H$\beta$ FWHM method, we find that the average mass of our sources (based on the broad line sources) is $\log M/M_{\sun} =$ 7.87 with a standard deviation of 0.66.    The range of masses, as shown in Figure~\ref{fig-masses}, is consistent with those found in other AGN surveys.  For example, our values are consistent with the range, $10^6$--$7 \times 10^9 M_{\sun}$, found by \citet{2002ApJ...579..530W} in a sample of 377 AGNs.  Our values are also similar to those of nearby PG QSOs derived from H-band host magnitudes \citep{2009ApJ...701..587V}.



\begin{figure}
\centering
\plotone{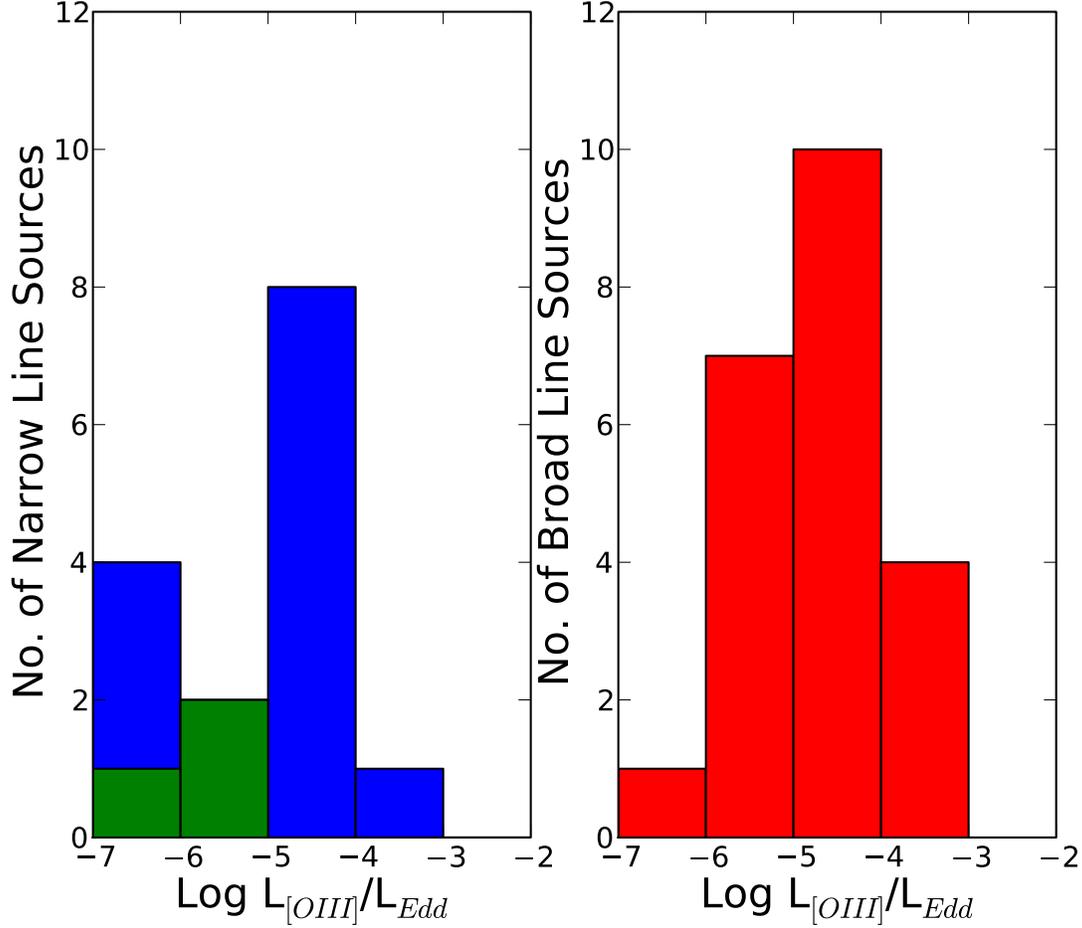}
\caption{Plotted are the distributions of L$_{[OIII]}/$L$_{Edd}$ for the narrow line (Seyferts = blue, LINERs = green) and broad line (red) sources.  The [\ion{O}{3}] 5007\AA~luminosity scales with the bolometric luminosity, making the ratio L$_{[OIII]}/$L$_{Edd}$ an indicator of the accretion rate.  While the ratio of L$_{[OIII]}/$L$_{Edd}$ is lower for the narrow line sources, a comparison of the accretion rates depends greatly on the bolometric corrections, which are determined from the spectral energy distributions and are not well known, particularly for the narrow line sources.  Only two of the \ion{H}{2}/ambiguous/other sources have masses available to calculate L$_{[OIII]}/$L$_{Edd}$.  The average value for these sources is low, with $\log $L$_{[OIII]}/$L$_{Edd} = -5.4$, but not as low as the value for the three LINERs with available mass measurements (-5.9).
\label{fig-lledd}}
\end{figure}

Since the masses of our narrow and broad line sources are similar while the average narrow line source luminosities are lower, we expect the values of L$_{[OIII]}/$L$_{Edd}$ to differ.  The [\ion{O}{3}] $\lambda 5007$\AA~luminosity is often used as an estimate of the bolometric luminosity of AGN, particularly for sources detected in the SDSS (see \citet{2004ApJ...613..109H}).  Typical bolometric corrections for extinction corrected [\ion{O}{3}] luminosities are expected to be between 600--800 for Seyfert 1s \citep{2008arXiv0812.1224K}.  There are, however, problems with using L$_{[OIII]}$ as an estimate of bolometric luminosity.  In the previous section, we showed that the hard X-ray luminosities, which are less affected by contamination from star formation and extinction, are not well-correlated with L$_{[OIII]}$, particularly for the narrow line sources.  Despite these problems, the ratio of L$_{[OIII]}/$L$_{Edd}$ allows us to compare a rough estimate of the accretion rates of our broad and narrow sources, which we can also compare with the more robust values we obtained in our X-ray study.  In Figure~\ref{fig-lledd}, we plot the results (where L$_{Edd}$ is defined as $1.38 \times 10^{38} (M/M_{\sun})$ and the mass is obtained from the 2MASS measurements).  We find, as expected, that the ratio of  L$_{[OIII]}/$L$_{Edd}$ is lower for the narrow line sources, with the average and standard deviations corresponding to $10^{-5.25 \pm 0.81}$ (narrow) and $10^{-4.61 \pm 0.85}$ (broad).  Since only three LINERs and two \ion{H}{2}/composite/ambiguous sources have available 2MASS-derived masses, we can not test whether sources in these categories have lower L$_{[OIII]}/$L$_{Edd}$ values than Seyferts.

For the broad line sources, our estimate of the average accretion rate ($L_{bol}/L_{Edd}$), assuming a bolometric correction of 600, is 0.015 with the 2MASS derived masses or 0.034 with the H$\beta$ FWHM derived masses.  Based on our X-ray analysis \citep{2009ApJ...690.1322W}, the 2MASS derived masses, and an assumed 2--10 keV bolometric correction of 35 for unabsorbed sources \citep{2005AJ....129..578B}, we estimate an X-ray derived accretion rate of 0.040.  Therefore, there is very good agreement between the optical and X-ray derived accretion rates, in an average sense.  Unfortunately, with increased uncertainty in the bolometric corrections, it is more difficult to determine these values for the narrow line/Sy 2 sources.

\section{Host Galaxy Properties}~\label{gal}
Since the Swift BAT-detected AGN are relatively close ($<z> \approx 0.03$) and bright, intrinsic stellar absorption features are seen in the majority of the spectra we analyzed.  This allows us to determine some of the properties of the host galaxies of our target AGN.  To do this, 
we have employed two particular methods to analyze the intrinsic stellar absorption features -- one using continuum fits and the other measuring stellar absorption indices.  We note, however, that the sampling of the host galaxy populations for the BAT-detected AGN is not uniform, but a function of the aperture size (2--3$\arcsec$), distance to the source, and both the size and orientation of the host galaxy within the slit.  It is our intent, in this paper, to determine basic conclusions about the stellar populations from the optical spectra.  More detailed information on the host galaxies of this sample, including star formation rates from Spitzer follow-ups and colors from an analysis of ground-based optical imaging data, will be presented by our collaborators.

The first method we used to obtain information about the AGN host galaxy populations was the continuum model fitting described in \S~\ref{continuum}.  For each of our sources, we fit the continuum with a combination of a power law (representing the non-thermal continuum) and a combination of a young, intermediate, and old single stellar population model, utilizing three different metallicities.  We then created a grid of test cases to test the ability of the continuum models to accurately describe the host galaxy spectrum, finding that metallicities could not be determined but that young stellar populations are clearly distinguished between both the intermediate and old stellar populations (see \S~\ref{testspectra}).  There is a degree of degeneracy between the intermediate and old populations as well which occurs when these populations are in combination with other populations (for example, a model of 50\% intermediate and 50\% young populations can be equally well modeled with a best fit continuum model that is a combination of young, intermediate, and old populations).

The main result of our continuum model fits is that the majority of the Swift BAT AGN in our sample have either a weak or no contribution from young stellar populations and are dominated by intermediate/old populations.  Of the sources with continuum light dominated by stellar populations, only one source has 50\% or more of their light dominated by a young (25\,Myr) population -- NGC 4151, whose host galaxy is a barred spiral (Sab) with a ring of star formation.  This is in contrast with the 15 sources with 50\% dominated by intermediate (2500\,Myr) populations and the 23 with 50\% or more dominated by old (10000\,Myr) populations.  To these results, however, we must add the caveat that we measured the continua with very simplified models.  It is also possible that degeneracies between the power law and young stellar component exist. 

Still, the result of the BAT AGN hosts being largely composed of intermediate to old populations, is supported further through an analysis of the H$\delta_A$ and D$_n(4000)$ stellar absorption indices.  These age sensitive indicators, the former sensitive to recent star bursts and the latter to an indicator of old populations through measurement of the \ion{Ca}{2} break, reveal few sources (6 total) in the region of the H$\delta_A$-D$_n(4000)$ parameter space occupied by systems with significant contributions from young stellar populations ($\ga 30$\%).  Due to contamination of the absorption features from AGN emission lines (i.e. [\ion{Ne}{3}] $\lambda$ 3869\AA~and H$\delta$), this result is based largely on the obscured sources.

Based on an SDSS study by \citet{2003MNRAS.346.1055K}, low luminosity narrow line AGN are hosted in old galaxies (as indicated by D$_n(4000)$).  This is consistent with the results of our study.  Additionally, we find that the distribution of our narrow line sources in the H$\delta_A$--D$_n(4000)$ plot is consistent with the location of `strong' AGN in the SDSS sample (Figure 17 in \citet{2003MNRAS.346.1055K}).  Since their definition of strong ($3.85 \times 10^{40}$\,ergs\,s$^{-1}$ in [\ion{O}{3}] $\lambda 5007$\,\AA) includes the majority of our sources, this shows that our results are consistent with the SDSS results.  As shown in \citet{2003MNRAS.346.1055K}, the values of D$_n(4000)$ for our narrow line sources are indicative of normal late-type galaxies.  This is also consistent with our analysis of the morphologies of the 9-month sample AGNs, as listed in NED.  In \citet{2009ApJ...690.1322W}, we had shown that the hosts of our sources (both Sy\,1 and Sy\,2 sources) were mostly spirals and irregulars.

\begin{figure}
\centering
\includegraphics[height=6.5cm]{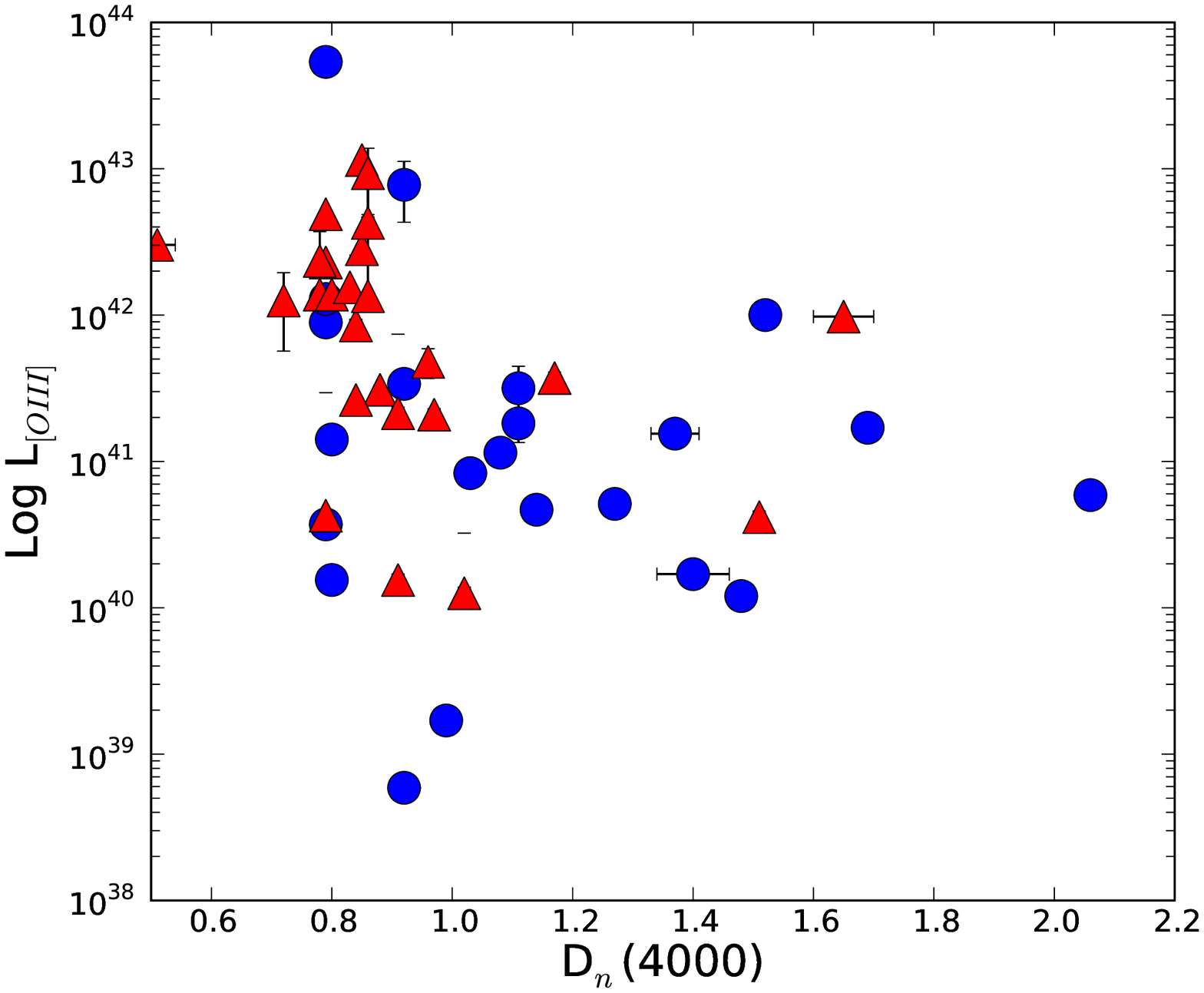}
\hspace{-1cm}
\includegraphics[height=6.5cm]{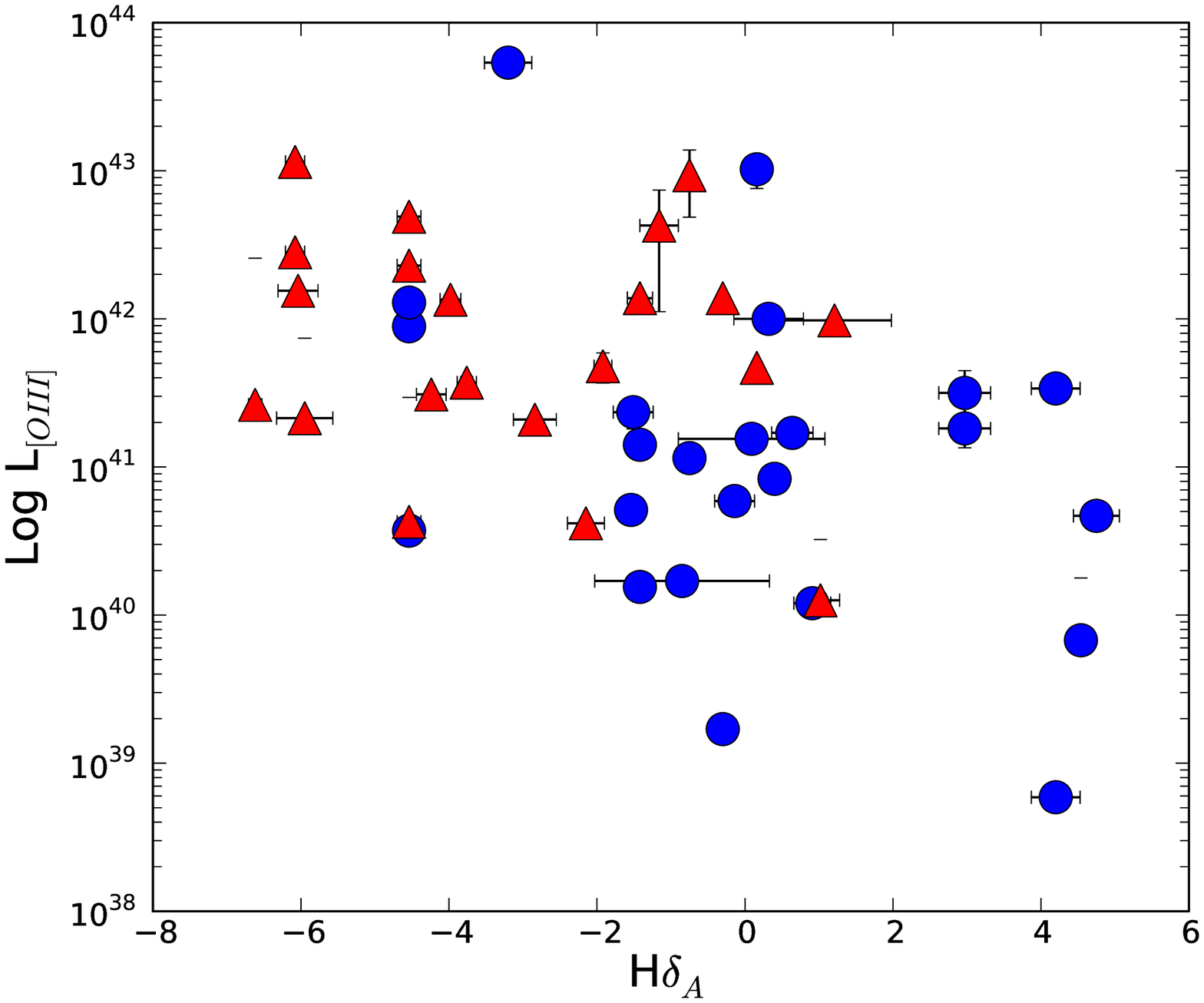} \\

\includegraphics[height=6.5cm]{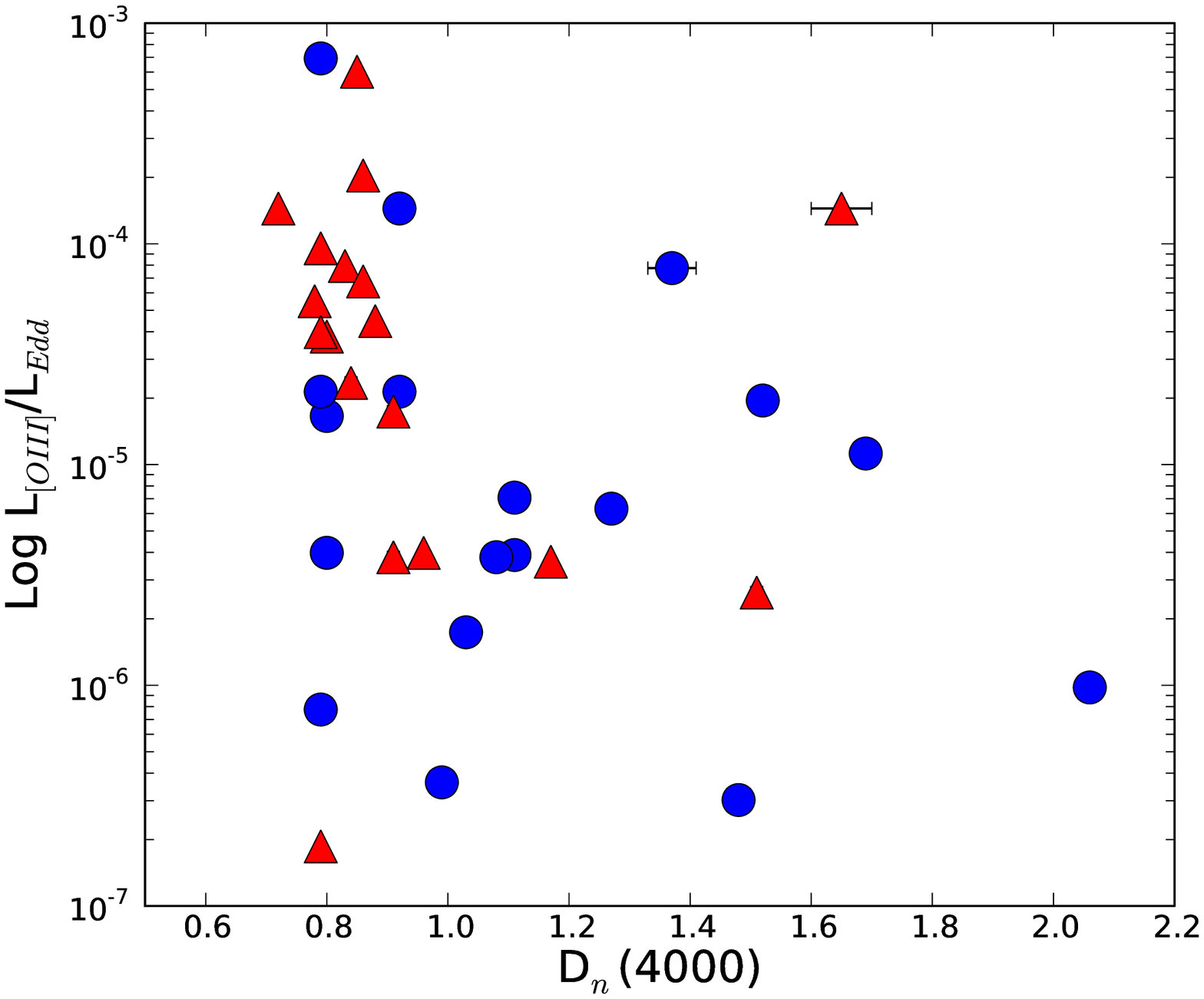}
\hspace{-1cm}
\includegraphics[height=6.5cm]{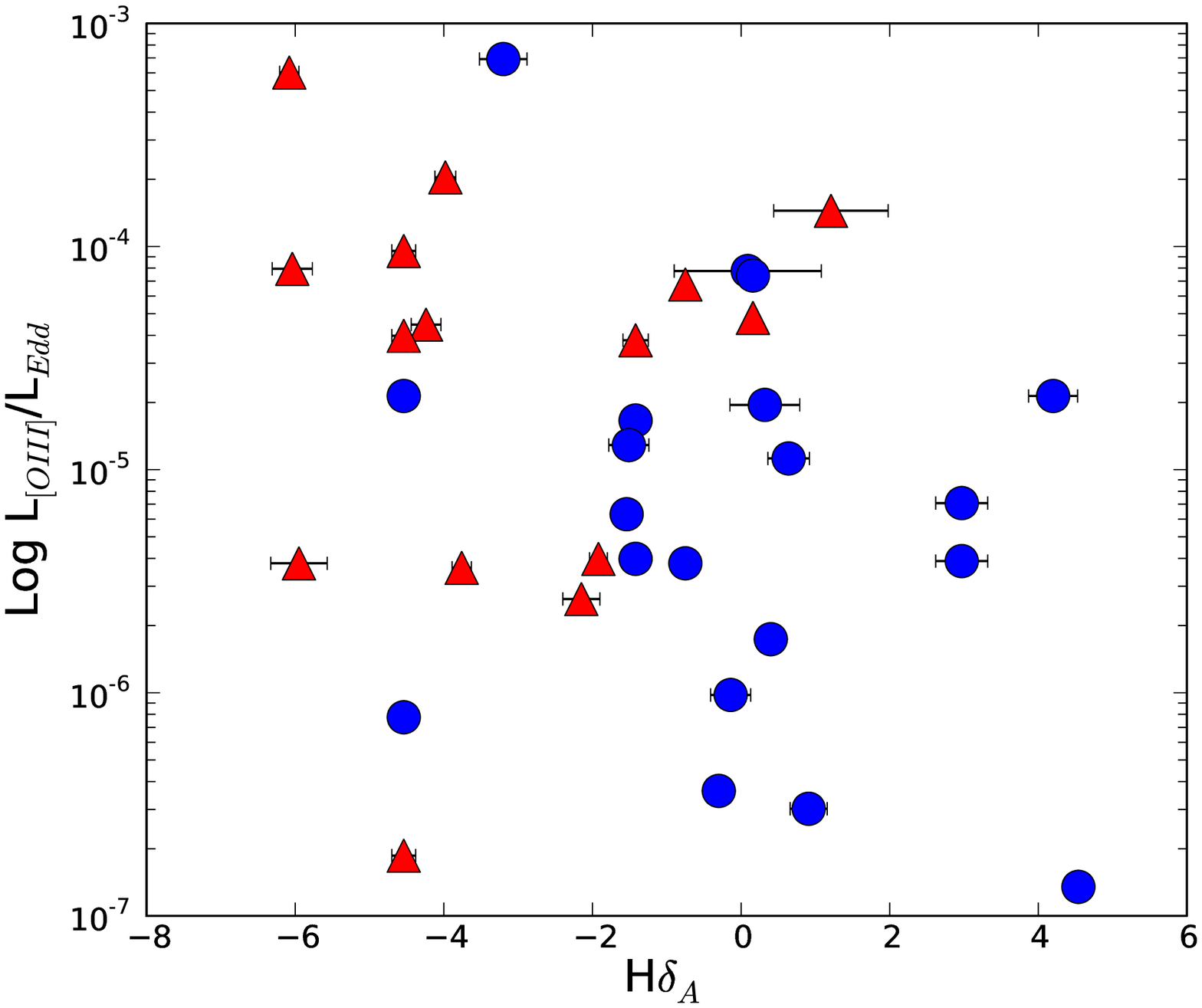}

\caption{Plotted are two age indicators, H$\delta_A$ which measures recent bursts of star formation and D$_n(4000)$ which measures the \ion{Ca}{2} break and is sensitive to old stellar populations versus the reddening corrected [\ion{O}{3}] 5007~\AA~luminosity and L$_{[OIII]}/$L$_{Edd}$ for the narrow line (circles) and broad line (triangles) sources.  
\label{fig-agelicklo3}}
\end{figure}

Another result from the \citet{2003MNRAS.346.1055K} study, is a connection between the age distribution of host galaxies and the [\ion{O}{3}] luminosity of the AGN.  In Figure~\ref{fig-agelicklo3}, we plot each of the stellar age indicators (H$\delta_A$ and D$_n(4000)$) versus the extinction-corrected [\ion{O}{3}] luminosity and the ratio  L$_{[OIII]}/$L$_{Edd}$ for both the narrow and broad line AGN.  The top plots of this figure are comparable to Figure 12 of \citet{2003MNRAS.346.1055K} (whose L$_{[OIII]}$ measurements are in units of L$_{\sun}$).  We find no direct correlation between either of these stellar absorption indices and either [\ion{O}{3}] luminosity or accretion rate ($R^2 \la 0.1$).  Since our sources include only the equivalents of SDSS `strong' AGN, it is not surprising that we do not see a correlation.  Our sample does not include weak AGN, which tend to have older populations (associated with early type galaxies).

Finally, we find a possible indication that the host galaxies of broad and narrow line sources may be different.  Namely, we see differences in the metallicity indicator Mgb.  Applying the Kolmogorov-Smirnov test, we find a $P$-value of 0.01, indicating that the populations are likely different.  For the broad line sources, we find an average value of 0.84 with a standard deviation of 1.65 in Mgb.  The narrow line sources have a much higher average Mgb measurement of 1.96 with a standard deviation of 2.27.  Based on our simulations, lower values of Mgb also correspond to younger populations (see the top left panel of Figure~\ref{fig-Licktest}).  Therefore, there is a degeneracy between age and metallicity such that the result of broad line sources having lower values of Mgb could indicate their hosts as either having a larger contribution from a younger population or from a lower metallicity than the hosts of narrow line sources.

\section{Conclusions}
AGN surveys are typically dominated by two selection effects: (1) dilution by starlight from the host galaxy and (2) obscuration by dust and gas in the host galaxy and/or the AGN itself (see 
\citet{1994PASP..106..113H,2004ASSL..308...53M}).  For these reasons, an unbiased AGN sample is difficult to define.  The Swift's BAT AGN survey provides one of the first truly unbiased (to all but the highest column densities) samples of local AGNs.

Since the BAT-detected sources are nearby, $<z> = 0.03$ \citep{2008ApJ...681..113T}, they are excellent targets for multi-wavelength follow-ups. In this paper, we presented the optical spectral properties from sources detected in the first 9-months of the survey.  Our analysis includes both the emission line properties of the AGN as well as the host properties revealed from intrinsic stellar absorption features.  The sample includes 40 spectra taken at the 2.1-m KPNO telescope, 24 archived SDSS spectra, and the emission line properties of 13 sources presented in the literature.  In total, this sample covers 81\% of the Swift BAT AGN sources viewable from KPNO.  It is comprised of 55\% broad line sources and 45\% narrow line sources, in the same ratio as the total Swift sample.

With our unbiased AGN sample, it is important to compile the fundamental properties of the sources both as a test to our current understanding of AGN and as a comparison to more biased methods of detection (e.g. optical surveys).  Using standard emission line diagnostic plots, we find that the majority of our hard X-ray detected sources are optically Seyferts (66\% of narrow line and 75\% of broad line sources).  This contrasts with the optically selected SDSS sample examined by \citet{2006MNRAS.372..961K}, which includes a large (75\%) fraction of \ion{H}{2} galaxies with few Seyferts (3\%).  Since \ion{H}{2} galaxies are less luminous than Seyferts in the X-ray band \citep{2003AA...399...39R}, it is not surprising that our hard X-ray flux limited sample detects the more luminous local sources, which are Seyferts.  In the same sense, the optical SDSS sample detects more LINERs, which are also less luminous sources than Seyferts, than we find in the Swift BAT sample.  In particular, we classify 16\% of the narrow line sources as LINERs and none of the broad line sources.  

One of the most fundamental properties of a black hole is its mass.  Under the unified AGN model, we expect to find no difference in the mass distribution between the broad and narrow line sources.  Indeed, we find the distributions of our 2MASS derived masses statistically consistent with being drawn from the same population.  Comparing 2MASS derived masses with a more accurate determination from the FWHM of H$\beta$ in broad line sources, we find the masses from both methods are well correlated.  The average value of our hard X-ray detected sources is  
$<M/M_{\sun}> = 10^{7.87 \pm 0.66}$, with a range of values consistent with those found in previous studies of AGNs \citep{2002ApJ...579..530W} and nearby PG QSOs \citep{2009ApJ...701..587V}.  

Determinations of the reddening from the ratio of narrow H$\alpha$/H$\beta$, as well as gas densities and temperatures in the narrow line regions from diagnostic emission lines of the same ion, are also consistent with both the unified model and previous results from optical studies.  Under the unified model, we expect narrow line sources to have heavier extinction (assuming the extinction is on the nuclear/galactic scale and not simply from the torus), while other narrow line region properties like density and temperature to be the same for narrow and broad line sources.
As expected, we find the average distribution of reddening values [E(B-V)] higher in the narrow line sources.  In our calculations of the gas density in the $S^+$ emission region, we find the same electron densities of $N_e \approx 10^3$\,cm$^{-3}$ for broad and narrow line sources.  Superficially, the $O^{+2}$ region appears at a higher temperature for the broad line sources.  However, as discussed in \citet{1978PhyS...17..285O}, a likely explanation is that the narrow and broad line sources both have similar temperatures (we find $T_e \approx 10000$\,K), but in the broad line sources we are able to probe  [\ion{O}{3}] $\lambda 4363$ emitting gas into denser regions of the narrow line region.

Based on the results of our X-ray study of our unbiased AGN sample, we suspect that the distributions of luminosities of the Swift AGN conflict with the unified model.  Namely, our X-ray results  \citep{2009ApJ...690.1322W} showed that the absorbed/type 2 AGNs (X-ray absorbed/optically narrow line sources, including optically classified Sy\,2s, LINERs, and \ion{H}{2} galaxies) have lower absorption corrected 2--10\,keV luminosities and accretion rates.  These same trends are found among the optically derived luminosities and accretion rates.  Specifically, we find average extinction-corrected 5007\AA~[\ion{O}{3}] luminosities of $10^{41.74\pm0.93}$\,ergs\,s$^{-1}$ and $10^{40.94\pm1.00}$\,ergs\,s$^{-1}$ and ratios of 
L$_{[OIII]}/$L$_{Edd}$ of $10^{-4.61\pm0.85}$ and $10^{-5.25\pm0.81}$, respectively for broad and narrow line sources.  Contrary to the results of \citet{2005ApJ...634..161H}, but in agreement with 
\citet{2008ApJ...682...94M}, we find that the 14--195\,keV BAT luminosities are only weakly correlated with [\ion{O}{3}] luminosity for broad and narrow line sources.  


Seemingly, the result of narrow line sources having lower luminosities and possibly accretion rates (depending on the bolometric corrections) poses a challenge to the unified model.  On closer inspection, we find that the narrow line sources with optical classifications as Seyferts have similar X-ray and optical luminosities to their broad line, Seyfert 1 counterparts.  Instead, it is the sources optically classified as LINERs and \ion{H}{2}/composite/ambiguous sources which have lower luminosities.  While these sources are clearly detected AGN based on their X-ray properties, modification of the unified model to include a luminosity dependence is clearly required to link these fainter non-Seyfert sources with the Seyfert 1s and 2s. 

Finally, through our continuum model fits and measurements of stellar absorption indices, we can make a few general comments on the host galaxy properties of our sources.  We find that the stellar ages of the hosts include small contributions from young populations (0.25\,Gyr).  The populations are more consistent with intermediate/old (2.5--10\,Gyr) populations.  Comparing with the results drawn from the SDSS survey, we find that our narrow line sources have the same properties as the `strong' narrow line AGN from \citet{2003MNRAS.346.1055K}.  Therefore, their stellar absorption properties (from the \ion{Ca}{2} break and H$\delta$ absorption) are like those of late type galaxies.  This is also consistent with the NED morphologies of our sources (both Sy 1s and Sy 2s), which are mostly spirals and irregulars \citep{2009ApJ...690.1322W}.


\acknowledgments
The authors would like to greatly thank Christy Tremonti for use of her analysis codes and discussions on how to modify them for application to our AGN sources.  L.W.\ acknowledges support by NASA grant NNX08AC14G. Also, she acknowledges support through NASA grant HST-HF-51263.01-A, through a Hubble Fellowship from the Space Telescope Science Institute, which is 
operated by the Association of Universities for Research in Astronomy, 
Incorporated, under NASA contract NAS5-26555.  S.V.\ acknowledges support from a Senior Award from the Alexander von Humboldt Foundation and thanks the host institution, MPE
Garching, where some of this work was performed. K.T.L.\ acknowledges support from the NASA Postdoctoral Program Fellowship (NNH 06CC03B).  The Kitt Peak National Observatory observations were obtained using MD-TAC time as part of the thesis of L.W.\ at the University of Maryland (for programs 0322 and 0107) and M.K.\ (program 0295). Kitt Peak National Observatory, National Optical Astronomy Observatory, is operated by the Association of Universities for Research in Astronomy (AURA) under cooperative agreement with the National Science Foundation.

{\it Facilities:} \facility{Swift ()}, \facility{KPNO:2.1m ()}, \facility{Sloan ()}

\appendix
\section{Galaxy Continuum Spectral Fits}\label{testspectra}
\begin{figure}
\plotone{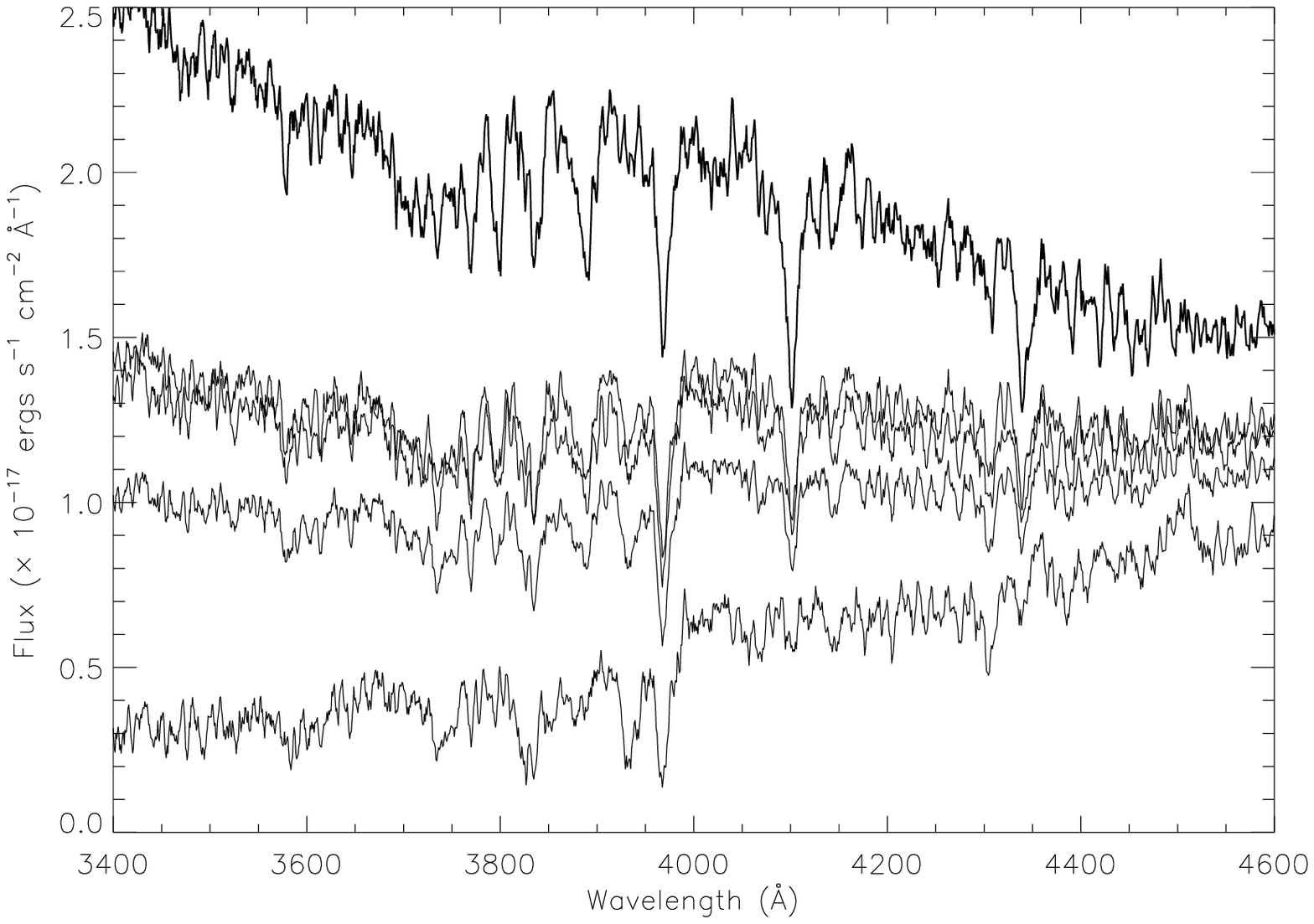}
\caption{Plotted are several of the test spectra created by broadening combinations of the stellar population models to a velocity dispersion of 300 km s$^{-1}$, adding random noise, and the effects of reddening.  From top to bottom, plotted are a young, 50\% young + 50\% intermediate, 50\% young + 50\% old, 33\% young + 33\% intermediate + 33\% old, and 50\% intermediate + 50\% old population.  Notice, there is very little difference between the 50\% young + 50\% intermediate and 50\% young + 50\% old populations.  We specifically plot the region surrounding the 4000\AA break, a region with prominent intrinsic absorption features.  
\label{fig-testspectra}}
\end{figure}

\begin{figure}
\plotone{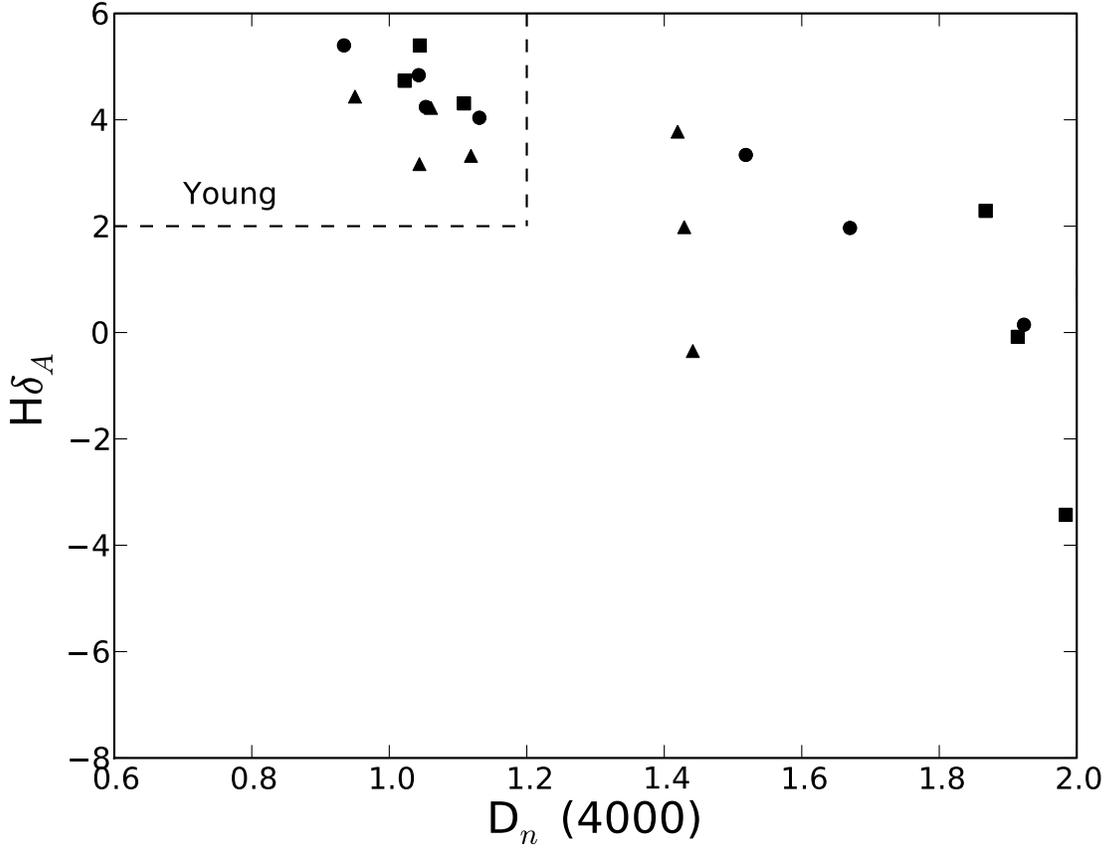}
\caption{Plotted is the stellar absorption index H$\delta_A$ versus the D$_n$(4000) index for the test spectra.  Both are commonly used as age indicators of a stellar population.  In the plot, our test spectra, consisting of combinations of single stellar population models, are shown for three metallicities: 0.2\,Z$_{\sun}$ (triangle),  Z$_{\sun}$ (circle), and 2.5 Z$_{\sun}$ (square).  We find that metallicity has little effect on the values of these stellar absorption indices, as expected.  We also find that populations with significant contributions from young populations (33\% or higher) fall within a small parameter space on the plot, towards the upper left hand corner.
\label{fig-hdeltatest}}
\end{figure}

\begin{figure}
\plotone{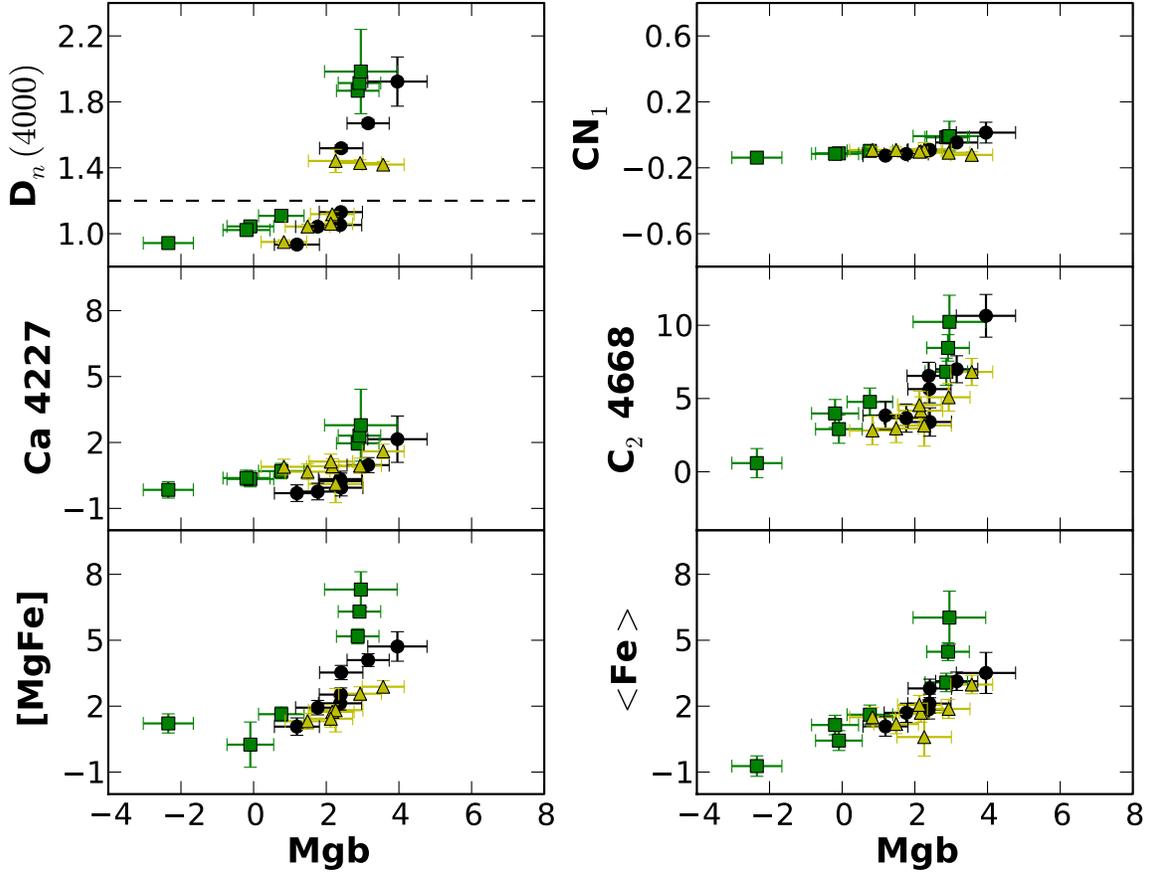}
\caption{Plotted are measured stellar absorption indices versus the Mgb stellar absorption index for the test spectra.  With the exception of D$_n(4000)$, which is an age indicator, the remaining plotted indices are sensitive to abundances of metals in the population.  In the plot, our test spectra, consisting of combinations of single stellar population models, are shown for three metallicities: 0.2\,Z$_{\sun}$ (yellow),  Z$_{\sun}$ (black), and 2.5 Z$_{\sun}$ (green).  The line in the top left plot indicates the division between young populations (below the line) and older populations (see Figure~\ref{fig-hdeltatest}).
\label{fig-Licktest}}
\end{figure}

In this section, we detail additional tests that we conducted to test the accuracy of the galaxy continuum fits.  As explained in \S~\ref{continuum}, we used a grid of single stellar age population models \citep{2003MNRAS.344.1000B} with 3 different ages (25, 2500, and 10000 Myr)  and at 3 different metallicities (2.5\,Z$_{\sun}$, Z$_{\sun}$, and 0.2 Z$_{\sun}$) to fit the continua of our AGN and template galaxy spectra.  In order to test the accuracy of these models, we constructed a grid of test spectra, broadening the sources by assuming  $FWHM = 300$\,km\,s$^{-1}$ ($\sigma \approx 128$\,km\,s$^{-1}$) and adding both random noise and reddening ($\tau = 1.5$) to the stellar population models used in the continuum fits.  This grid of models includes a young, intermediate, and old population, as well as the following combinations: 50\% young + 50\% intermediate, 50\% young + 50\% old, 50\% intermediate + 50\% old, and 33\% young + 33\% intermediate + 33\% old.  In Figure~\ref{fig-testspectra}, we plot several of these test spectra.

Assuming an error of 10\% in the flux, we fit each of the test spectra with the model spectra used to fit the continua of our target spectra.  The results of these fits are shown in Table~\ref{tbl-testspectra}.  Our results show that the velocity dispersions are accurately measured by the models in all cases.  The metallicities, however, are not since all of our test spectra have solar metallicity but a range of values are found from the fitting process.  We also find that the young stellar population component is measured well, though its contribution is underestimated by up to about 20\%.  Finally, we find that there is a degeneracy between the intermediate and old populations when they are found in combination with the young population.  This is well illustrated in Figure~\ref{fig-testspectra}, where there is little difference between the 50\% young + 50\% intermediate and 50\% young + 50\% old spectra.  We do, however, find that a 100\% intermediate population is distinguishable from a 100\% old population.  Therefore, the main conclusion that we draw from our test spectra is that our continuum model fits can clearly distinguish between young and intermediate/old populations.  See \citet{2003MNRAS.341...33K} for more detailed investigations used to study the SDSS host galaxies.

In a similar manner, we also created test spectra to look for degeneracies between the stellar continuum and power law component.  To accomplish this, we used the same set of test galaxy models as above.  For each of these test spectra, we added a power law component with an index ($p_1$) set to the average value determined from fits to our AGN sources (0.67).  We constrained the values such that the light fraction from both the stellar light and power law contributed 50\% of the light at 5500\AA.  Results of these fits are presented in Table~\ref{tbl-sims2}. We find that there is no obvious degeneracy between any of the stellar population models and the power law component (at least at this power law index).  The average fitted power law index, 0.74, is slightly higher than the true value while the fitted fraction of light contributed from the power law tends to be slightly lower than the true value (most of the values are between 0.41--0.45 instead of 0.50).  Generally, we find that the fitted values are consistent with the input parameters.

In addition to testing the accuracy of the continuum fits, we also used our grid of test galaxy spectra (excluding a power law contribution) to interpret the results of measurements of stellar absorption indices in our target spectra.  In addition to using the grid of solar models we described above, we created grids of populations with metallicities 2.5 and 0.20 times solar abundance.  For all of these sources, we measured the stellar absorption indices in the same manner as for our target spectra (see \S~\ref{gal-absorption}).  In this way, we can use our test spectra, which are of approximately the same signal-to-noise as our target spectra, to understand the results of an analysis of the stellar absorption indices.

In Figure~\ref{fig-hdeltatest}, we plot the H$\delta_A$ index versus D$_n$(4000) for our test spectra.  As we described in \S~\ref{gal-absorption}, these two indices are commonly used as indicators of the age of stellar populations.  As the plot shows, metallicity of the stellar population models does not have a large effect on these stellar absorption indices.  Further, as expected, there is a clear dependence on age, where populations with a significant (33\% or higher) contribution from a young population have both the highest values of H$\delta_A$, associated with recent bursts of star formation, and the lowest values of D$_n$(4000), which indicates the strength of the \ion{Ca}{2} break.  We find that the populations with significant contributions of young populations have H$\delta_A > 2$ and D$_n$(4000) $< 1.2$. 

In Figure~\ref{fig-Licktest}, we plot additional stellar absorption indices often used as metallicity indicators (as well as the \ion{Ca}{2} break age indicator) for the test galaxy spectra.  Each metallicity is represented with a different color, with the same grid of different stellar population components mentioned above.  We point out that, as shown in the D$_n(4000)$ versus Mgb plot, that the populations with significant contributions from young stars ($< 1.2$) tend to have lower values of Mgb ($\la 3$).  Distinctions between intermediate/old higher metallicity (Z$_{\sun}$ and 2.5 Z$_{\sun}$) and low metallicity (0.2 Z$_{\sun}$) populations are seen in the C$_2 4668$ vs. Mgb, [MgFe] vs. Mgb and $<$Fe$>$ vs. Mgb plots -- where higher values in x and y parameters are seen for the higher metallicity populations.  For young populations of any metallicity, it is more difficult to distinguish between different metallicity populations.

\section{Notes on Individual Spectra}
In this section, we include notes on the emission line spectra of the sources examined.  These notes particularly relate to peculiarities in the spectra or the fitting procedure for sources indicated.

For 10 broad line sources (or $\approx 1/3$ of the broad line sources), absorption lines from the \ion{Na}{1}D doublet~$\lambda5890, 5896$\AA~are seen.  These absorption features are seen in Mkn 1018, Mkn 590, MCG -01-13-025, Mrk 6, SDSS J090432.19+553830.1, NGC 3227, NGC 3516,  NGC 4593, MCG +09-21-096, NGC 5548, and RX J2135.9+4728.  In the spectrum of MCG +09-21-096, the absorption line is embedded in a broad \ion{He}{1} (FWHM $\approx 2270$\,km\,s$^{-1}$) emission line.  The \ion{Na}{1}D doublet was also detected (by eye) in 9 narrow line sources: NGC 788, Mkn 18, SDSS J091129.97+452806.0, SDSS J091800.25+042506.2, Ark 347, NGC 4102, NGC 6240, UGC 11871, and NGC 7319.

Additionally:
\par{\bf BROAD LINE SOURCES:}\\

\par{\bf LEDA 138501:} H$\beta$ has a ``red'' wing.

\par{\bf MCG -01-13-025, Mrk 1018, NGC 3227:}  Strong intrinsic absorption lines are seen in the spectra of these broad line sources.  \ion{He}{1} is seen in absorption for both sources.

\par{\bf IRAS 05589+2828:} There is a clear broad component to \ion{He}{2} $\lambda 4686$\AA.  H$\beta$ has a red wing.

\par{\bf MCG +04-22-042:} There is a clear broad component to \ion{He}{2} $\lambda 4686$\AA.

\par{\bf SBS 1136+594:} There is a clear broad component to \ion{He}{2} $\lambda 4686$\AA. 

\par{\bf UGC 6728:} Two narrow emission lines are present for each of the [\ion{O}{3}] emission lines (at $\lambda 4959$\AA and $\lambda 5007$\AA).

\par{\bf NGC 4593:} Two narrow emission lines are present for each of the [\ion{O}{3}] emission lines (at $\lambda 4959$\AA and $\lambda 5007$\AA).

\par{\bf MCG +09-21-096:} The profiles of the broad Balmer lines are complex, with broad ``boxy'' shapes (including H$\delta$, H$\gamma$, H$\beta$, and H$\alpha$).

\par{\bf Mrk 813:} H$\beta$ is blended with the nearby [\ion{O}{3}] emission lines.

\par{\bf 4C +74.26}: H$\beta$ is extremely broad and blended with the nearby [\ion{O}{3}] emission lines at this resolution.

\par{\bf NARROW LINE SOURCES:}\\

\par{\bf Mkn 18:}  Both the KPNO and SDSS spectra show an additional broad component to H$\alpha$ of approximately 370\,km\,s$^{-1}$.

\par{\bf Ark 347:} The H$\alpha$ region is quite complex.  Six distinct narrow lines are seen in the region including [\ion{N}{2}] $\lambda 6548$\AA, H$\alpha$, and [\ion{N}{2}] $\lambda 6583$\AA.  The measured wavelengths of these lines are: $6546.72 \pm 0.17, 6555.7 \pm 0.34, 6563.25 \pm 0.19, 6570.23 \pm 0.08, 6582.42 \pm 0.20$, and $6591.70 \pm 0.09$\AA.

\bibliography{ms.bib}

\clearpage
\begin{deluxetable}{llllllll}
\tabletypesize{\scriptsize}
\tablecaption{SWIFT BAT-detected AGN\label{tbl-1}}
\tablewidth{0pt}
\tablehead{
\colhead{Source} & \colhead{RA (h m s)}  & \colhead{Dec (d m s)} &
\colhead{Redshift} & \colhead{E(B-V)\tablenotemark{1}} &
\colhead{Type\tablenotemark{2}} & \colhead{Host Galaxy\tablenotemark{2}} & \colhead{Obs.\tablenotemark{3}}
}

\startdata
\object{Mkn 352} & 00:59:53.3 & +31:49:36.8 & 0.015 & 0.06 & Sy 1 & SA0 & Lit. \\

\object{NGC 788} & 02:01:06.5 & -- 06:48:55.8 & 0.014 & 0.03 & Sy 2 & SA(s)0/a & KPNO \\

\object{Mkn 1018} & 02:06:16.0 & -- 00:17:29.2 & 0.043 & 0.03 & Sy 1.5 & S0; merger & SDSS\\

\object{LEDA 138501} & 02:09:34.3  & +52:26:33.0  & 0.049 & 0.16 & Sy 1 & & KPNO\\

\object{Mkn 590} & 02:14:33.6 & -- 00:46:00.3 & 0.026 & 0.04 & Sy 1.2 & SA(s)a & SDSS\\

\object{NGC 931} & 02:28:14.5 & +31:18:42.1 &  0.017 & 0.10 & Sy 1.5 & Sbc & Lit.\\



\object{2MASX J03181899+6829322} & 03:18:19.0 & +68:29:31.6 & 0.090 & 0.72 & Sy 1.9 & & KPNO \\ 

\object{NGC 1275} & 03:19:48.2 & +41:30:42.1 & 0.018 & 0.16 & Sy 2& NLRG & Lit. \\
\enddata

\tablenotetext{1}{Milky Way reddening values, E(B-V), obtained from NED.\\}
\tablenotetext{2}{AGN type and host galaxy type from NED, \citet{2008ApJ...681..113T}, and the results of this paper.  For AGN types, optical identifications
are listed, where available.  Values in parentheses indicate classifications from this paper, where 'Sy 1' is a source with broad emission lines and narrow line emission consistent with a Seyfert and 'Sy 2' is a source without broad emission lines and with narrow line emission consistent with a Seyfert. Sub-classifications were made (i.e. Sy\,1.5) following the criteria of  \citet{1981ApJ...249..462O} (based upon the ratio of the broad to narrow components of H$\alpha$ and H$\beta$.
Where ``Gal'' is indicated, there are no optical emission lines indicative
of the presence of an AGN.  The optical spectrum looks like a galaxy spectrum.  Additional host
galaxy classifications were obtained from the LEDA database.  Where ``?'' is indicated, there is no available
classification.\\}
\tablenotetext{3}{Observation type from Sloan Digital Sky Survey archive (SDSS) or our Kitt Peak Observations (KPNO).  Sources with line ratios that we have obtained in the literature are indicated by (Lit.).\\}
\tablenotetext{4}{This source was initially included in the 9-month catalog based on an earlier
method of source selection.  However, with subsequent analysis it fell below the 4.8$\sigma$
detection threshold.  It is detected above 5$\sigma$ and included in the 22-month BAT survey.}
\end{deluxetable}

\begin{deluxetable}{llllccl}
\tabletypesize{\small}
\tablecaption{Details of KPNO Observations\label{tbl-kpno}}
\tablewidth{0pt}
\tablehead{
\colhead{Source} & \colhead{Grating} &
\colhead{UT Date} & \colhead{Exp.~(s)}  & 
\colhead{x (kpc)\tablenotemark{\dagger}} & \colhead{y (kpc)\tablenotemark{\dagger}} &
\colhead{B?\tablenotemark{*}}
}

\startdata
NGC 788 & 35 & 2006-11-20 & 3600 & 0.53 & 4.49\\
LEDA 138501 & 26new & 2006-11-17 & 1800 & 1.93 & 5.53 & B\\
LEDA 138501 & 35 & 2006-11-19 & 1800 & 1.93 & 7.53 & B\\
3C 111 & 35 & 2006-11-20 & 2700 & 1.90 & 7.42 & B \\
2MASX J04440903+2813003 & 26new & 2006-11-17 & 3380 & 0.44 & 2.25\\
2MASX J04440903+2813003 & 35 & 2006-11-19 & 3601 & 0.44 & 1.71 \\
MCG -01-13-025 & 26new & 2006-11-18 & 3600 & 0.62 & 2.79& B \\
MCG -01-13-025 & 35 & 2006-11-19 & 3601 & 0.62 & 2.79 & B \\
\enddata
\tablenotetext{\dagger}{An estimate of the extraction aperture along the slit in both
x and y is given in units of kpc.  The x value is calculated as the fixed 2$\arcsec$ slit size,
converted to kpc using the redshift to the source.  The y value is obtained from the aperture
size used to extract the individual spectrum.\\}
\tablenotetext{*}{B indicates the presence of broad lines (particularly \ion{H}{1} Balmer lines) from a
visual inspection of the spectra.}
\end{deluxetable}

\begin{deluxetable}{lllccc}
\tabletypesize{\small}
\tablecaption{Details of KPNO Observations of Template Galaxies\label{tbl-kpnotemplates}}
\tablewidth{0pt}
\tablehead{
\colhead{Source} & \colhead{Grating} &
\colhead{UT Date} & \colhead{Exp.~(s)} &
\colhead{x (kpc)\tablenotemark{\dagger}} & \colhead{y (kpc)\tablenotemark{\dagger}}
}

\startdata
NGC 205 & 26new & 2006-11-17 & 2101 & 0.03 & 0.09\\
NGC 205 & 35 & 2006-11-19 & 1800 & 0.03 & 0.19 \\
NGC 221 & 26new & 2006-11-18 & 1800 & 0.03 & 0.16 \\
NGC 221 & 35 & 2006-11-20 & 1799 & 0.03 & 0.15\\
NGC 628 & 26new & 2006-11-17 & 3600 & 0.09 & 0.65\\
NGC 628 & 35 & 2006-11-19 & 1800 & 0.09 & 1.01\\
NGC 1023 & 26new & 2006-11-17 & 1799& 0.08 & 0.67 \\
NGC 1023 & 35 & 2006-11-20 & 1800 & 0.08 & 0.68\\
NGC 3384 & 26new & 2006-11-17 & 1799 & 0.09 & 0.66\\
NGC 3884 & 35 & 2006-11-20 & 1801 & 0.09 & 0.82\\
NGC 3640 & 26new & 2006-11-17 & 1800 & 0.16 & 1.52\\
NGC 3640 & 35 & 2006-11-20 & 1800 & 0.16 & 1.91\\
NGC 4914 & 26new & 2007-04-14 & 1200 & 0.61 & 3.01\\
NGC 4914 & 35 & 2007-04-15 & 1200 & 0.61 & 3.71\\
NGC 5308 & 26new & 2007-06-16 & 1801 & 0.26 & 1.83\\
NGC 5308 & 35 & 2007-06-18 & 1800 & 0.26 & 5.50\\
NGC 5557 & 26new & 2007-04-14 & 1200 & 0.42 & 3.36\\
NGC 5557 & 35 & 2007-04-15 & 1200 & 0.42 & 4.21\\
NGC 5638 & 26new & 2007-06-16 & 1800 & 0.22 & 1.77\\
NGC 5638 & 35 & 2007-06-18 & 1800 & 0.22 & 5.57\\
NGC 6654 & 26new & 2007-06-16 & 1800 & 0.24 & 1.26\\
NGC 6654 & 35 & 2007-06-18 & 1799 & 0.24 & 3.53\\
\enddata
\tablenotetext{\dagger}{An estimate of the extraction aperture along the slit in both
x and y is given in units of kpc.}
\end{deluxetable}

\begin{deluxetable}{llcllcl}
\tabletypesize{\small}
\tablecaption{Details of SDSS Observations\label{tbl-sdss}}
\tablewidth{0pt}
\tablehead{
\colhead{Source} & \colhead{UT Date} &
 \colhead{Exp.~(s)} & \colhead{Plate} & \colhead{Tile} & \colhead{d (kpc)\tablenotemark{\dagger}} & \colhead{B?\tablenotemark{*}}
}

\startdata
Mkn 1018 & 2000-09-25 & 2700 & 404 & 193 & 2.53 & B \\
Mkn 590 & 2003-01-08 & 4200 & 1073 & 9328 & 1.52 & B \\
Mkn 18 & 2007-12-05 & 4204 & 1785 & 1290 & 0.64 \\
SDSS J090432.19+553830.1 & 2000-12-29 & 9000 & 450 & 238 & 2.17 &B \\
SDSS J091129.97+452806.0 & 2002-02-07 & 4803 & 832 & 603 & 1.58 \\
SDSS J091800.25+042506.2 & 2003-03-09 & 3000 & 991 & 763 & 9.40\\
MCG +04-22-042 & 2005-12-23 & 4800 & 2290 & 1658 & 1.94 & B\\
Mkn 110 & 2001-12-09 & 4803 & 767 & 553 & 2.05& B \\
\enddata
\tablenotetext{\dagger}{The diameter of the aperture size for SDSS (3\arcsec) in kpc.\\}
\tablenotetext{*}{B indicates the presence of broad lines (particularly \ion{H}{1} Balmer lines) from a
visual inspection of the spectra.}

\end{deluxetable}

\begin{deluxetable}{lllclccc}
\tabletypesize{\small}
\tablecaption{Stellar Light Fits to the Galaxy Templates\label{tbl-templates}}
\tablewidth{0pt}
\tablehead{
\colhead{Galaxy} & \colhead{Type\tablenotemark{*}}  & \colhead{$v_{disp}$\tablenotemark{*}} &
\colhead{$FWHM$\tablenotemark{\dagger}} & \colhead{$Z$\tablenotemark{\dagger}} &
\colhead{$Lf_{young}$\tablenotemark{\dagger}} & \colhead{$Lf_{interm}$\tablenotemark{\dagger}} & 
\colhead{$Lf_{old}$\tablenotemark{\dagger}} 
}

\startdata
\object{NGC 205} & E5 pec & 40.8 & 300 & $0.2 Z_{\sun}$ & 0.02 & 0.98 & -- \\
\object{NGC 221} & cE2 & 71.8 & 170 & $Z_{\sun}$ & 0.05 & -- & 0.95\\
\object{NGC 628} & SA(s)c & 72.2 & 270 & $Z_{\sun}$ & 0.07 & 0.40 & 0.53  \\
\object{NGC 1023} & SB(rs)0- & 204.5 & 300 & $2.5 Z_{\sun}$ & -- & 0.02 & 0.98\\
\object{NGC 3384} & SB(s)0- & 148.4 & 300 & $Z_{\sun}$ & -- & -- & 1.00 \\
\object{NGC 3640} & E3 & 181.6 & 430 & $2.5 Z_{\sun}$ & -- & 0.10 & 0.90\\
\object{NGC 4914} & E+ & 223.6 & 330 & $2.5 Z_{\sun}$ & -- & 0.46 & 0.53\\
\object{NGC 5308} & S0- & 227.2 & 370 & $Z_{\sun}$ & -- & -- & 1.00 \\
\enddata
\tablenotetext{*}{The galaxy type was obtained from NED while the central velocity dispersion was found in LEDA.  Typical errors on the central velocity dispersion are of the order 5\,km\,s$^{-1}$.  These templates were selected from the non-active galaxy templates listed in \citet{1997ApJS..112..315H}\\}
\tablenotetext{\dagger}{The fitted values using the stellar population models of \citet{2003MNRAS.344.1000B} include the FWHM (km\,s$^{-1}$), metallicity ($Z$), and light fractions ($Lf$) at 5500\AA ~ using populations at 25 (young), 2500 (interm), and 10000 (old) Myr.  A dash indicates no contribution from the indicated component.}
\end{deluxetable}

\begin{deluxetable}{llllcccccl}
\tabletypesize{\small}
\tablecaption{Stellar Light Fits to the AGN Sources\label{tbl-continuum}}
\tablewidth{0pt}
\tablehead{
\colhead{Source} & 
\colhead{$FWHM$\tablenotemark{\dagger}} & \colhead{$Z$\tablenotemark{\dagger}} &
\colhead{$p_0$\tablenotemark{\dagger}} & \colhead{$p_1$\tablenotemark{\dagger}} &
\colhead{$Lf_{pow}$\tablenotemark{\dagger}} &
\colhead{$Lf_{young}$\tablenotemark{\dagger}} & \colhead{$Lf_{interm}$\tablenotemark{\dagger}} & 
\colhead{$Lf_{old}$\tablenotemark{\dagger}} & \colhead{$\chi^2/dof$}
}

\startdata
\multicolumn{8}{c}{\bf KPNO Spectra} \\
\hline
NGC 788 & 200 & 2.5 $Z_{\sun}$ & 1.00 & 0.77 & 0.73 & 0.05 & 0.09 & 0.12 & 10.3 \\
LEDA 138501 & 400 & 0.2 $Z_{\sun}$ & 0.56 & 0.37 & 1.00 & -- & -- & -- & 1.0 \\
2MASX J03181899+6829322 & 200 & 0.2 $Z_{\sun}$ & -- & -- & -- & -- & 1.00 & -- & 17.9 \\
3C 105 & 170 & 0.2 $Z_{\sun}$ & -- & -- & -- & -- & 1.00 & -- & 25.3 \\
3C 111 & 200 & 2.5 $Z_{\sun}$ & 0.32 & 0.96 & 1.00 & -- & -- & -- & 2840 \\
2MASX J04440903+2813003 & 460 & 2.5 $Z_{\sun}$ & 0.01 & 1.58 & 1.00 & -- & -- & -- & 101 \\
MCG -01-13-025 & 330 & 2.5 $Z_{\sun}$ & -- & -- & -- & 0.06 & -- & 0.94 & 3.0 \\
\enddata
\tablenotetext{\dagger}{The fitted values using the stellar population models of \citet{2003MNRAS.344.1000B} include FWHM (km\,s$^{-1}$), metallicity ($Z$), and light fractions ($Lf$) at 5500\AA~ using both a power law and stellar population models with ages of: 25 (young), 2500 (interm), and 10000 (old) Myr.  The values $p_0$ and $p_1$ are the power law components, defined as $p_0 \times \lambda^{p_1}$.  The constant factor, $p_0$, is constrained to range from 0 to 1 and is the specific flux at 1\AA with units of $10^{-17}$\,\flux\,\AA$^{-1}$.  Where a component's contribution (e.g. power law) was not required in the best-fit, a dash is indicated.\\}  
\tablenotetext{\ddagger}{For the indicated sources, the value of $p_0 < 0.01$ but non-negligible.  The parameter $p_0$ for the marked sources is: $9.9 \times 10^{-5}$ (IRAS 05589+2828), $4.2 \times 10^{-12}$ (1RXS J174538.1+290823), $2 \times 10^{-4}$ (UGC 11871), and $1.6 \times 10^{-7}$ (NGC 5506). }
\end{deluxetable}

\begin{deluxetable}{lclllllll}
\tabletypesize{\scriptsize}
\rotate
\tablecaption{Emission Line Properties For Strong Lines (Narrow Line Sources)\label{tbl-stronghbeta}}
\tablewidth{0pt}
\tablehead{
\colhead{Source} & 
\colhead{FWHM$_{\rm {blue}}$\tablenotemark{\dagger}} & 
  \colhead{[\ion{O}{2}]~$\lambda 3727$\tablenotemark{*}} &  \colhead{H$\gamma$~$\lambda 4340$\tablenotemark{*} } &
   \colhead{H$\beta$~$\lambda 4861.3$\tablenotemark{*} } & \colhead{[\ion{O}{3}]~$\lambda 4959$\tablenotemark{*}} &
\colhead{[\ion{O}{3}]~$\lambda 5007$\tablenotemark{*}}  \\
\\
\colhead{} & 
\colhead{FWHM$_{\rm {red}}$\tablenotemark{\dagger}} &   \colhead{[\ion{O}{1}]~$\lambda 6300$\tablenotemark{*}} &
\colhead{ [\ion{N}{2}]~$\lambda6548$\tablenotemark{*}} &
\colhead{[\ion{N}{2}]~$\lambda6583$\tablenotemark{*}} & \colhead{[\ion{S}{2}]~$\lambda6716$\tablenotemark{*}} & 
\colhead{[\ion{S}{2}]~$\lambda6731$\tablenotemark{*}} & \colhead{log F(H$\alpha$)} 

}

\startdata
\multicolumn{8}{c}{\bf KPNO Spectra}\\
\hline
NGC 788 &  674.2 $\pm$ 30.6 & \nodata & \nodata & 1.82 $\pm$ 0.65 & 2.31 $\pm$ 0.76 & 3.79 $\pm$ 1.53 \\
 &  177.7 $\pm$ 5.0 & 0.81 $\pm$ 0.22 & 0.44  & 1.33  & 0.95  & 0.92  & -13.29  
 \\ \\

2MASX J03181899+6829322 &  144.8 $\pm$ 1.5 & \nodata & 0.14 $\pm$ 0.01 & 0.37 $\pm$ 0.01 & 1.06 $\pm$ 0.01 & 2.93 $\pm$ 0.03 \\ 
 &  50.0\tablenotemark{\ddagger} & 0.09 $\pm$ 0.01 & 0.14  & 0.43 $\pm$ 0.01 & 0.35 $\pm$ 0.02 & 0.26 $\pm$ 0.02 & -14.09  
 \\ \\

3C 105 &  243.6 $\pm$ 2.1 & \nodata & \nodata & 0.13 $\pm$ 0.01 & 0.74 $\pm$ 0.01 & 2.49 $\pm$ 0.04 \\
\enddata
\tablenotetext{\dagger}{The  FWHM of the lines, in km\,s$^{-1}$, are tied together for all of the narrow emission lines listed in this table.
\\}
\tablenotetext{*}{Ratio of the intensity of the indicated line to the intensity of H$\alpha$.  The units of the H$\alpha$ flux are ergs\,s$^{-1}$\,cm$^{-2}$. Where errors are not indicated, the errors are on the order of $10^{-3}$.
\\}
\tablenotetext{\ddagger}{Indicated FWHM of the lines was fixed to the narrow velocity value of 50\,km\,s$^{-1}$.\\}
\end{deluxetable}

\begin{deluxetable}{lccccc}
\tabletypesize{\scriptsize}
\tablecaption{Emission Line Fluxes For Weaker Lines (Narrow Line Sources)\label{tbl-weakblue}}
\tablewidth{0pt}
\tablehead{
\colhead{Source} &
\colhead{[\ion{Ne}{3}]~$\lambda 3869$\tablenotemark{*}}  & \colhead{H$\delta$} &  \colhead{[\ion{O}{3}]~$\lambda 4363$\tablenotemark{*}} & 
\colhead{\ion{He}{2}~$\lambda 4686$\tablenotemark{*}} & \colhead{[\ion{N}{1}]~$\lambda 5199$\tablenotemark{*}} \\ \\

& \colhead{ \ion{He}{1}~$\lambda 5876$\tablenotemark{*}} &
\colhead{[\ion{Fe}{7}]~$\lambda6087$\tablenotemark{*}} & \colhead{[\ion{O}{1}]~$\lambda6363$\tablenotemark{*}} & 
\colhead{[\ion{Fe}{10}]~$\lambda6375$\tablenotemark{*}} & \colhead{[\ion{Ar}{3}]~$\lambda 7136$\tablenotemark{*}} 
}
\startdata
NGC 788 & \nodata & \nodata & \nodata & -14.13 $\pm$ 0.16 & -14.09 $\pm$ 0.12 \\
&  -14.28 $\pm$ 0.14 & -14.72 $\pm$ 0.22 & -15.03 $\pm$ 0.31 & \nodata & -14.71 $\pm$ 0.24 \\ \\

2MASX J03181899+6829322 & \nodata & -14.02 $\pm$ 0.16 & -14.87 $\pm$ 0.18 & -15.69 $\pm$ 0.34 & -15.75 $\pm$ 0.39 \\
& \nodata & -15.62 $\pm$ 0.32 & -16.50 $\pm$ 0.78 & -14.96 $\pm$ 0.20 & -15.87 $\pm$ 0.54 \\ \\

3C 105 & \nodata & \nodata & \nodata & \nodata & -15.84 $\pm$ 0.43 \\
&  \nodata & -15.29 $\pm$ 0.24 & \nodata & -15.24 $\pm$ 0.27 & -14.60 $\pm$ 0.15 \\
\enddata
\tablenotetext{*}{ ~Logarithm of the intensity of the indicated line.  \\}
\end{deluxetable}

\begin{deluxetable}{lcccccccc}
\tabletypesize{\scriptsize}
\rotate
\tablecaption{Emission Line Properties For Strong Blue Lines (Broad Line Sources)\label{tbl-bluebroad}}
\tablewidth{0pt}
\tablehead{
\colhead{Source} &

\colhead{FWHM (km s$^{-1}$)}  & 

\colhead{H$\beta_N$\tablenotemark{*}}  &

\colhead{[\ion{O}{3}]~$\lambda 4959$\tablenotemark{*}} &  

\colhead{[\ion{O}{3}]~$\lambda 5007$\tablenotemark{*}} &

\colhead{H$\beta_B$ FWHM (km s$^{-1}$)} &
 
\colhead{H$\beta_B$\tablenotemark{*}} &
\colhead{F$_{5100\AA}$}
}
\startdata
\multicolumn{8}{c}{\bf KPNO Spectra}\\
\hline

3C 111 &  214.6 $\pm$ 0.4 & -12.70  & -11.97  & -11.54  & 4960.5 $\pm$ 1.6 & -11.55  & -13.90 $\pm$ 0.05 \\ 
MCG -01-13-025 &  656.3 $\pm$ 666.1 & -13.76  & -13.32  & -13.02  & 8162.8 $\pm$ 288.6 & -13.11 $\pm$ 0.01 & -14.63 $\pm$ 0.02 \\ 
1RXS J045205.0+493248 &  374.1 $\pm$ 4.8 & -13.01 $\pm$ 0.01 & -12.48 $\pm$ 0.01 & -12.04 $\pm$ 0.01 & 7402.1 $\pm$ 21.7 & -12.36  & -14.33 $\pm$ 0.01 \\  
MCG +08-11-011 &  986.1 $\pm$ 20.5 & -12.15 $\pm$ 0.07 & -11.71 $\pm$ 0.07 & -11.24 $\pm$ 0.07 & 3762.3 $\pm$ 28.6 & -11.61  & -13.60 $\pm$ 0.02 \\ 
IRAS 05589+2828 &  563.2 $\pm$ 33.5 & -12.87 $\pm$ 0.33 & -12.81 $\pm$ 0.33 & -12.36 $\pm$ 0.33 & 5564.9 $\pm$ 15.1 & -12.34  & -14.40 $\pm$ 0.01 \\ 
Mkn 6 &  750.1 $\pm$ 212.0 & -12.41 $\pm$ 0.53 & -11.97  & -11.52 $\pm$ 0.00 & 4757.8 $\pm$ 69.5 & -12.19 $\pm$ 0.01 & -13.89 $\pm$ 0.02 \\ 
Mkn 79 &  1078.6 $\pm$ 96.9 & -13.11 $\pm$ 0.04 & -12.56 $\pm$ 0.02 & -12.09 $\pm$ 0.02 & 3940.9 $\pm$ 54.4 & -12.55 $\pm$ 0.01 & -14.66 $\pm$ 0.03 \\
\enddata
\\

\tablenotemark{*}{The logarithm of the indicated lines are given in ergs\,s$^{-1}$\,cm$^{-2}$, where H$\beta_N$ indicates the narrow component of H$\beta$ and H$\beta_B$ indicates the broad component.  The limits on the velocity offsets of the lines were $\pm 1000$\,km\,s$^{-1}$.  Where error-bars are not listed, they are on the order of $10^{-3}$.\\}

\end{deluxetable}

\begin{deluxetable}{lccccccl}
\tabletypesize{\scriptsize}
\rotate
\tablecaption{Emission Line Properties For Strong Red Lines (Broad Line Sources)\label{tbl-redbroad}}
\tablewidth{0pt}
\tablehead{
\colhead{Source} &

\colhead{FWHM (km s$^{-1}$)}  & 


\colhead{H$\alpha_N$\tablenotemark{*}}  &

\colhead{[\ion{N}{2}]~$\lambda 6583$\tablenotemark{*}} & 

\colhead{[\ion{S}{2}]~$\lambda 6716$\tablenotemark{*}} &  

\colhead{[\ion{S}{2}]~$\lambda 6731$\tablenotemark{*}} &

\colhead{H$\alpha_B$ FWHM (km s$^{-1}$)} &
 
\colhead{H$\alpha_B$\tablenotemark{*}} 
}
\startdata
\multicolumn{8}{c}{\bf KPNO Spectra}\\
\hline

3C111 & 481.1 $\pm$ 2.4 & -12.14  & -12.81  & -12.93  & -12.98 & 4589.8 $\pm$ 0.5 & -11.23  \\ 
MCG-01-13-025 &  888.8 $\pm$ 70.9 & -13.21 $\pm$ 0.01 & -13.20  & -13.58 $\pm$ 0.03 & -13.48$\pm$ 0.02 & 6381.7 $\pm$ 27.7 & -12.55  \\ 
1RXS J045205.0+493248 &  310.2 $\pm$ 2.0 & -12.48  & -12.68  & -13.08  & -13.09 & 5706.2 $\pm$ 2.1 & -11.88  \\ 
MCG +08-11-011 &  766.1 $\pm$ 19.6 & -11.51  & -11.60  & -12.35  & -12.19 & 4214.3 $\pm$ 9.3 & -11.20  \\ 
IRAS 05589+2828&  785.1 $\pm$ 67.6 & -12.49  & -12.85  & -13.85 $\pm$ 0.02 & -13.86$\pm$ 0.03 & 5416.2 $\pm$ 7.1 & -12.11  \\ 
Mkn 6 &  848.2 $\pm$ 25.5 & -11.97  & -12.26  & -12.60  & -12.45 & 6800.9 $\pm$ 15.1 & -11.51  \\ 
Mkn 79 &  395.5 $\pm$ 38.5 & -12.66  & -12.74  & -13.36 $\pm$ 0.01 & -13.45$\pm$ 0.01 & 3660.3 $\pm$ 5.5 & -12.08  \\
\enddata

\tablenotemark{*}{The logarithm of the indicated lines are given in ergs\,s$^{-1}$\,cm$^{-2}$, where H$\alpha_N$ indicates the narrow component of H$\alpha$ and H$\alpha_B$ indicates the broad component.  The [\ion{N}{2}] $\lambda 6548$ line, not shown in the table, was fixed to a ratio of 1:2.98 with [\ion{N}{2}] $\lambda 6583$.  Where error-bars are not included, they are on the order of $10^{-3}$.\\}
\tablenotetext{\dagger}{The indicated value is a lower limit.}
\end{deluxetable}

\begin{deluxetable}{lccccccc}
\tabletypesize{\scriptsize}
\rotate
\tablecaption{Emission Line Properties For Weaker Narrow Lines (Broad Line Sources)\label{tbl-weakbluebroad}}
\tablewidth{0pt}
\tablehead{
\colhead{Source} &
\colhead{[\ion{O}{2}]~$\lambda 3727$\tablenotemark{*}}  & 

\colhead{[\ion{Ne}{3}]~$\lambda 3869$\tablenotemark{*}}  & 

\colhead{[\ion{Ne}{3}]~$\lambda 3968$\tablenotemark{*}}  &

\colhead{H$\delta$ 4101\tablenotemark{*}} &  

\colhead{H$\gamma$ 4340\tablenotemark{*}} &

\colhead{[\ion{O}{3}]~$\lambda 4363$\tablenotemark{*}} & 

\colhead{\ion{He}{2}~$\lambda 4686$\tablenotemark{*}}  \\
\\
&
\colhead{[\ion{N}{1}]~$\lambda 5199$\tablenotemark{*}} &

\colhead{\ion{He}{1}~$\lambda 5876$\tablenotemark{*}}  & 

\colhead{[\ion{Fe}{7}]~$\lambda 6087$\tablenotemark{*}}  & 

\colhead{[\ion{O}{1}]~$\lambda 6300$\tablenotemark{*}}  &

\colhead{[\ion{O}{1}]~$\lambda 6363$\tablenotemark{*}} &  

\colhead{[\ion{Fe}{10}]~$\lambda 6375$\tablenotemark{*}} &

\colhead{[\ion{Ar}{3}]~$\lambda 7136$\tablenotemark{*}} 
}
\startdata
\multicolumn{8}{c}{\bf KPNO Spectra}\\
\hline
LEDA 138501&  -13.86 $\pm$  0.08 &  -13.57 $\pm$  0.08 &  -13.66 $\pm$  0.12 &  -12.76 $\pm$  0.10 &  -13.39 $\pm$  0.06 &  -13.36 $\pm$  0.05 &  -13.96 $\pm$  0.09 \\
& \nodata & \nodata & \nodata & \nodata & \nodata & \nodata & \nodata \\ \\

3C 111 & \nodata & \nodata & \nodata & \nodata & \nodata & \nodata & \nodata \\
&  \nodata & \nodata & -12.85 $\pm$  0.14 &  -13.01 $\pm$  0.17 &  -13.78 $\pm$  0.18 &  \nodata \\ \\

MCG -01-13-025 &  -13.23 $\pm$  0.17 &  \nodata & \nodata & \nodata & \nodata & \nodata \nodata & \nodata \\
\enddata
\tablenotetext{*}{The logarithm of the flux for each indicated line is given in units of ergs\,s$^{-1}$\,cm$^{-2}$.}
\end{deluxetable}

\begin{deluxetable}{lcccccccc}
\tabletypesize{\scriptsize}
\rotate
\tablecaption{Measurements of Intrinsic Stellar Absorption\label{tbl-lick}}
\tablewidth{0pt}
\tablehead{
\colhead{Source} & \colhead{D$_n(4000)$}  & \colhead{H$\delta_A$ (\AA)} &
\colhead{CN$_1$ (mag)} & \colhead{Ca 4227 (\AA)} &
\colhead{C2 4668 (\AA)} & \colhead{Mgb (\AA)}  & \colhead{[MgFe]$^{\prime}$ (\AA)}  & \colhead{$<$Fe$>$ (\AA)} 
}
\startdata
\multicolumn{9}{c}{\bf KPNO Spectra}\\
\hline

NGC 788 &\nodata &\nodata &\nodata &\nodata & -24.12 $\pm$ 0.80 & -8.50 $\pm$ 0.14 & 13.61 $\pm$ 0.06 & -7.05 $\pm$ 0.11 \\
LEDA 138501 & 0.88 $\pm$ 0.00 & -3.25 $\pm$ 0.16 & 0.11 $\pm$ 0.00 & 0.24 $\pm$ 0.09 & -2.62 $\pm$ 0.19 & -0.09 $\pm$ 0.12 & 0.69 $\pm$ 0.17 & -0.25 $\pm$ 0.09 \\
2MASX J03181899+6829322 &\nodata &\nodata & -0.52 $\pm$ 0.02 & -0.36 $\pm$ 0.68 & -4.64 $\pm$ 0.45 & 2.33 $\pm$ 0.30 & \nodata & 2.19 $\pm$ 0.20 \\
3C 105 &\nodata &\nodata &\nodata &\nodata & 3.63 $\pm$ 0.48 & 3.89 $\pm$ 0.26 & 3.66 $\pm$ 0.16 & 3.52 $\pm$ 0.18 \\
3C 111 &\nodata &\nodata &\nodata &\nodata & 21.49 $\pm$ 0.08 & -210.88 $\pm$ 92.86 & \nodata & -98.29 $\pm$ 46.43 \\
2MASX J04440903+2813003 &\nodata & -0.22 $\pm$ 0.00 & 0.00 $\pm$ 0.00 & 0.04 $\pm$ 0.00 & -0.21 $\pm$ 0.00 & -0.14 $\pm$ 0.00 & 0.16  & -0.10 $\pm$ 0.00 \\
MCG -01-13-025 & 1.51 $\pm$ 0.01 & -2.15 $\pm$ 0.25 & 0.09 $\pm$ 0.01 & 1.05 $\pm$ 0.12 & 5.14 $\pm$ 0.17 & 4.21 $\pm$ 0.17 & 4.32 $\pm$ 0.09 & 3.20 $\pm$ 0.13 \\
\enddata

\end{deluxetable}

\begin{deluxetable}{lccccccc}
\tabletypesize{\scriptsize}
\tablecaption{De-reddened Emission Line Properties\label{tbl-literature}}
\tablewidth{0pt}
\tablehead{
\colhead{Source\tablenotemark{*}} & \colhead{H$\alpha$/H$\beta$}  & \colhead{E(B - V)$_{int}$} &
\colhead{[\ion{O}{3}]/H$\beta$} & \colhead{[\ion{O}{1}]/H$\alpha$} &
\colhead{[\ion{N}{2}]/H$\alpha$} & \colhead{[\ion{S}{2}]/H$\alpha$} & 
\colhead{[\ion{O}{3}]/[\ion{O}{2}]} 
}
\startdata
\multicolumn{8}{c}{\bf KPNO Spectra}\\
\hline
            NGC 788	& 0.55	& \nodata	& 2.08	& 0.81	& 1.33	& 1.87	& \nodata \\
2MASX J03181899+6829322	& 2.70	& \nodata	& 7.92	& 0.09	& 0.43	& 0.61	& \nodata \\
            3C 105	& 7.69	& 0.92	& 17.07	& 0.11	& 0.77	& 0.39	& \nodata \\
            3C 111 (B)	& 3.60	& 0.15	& 14.01	& 0.10	& 0.19	& 0.27	& \nodata \\
2MASX J04440903+2813003	& 6.67	& 0.77	& 3.63	& -0.07	& 0.68	& 0.48	& \nodata \\
MCG -01-13-025 (B)	& 3.50	& 0.12	& 5.38	& 0.43	& 0.93	& 0.86	& 1.39 \\
1RXS J045205.0+493248 (B)	& 3.43	& 0.10	& 9.22	& 0.17	& 0.57	& 0.45	& 1.89 \\
\enddata
\tablenotetext{*}{Sources with broad lines (approximately Sy 1 -- Sy 1.5) are indicated with a (B).
}
\tablenotetext{1}{Reference: \citet{1997ApJS..112..315H} \\}
\tablenotetext{2}{Reference: \citet{1987ApJS...63..295V} \\}
\tablenotetext{3}{Reference: \citet{2001ApJS..132...37K} \\}
\end{deluxetable}

\begin{deluxetable}{lccccc}
\tabletypesize{\scriptsize}
\tablecaption{Classification\label{tbl-class}}
\tablewidth{0pt}
\tablehead{
\colhead{Source} & 
\colhead{[\ion{N}{2}]/H$\alpha$} & \colhead{[\ion{S}{2}]/H$\alpha$} & \colhead{[\ion{O}{1}]/H$\alpha$} 
& \colhead{[\ion{O}{3}]/[\ion{O}{2}]} & \colhead{Class}
}
\startdata
\multicolumn{5}{c}{\bf KPNO Spectra}\\
\hline
            NGC 788 & AGN &   LINER &   LINER & \nodata &   LINER\\
2MASX J03181899+6829322 & AGN & Seyfert & Seyfert & \nodata & Seyfert\\
            3C 105 & AGN & Seyfert & Seyfert & \nodata & Seyfert\\
3C 111 (B)& AGN & Seyfert & Seyfert & \nodata & Seyfert\\
2MASX J04440903+2813003 & AGN & Seyfert & Seyfert & \nodata & Seyfert\\
     MCG -01-13-025 (B) & AGN & Seyfert &   LINER &   LINER &  Ambig.\\
   1RXS J045205.0+493248 (B) & AGN & Seyfert & Seyfert & Seyfert & Seyfert\\
\enddata
\\
{Parenthesis are placed around classifications where the probable
class is noted despite the fact that not all of the criteria point to the same class.  This is discussed within the text.  The symbol (B) indicates broad line sources (i.e. approximately Sy 1--1.5).}
\tablenotetext{1}{Reference: \citet{1997ApJS..112..315H} \\}
\tablenotetext{2}{Reference: \citet{1987ApJS...63..295V} \\}
\tablenotetext{3}{Reference: \citet{2001ApJS..132...37K} \\}
\end{deluxetable}

\begin{deluxetable}{lcccccc}
\tabletypesize{\scriptsize}
\tablecaption{Mass and Luminosity\label{tbl-masses}}
\tablewidth{0pt}
\tablehead{
\colhead{Source} & 
\colhead{$\lambda L_{\lambda}$\tablenotemark{1}} & \colhead{M$_{\rm{H}\beta}$\tablenotemark{2}} & 
\colhead{L$_{[OIII]}$(obs)\tablenotemark{3}} 
& \colhead{L$_{[OIII]}$(corr)\tablenotemark{3}}  &
\colhead{M$_{reverb}$\tablenotemark{2}} & 
\colhead{M$_{\rm{2MASS}}$\tablenotemark{2}}
}
\startdata
NGC 788 & & & 40.84$\pm$0.15&	41.50$\pm$0.15 & &  8.51\\

Mkn 1018 &	43.61$\pm$0.01 &	8.25$\pm$0.02 &	41.64$\pm$0.09&	41.68$\pm$0.09 & & 8.94\\ 
Mkn 590 &	43.12$\pm$0.01 &	7.94$\pm$0.03 &	41.66$\pm$0.04&	41.66$\pm$0.04 & 7.68 $\pm 0.06$& 8.87\\ 
2MASX J03181899+6829322 & & & 41.59&	41.64 & &  \\
3C 105 & & & 41.50$\pm$0.01&	41.50$\pm$0.01 & & 7.79 \\
3C 111 &	44.47$\pm$0.05 &	8.54$\pm$0.03 &	43.12&	43.12 \\ 
2MASX J04440903+2813003 & & & 40.00&	40.00 & &  \\

MCG -01-13-025 &	42.77$\pm$0.02 &	8.12$\pm$0.04 &	40.67&	40.67 & & 8.06\\
\enddata
\tablenotetext{1}{$\lambda L_{\lambda}$ is derived from the power law continuum flux at 5100\AA~and
is the logarithm of the luminosity in units of ergs\,s$^{-1}$.\\}
\tablenotetext{2}{Indicated masses are the logarithm of the mass in solar masses.\\}
\tablenotetext{3}{The logarithm of [\ion{O}{3}]\,5007\AA~luminosity is in units of ergs\,s$^{-1}$, including the observed luminosity (obs) and the extinction corrected values (corr) using values of E(B-V) recorded in Table~\ref{tbl-literature}.  Where errors are not indicated, they are of the order of $10^{-3}$.}
\end{deluxetable}

\clearpage

\clearpage

\clearpage

\clearpage

\clearpage

\clearpage

\clearpage

\clearpage

\clearpage

\clearpage
\clearpage

\clearpage 

\begin{deluxetable}{llllll}
\tabletypesize{\small}
\tablecaption{Stellar Light Fits to the Test Spectra\label{tbl-testspectra}}
\tablewidth{0pt}
\tablehead{
\colhead{Test Spectrum} & 
\colhead{$FWHM$\tablenotemark{\dagger}} & \colhead{$Z$\tablenotemark{\dagger}} &
\colhead{$Lf_{young}$\tablenotemark{\dagger}} & \colhead{$Lf_{interm}$\tablenotemark{\dagger}} & 
\colhead{$Lf_{old}$\tablenotemark{\dagger}} 
}

\startdata
25 Myr (Y) &  300 & $0.2 Z_{\sun}$ & 0.89 & 0.00 & 0.11 \\
2500 Myr (I) &  330 & 2.5 $Z_{\sun}$ & 0.00 & 1.00 & 0.00 \\
10000 Myr (O) & 300 & $Z_{\sun}$ & 0.00 & 0.00 & 1.00 \\

0.5 $\times$ (Y + I )&  300 & $0.2 Z_{\sun}$ & 0.41 & 0.28 & 0.30 \\
0.5 $\times$ (Y + O ) &  300 & $Z_{\sun}$ & 0.32 & 0.68 & 0.00 \\

0.5 $\times$ (I + O) &  330 & 2.5 $Z_{\sun}$ & 0.00 & 1.00 & 0.00 \\

0.33 $\times$ (Y + I + O) &  300 & $Z_{\sun}$ & 0.19 & 0.81 & 0.00 \\
\enddata
\tablenotetext{\dagger}{The fitted values using the stellar population models of \citet{2003MNRAS.344.1000B} include FWHM (km\,s$^{-1}$), metallicity ($Z$), and light fractions ($Lf$) at 5500\AA \,using populations at 25 (young or Y), 2500 (interm or I), and 10000 (old or O) Myr.}
\end{deluxetable}

\begin{deluxetable}{lcllllllll}
\tabletypesize{\small}
\tablecaption{Best-fit Power law + Stellar Light Fits to the Test Spectra\label{tbl-sims2}}
\tablewidth{0pt}
\tablehead{
\colhead{Source} & 
\colhead{$FWHM$\tablenotemark{\dagger}} & \colhead{$Z$\tablenotemark{\dagger}} &
\colhead{$p_0$\tablenotemark{\dagger}} & \colhead{$p_1$\tablenotemark{\dagger}} &
\colhead{$Lf_{pow}$\tablenotemark{\dagger}} &
\colhead{$Lf_{young}$\tablenotemark{\dagger}} & \colhead{$Lf_{interm}$\tablenotemark{\dagger}} & 
\colhead{$Lf_{old}$\tablenotemark{\dagger}} 
}
\startdata
25 Myr (Y) + pow & 300 &  $Z_{\sun}$ & $9.3 \times 10^{-4}$ & 0.77 & 0.42 & 0.58 & 0.00 & 0.00 \\
2500 Myr (I) + pow  & 300 &  $Z_{\sun}$ & $3.3 \times 10^{-3}$ & 0.66 & 0.44 & 0.00 & 0.56 & 0.00 \\
10000 Myr (O) + pow& 300 &  $Z_{\sun}$ & $4.6 \times 10^{-4}$ & 0.87 & 0.45 & 0.00 & 0.09 & 0.46 \\
0.5 $\times$ (Y + I ) + pow& 300 &  $Z_{\sun}$ & $8.5 \times 10^{-4}$ & 0.79 & 0.41 & 0.30 & 0.24 & 0.05 \\
0.5 $\times$ (Y + O ) + pow& 300 &  $Z_{\sun}$ & $4.2 \times 10^{-2}$ & 0.41 & 0.61 & 0.20 & 0.00 & 0.19 \\
0.5 $\times$ (I + O) + pow& 300 &  $Z_{\sun}$ & $6.2 \times 10^{-4}$ & 0.83 & 0.44 & 0.00 & 0.29 & 0.28 \\
0.33 $\times$ (Y + I + O) + pow& 300 &  $Z_{\sun}$ & $3.8 \times 10^{-4}$ & 0.87 & 0.41 & 0.19 & 0.20 & 0.20 \\
\enddata
\tablenotetext{\dagger}{The fitted values using the stellar population models of \citet{2003MNRAS.344.1000B} include FWHM (km\,s$^{-1}$), metallicity ($Z$), and light fractions ($Lf$) at 5500\AA~ using both a power law and stellar population models with ages of: 25 (young), 2500 (interm), and 10000 (old) Myr.  The values $p_0$ and $p_1$ are the power law components, defined as $p_0 \times \lambda^{p_1}$.  The constant factor, $p_0$, is constrained to range from 0 to 1.}
\end{deluxetable}

\clearpage 

\end{document}